\newif\ifemulateapj
\emulateapjtrue        % use emulateapj format

\ifemulateapj
    \documentclass[iop,numberedappendix]{emulateapj}
\else
    \documentclass[12pt,preprint]{aastex}
\fi

\usepackage{afterpage}
\usepackage{xcolor}
\definecolor{darkred}{rgb}{0.8,0.13,0.13}
\definecolor{darkblue}{rgb}{0.13,0.15,0.8}
\definecolor{medblue}{rgb}{0,0,0.5}
\usepackage{hyperref}
\hypersetup{
    pdfauthor={Larry Bradley},
    pdftitle={CLASH: A Census of Magnified Star-Forming Galaxies at z=6-8},
    pdfstartview=Fit,
    pdfpagelayout=SinglePage,
    pdfpagemode=UseNone,
    colorlinks=true,
    linkcolor={darkred},
    citecolor={darkblue},
    urlcolor={medblue},
    backref=true
}

\newcommand{\tnm}[1]{\tablenotemark{#1}}
\newcommand{\peras}{\ensuremath{\mathrm{arcsec}^{-1}}}
\newcommand{\kms}{\mbox{km\ s$^{-1}$}}
\newcommand{\kmsMpc}{\kms~\mbox{Mpc}$^{-1}$}

\newcommand{\Lstar}{\ensuremath{L_{*}}}

\newcommand{\mult}[2]{\mbox{#1$\,\times\,$#2}}
\newcommand{\dg}{\ensuremath{^\circ}}
\newcommand{\HST}{\textit{HST}}
\newcommand{\Spitzer}{\textit{Spitzer}}

\newcommand{\iband}{\ensuremath{i_{775}}}
\newcommand{\Iband}{\ensuremath{I_{814}}}
\newcommand{\zband}{\ensuremath{z_{850}}}
\newcommand{\Yband}{\ensuremath{Y_{105}}}
\newcommand{\YJband}{\ensuremath{J_{110}}}
\newcommand{\Jband}{\ensuremath{J_{125}}}
\newcommand{\JHband}{\ensuremath{JH_{140}}}
\newcommand{\Hband}{\ensuremath{H_{160}}}

\newcommand{\wl}{\mbox{$\lambda$}}
\newcommand{\ionrm}[1]{\mbox{\small\sc{\romannumeral #1}}}
\newcommand{\forbdw}[3]{\mbox{[#1~\ionrm{#2}]~\wl\wl#3}}
\newcommand{\forbww}[4]{\mbox{[#1~\ionrm{#2}]~\wl\wl#3,~#4}}
\newcommand{\OIIw}{\forbdw{O}{2}{3727}}
\newcommand{\OIIIww}{\forbww{O}{3}{4959}{5007}}
\newcommand{\Ha}{\mbox{H\ensuremath{\alpha}}}

\def\figcla {
    \begin{figure}[h!]
    \centerline{\includegraphics[width=1.0\columnwidth]{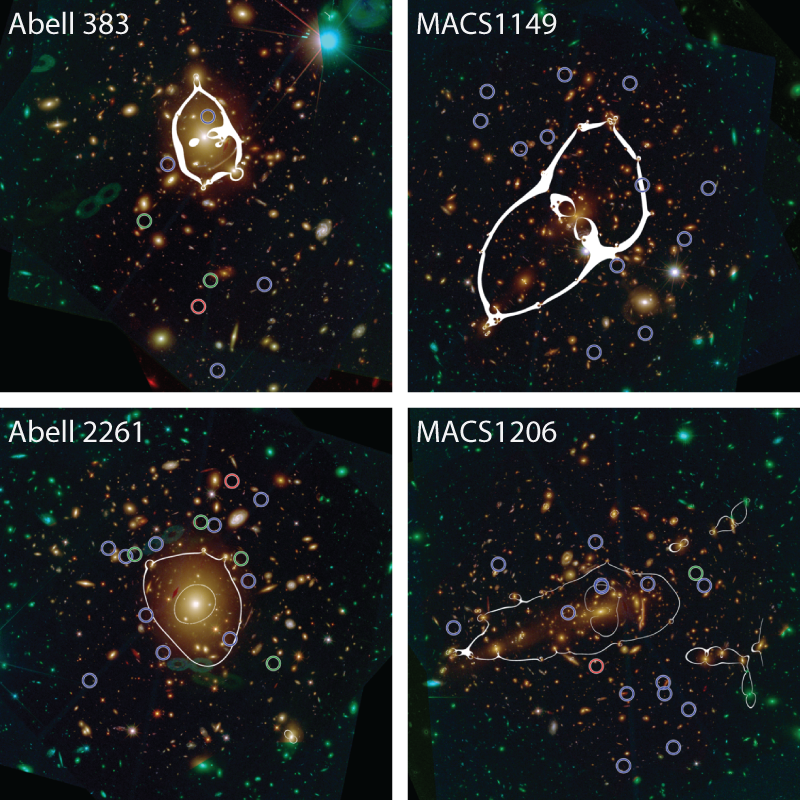}}
    \caption{ACS+IR color images of the galaxy clusters A383, MACS1149, A2261, and MACS1206.  The field of view of each image is \mult{3\farcm25}{3\farcm25} and is shown with North up and East left.  The locations of our high-redshift candidate galaxies at $z\sim6$, $z\sim7$, and $z\sim8$ are marked by the blue, green, and red circles, respectively.  The white contours denote the approximate location of the critical lines ($\mu > 200$) at $z\sim6$ calculated from our fiducial lensing models.  MACS1149 is part of the HFF program.}
    \label{fig:cl1}
    \end{figure}
}

\def\figclb {
    \begin{figure}[h!]
    \centerline{\includegraphics[width=1.0\columnwidth]{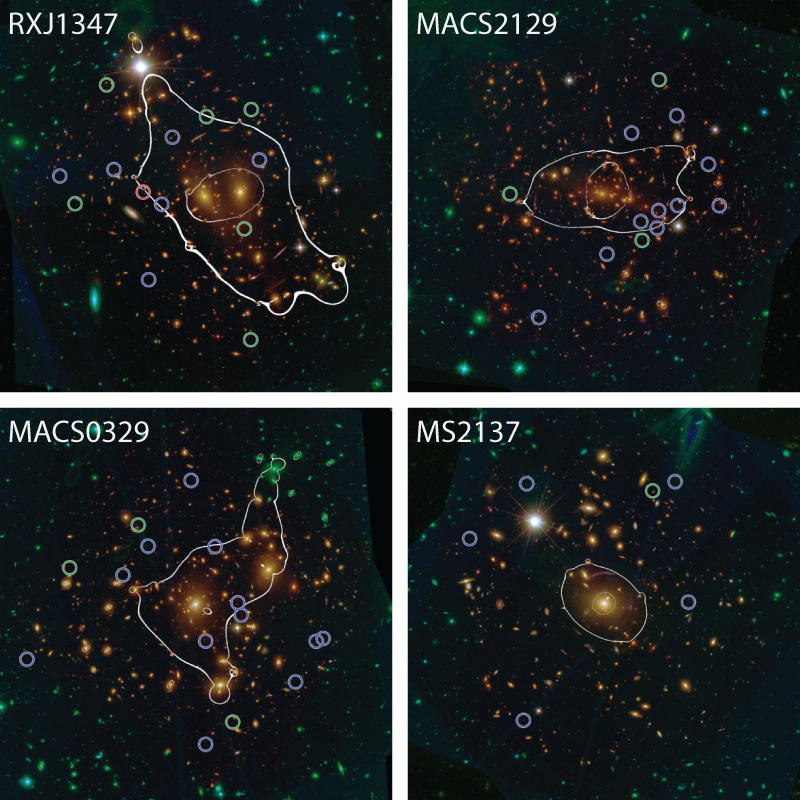}}
    \caption{Same as in Figure~\ref{fig:cl1}, but for the galaxy clusters RXJ1347, MACS2129, MACS0329, and MS2137.}
    \label{fig:cl2}
    \end{figure}
}

\def\figclc {
    \begin{figure}[h!]
    \centerline{\includegraphics[width=1.0\columnwidth]{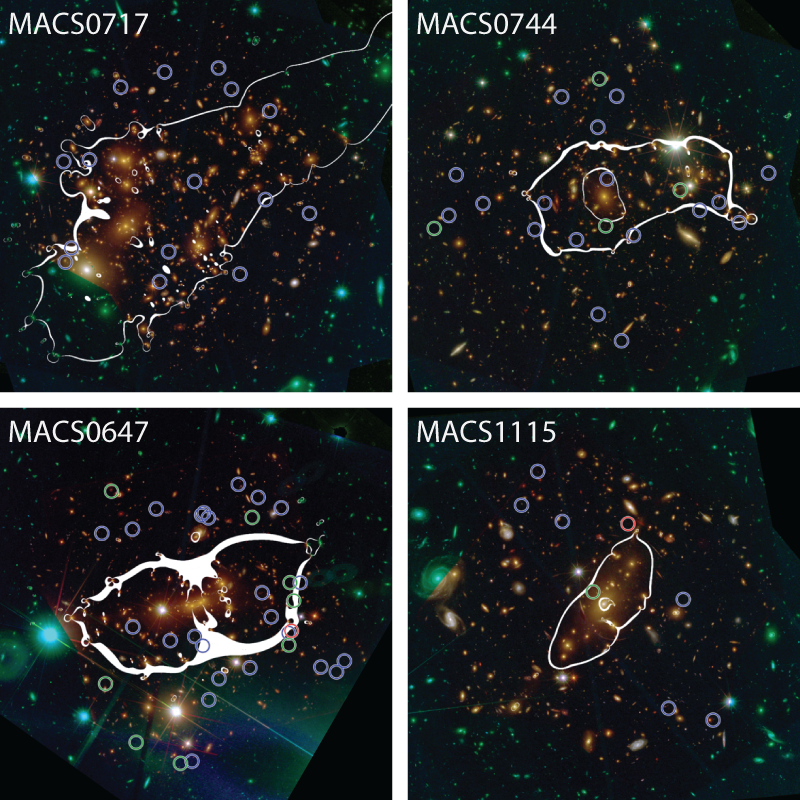}}
    \caption{Same as in Figure~\ref{fig:cl1}, but for the galaxy clusters MACS0717, MACS0744, MACS0647, and MACS1115.  MACS0717 is part of the HFF program.}
    \label{fig:cl3}
    \end{figure}
}

\def\figcld {
    \begin{figure}[h!]
    \centerline{\includegraphics[width=1.0\columnwidth]{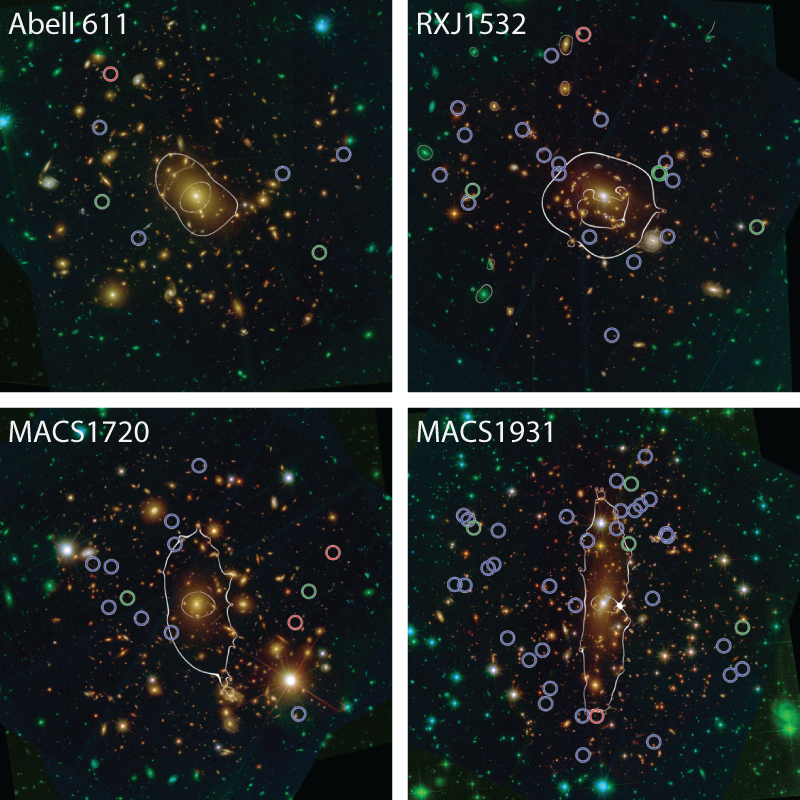}}
    \caption{Same as in Figure~\ref{fig:cl1}, but for the galaxy clusters A611, RXJ1532, MACS1720, and MACS1931.  For RXJ1532, we have not identified any multiply-imaged galaxies, and therefore the critical curve shown is from an approximate lens model based on the light distribution (see Appendix~\ref{sec:rxj1532}).}
    \label{fig:cl4}
    \end{figure}
}

\def\figcle {
    \begin{figure}[h!]
    \centerline{\includegraphics[width=1.0\columnwidth]{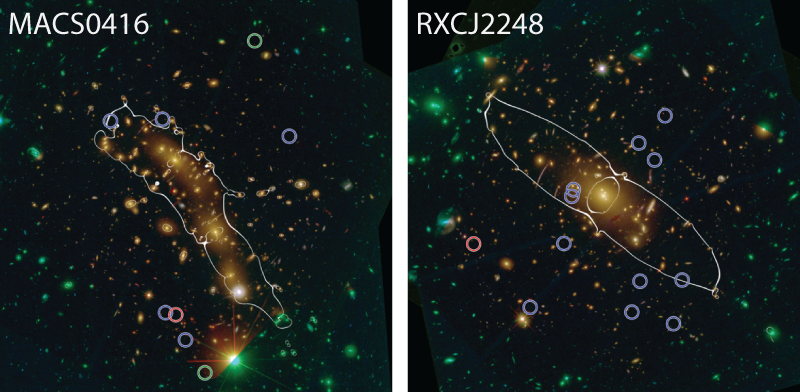}}
    \caption{Same as in Figure~\ref{fig:cl1}, but for the galaxy clusters MACS0416 and RXJC2248.  Both of these clusters are part of the HFF program.}
    \label{fig:cl5}
    \end{figure}
}

\def\figobjzsix {
    \begin{figure*}[htbp!]
    \includegraphics[width=0.5\textwidth]{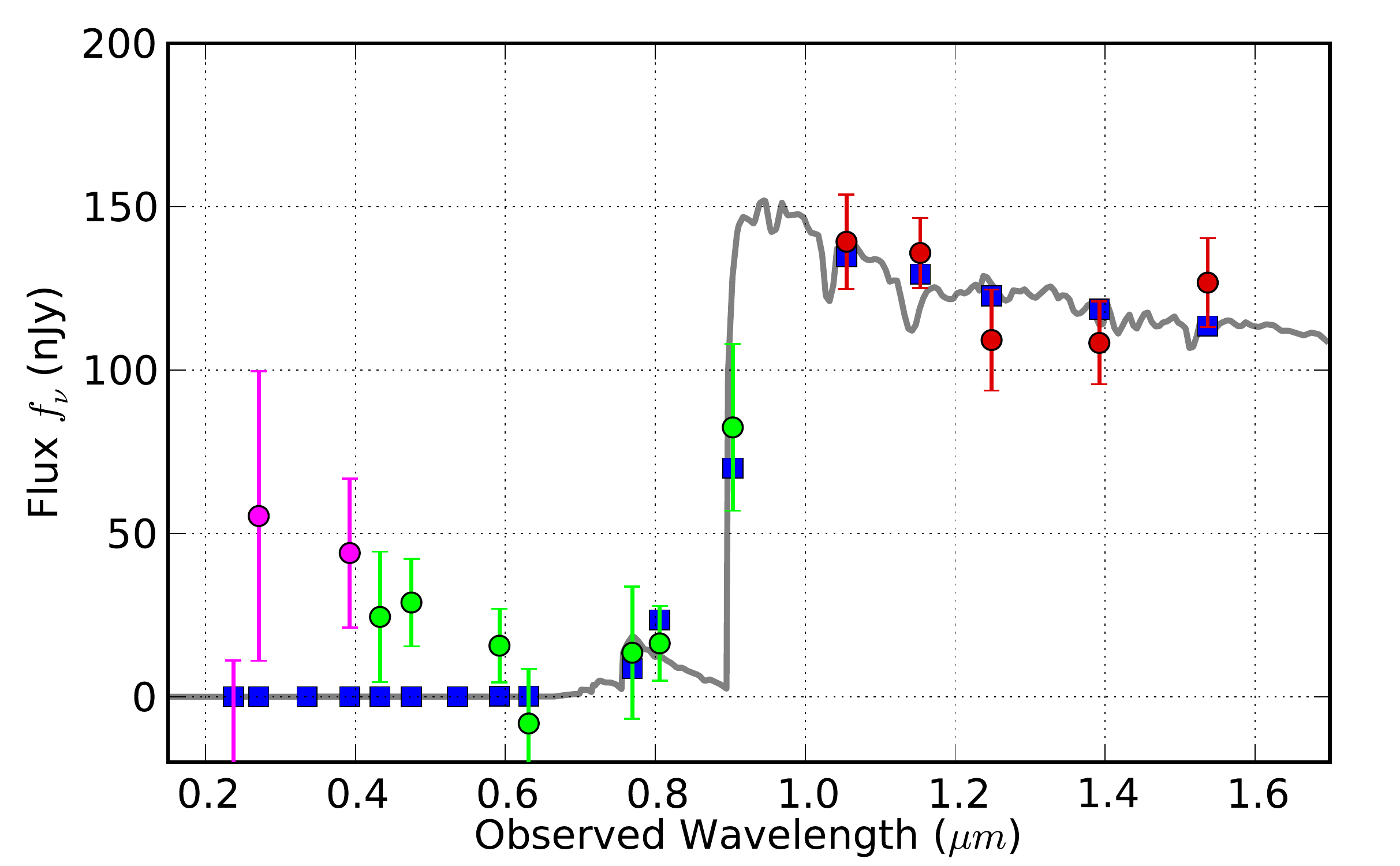}\includegraphics[width=0.5\textwidth]{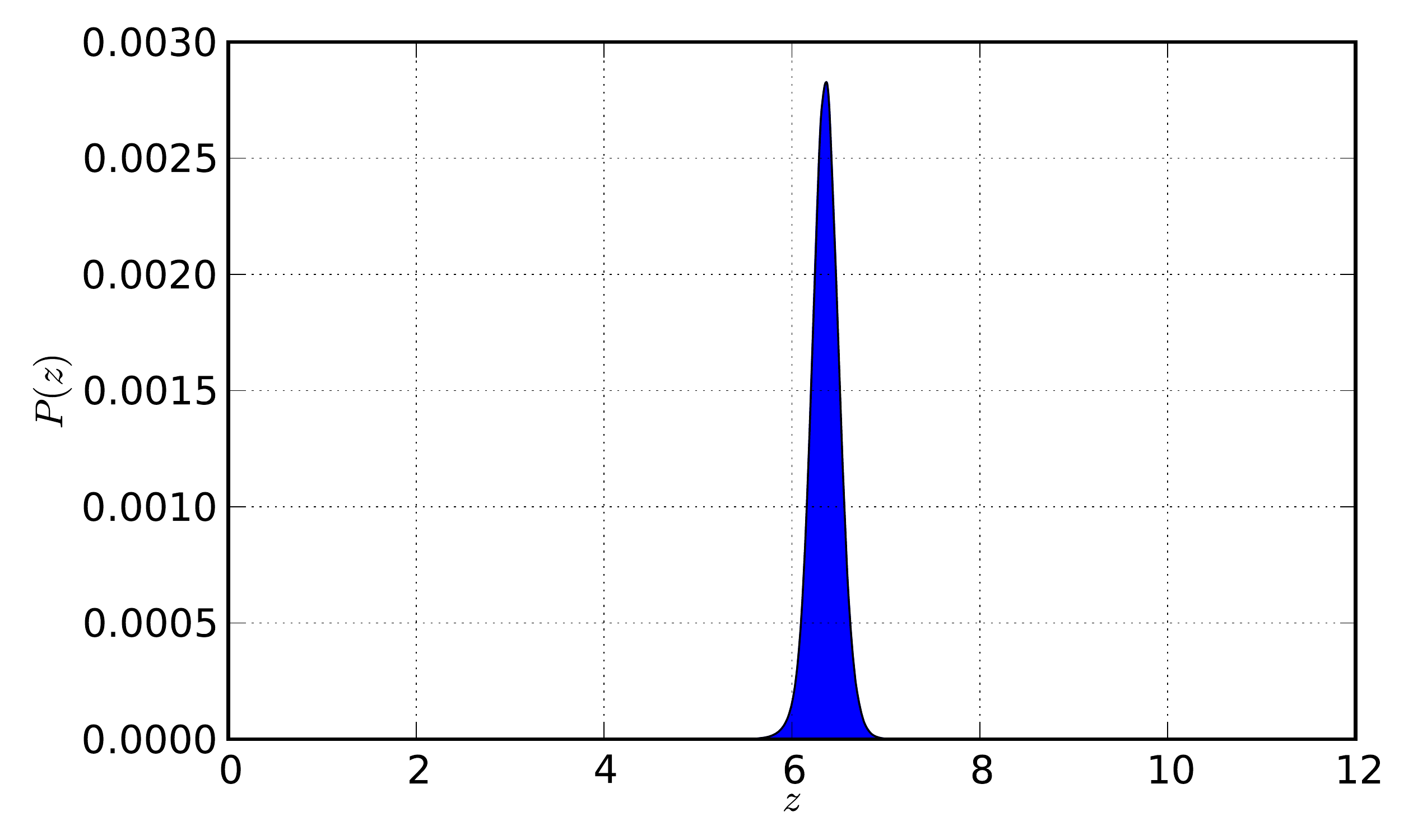}\\[3mm]
    \centerline{\includegraphics[width=0.8\textwidth]{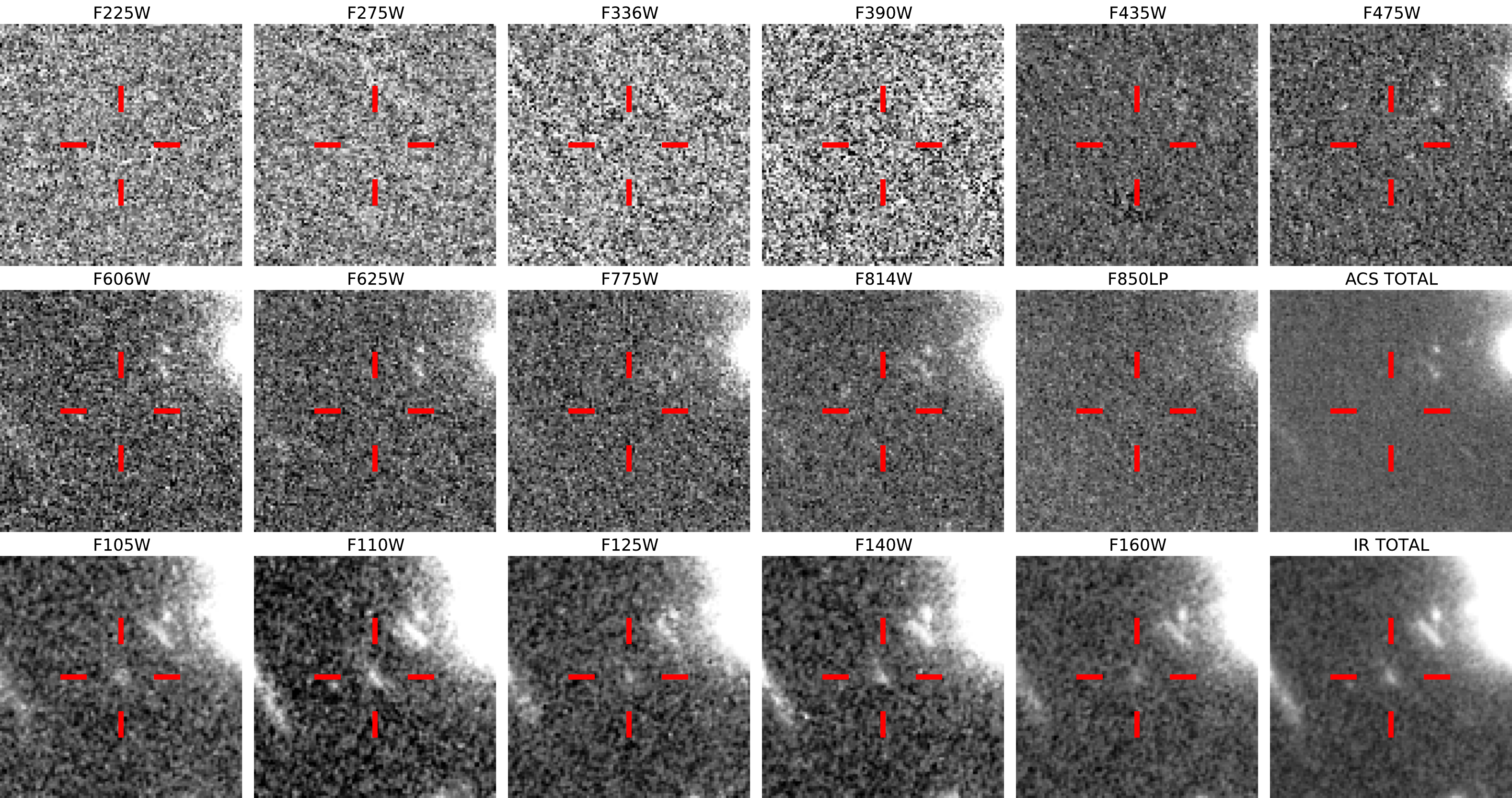}}
    \caption{Top left: Observed SED (magenta, green, and red data points with $1\sigma$ error bars) and BPZ SED template fit (gray line with blue data points). Top right: Posterior photometric redshift probability distribution. Bottom: multiband postage stamp images for the $z\sim6.4$ candidate A2261$-$0754.  The field of view of each stamp image is \mult{13\farcs1}{13\farcs1} and is shown at a position angle (E of N) of $0\dg$.  The stamps marked ACS and IR total represent the inverse-variance weighted sum of all the images taken with those two respective detectors.}
    \label{fig:z6obj}
    \end{figure*}
}

\def\figobjzseven {
    \begin{figure*}[htbp!]
    \includegraphics[width=0.5\textwidth]{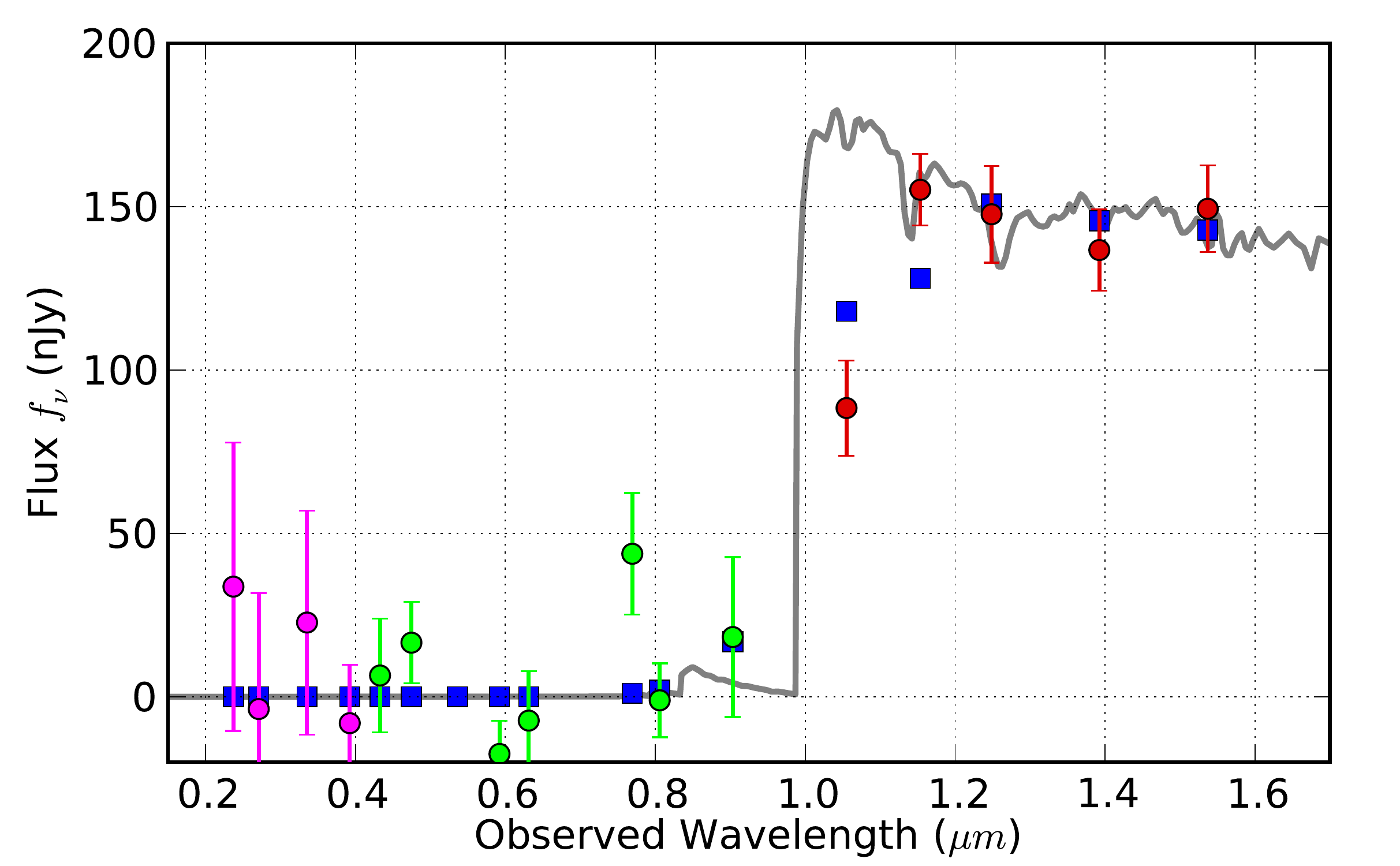}\includegraphics[width=0.5\textwidth]{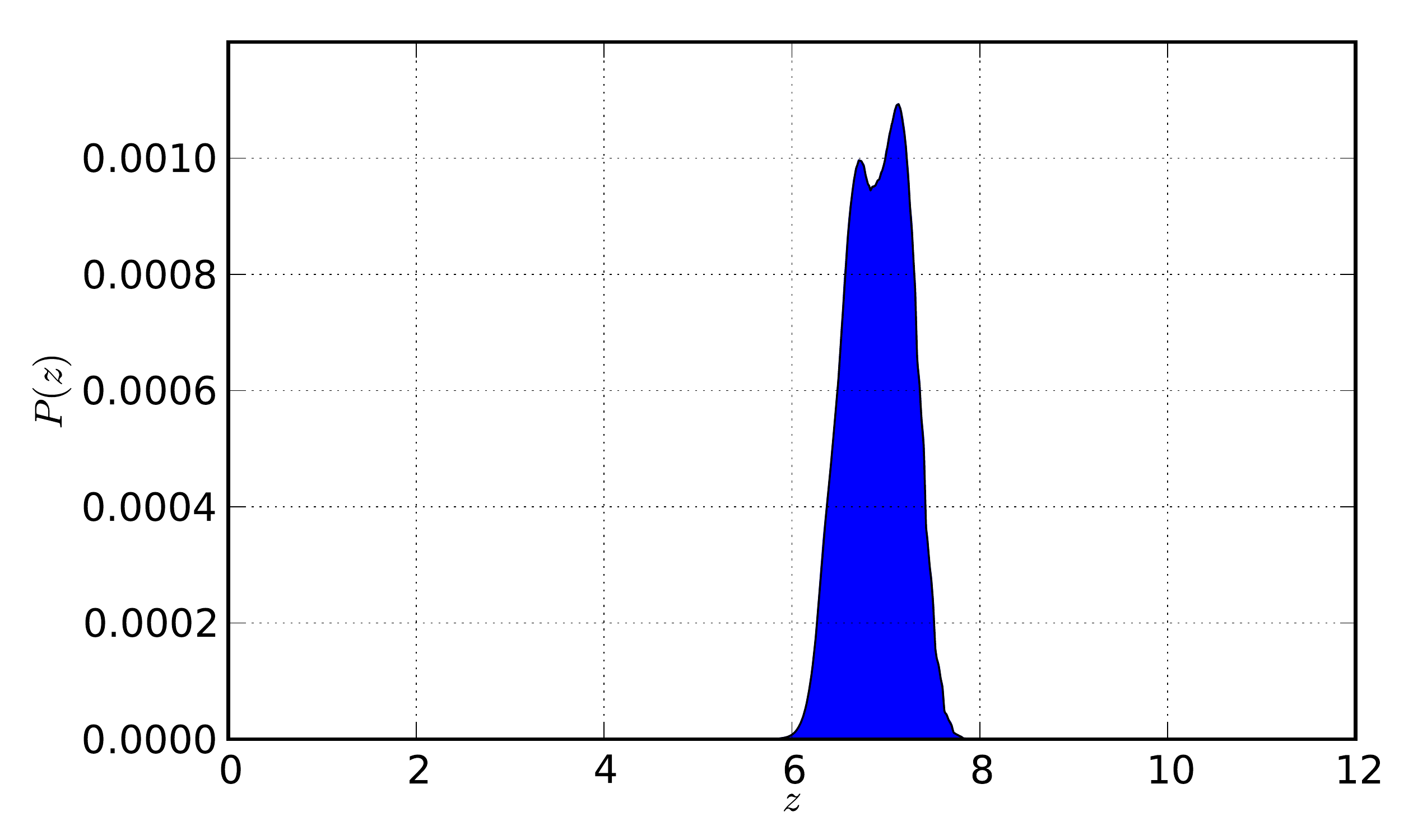}\\[3mm]
    \centerline{\includegraphics[width=0.8\textwidth]{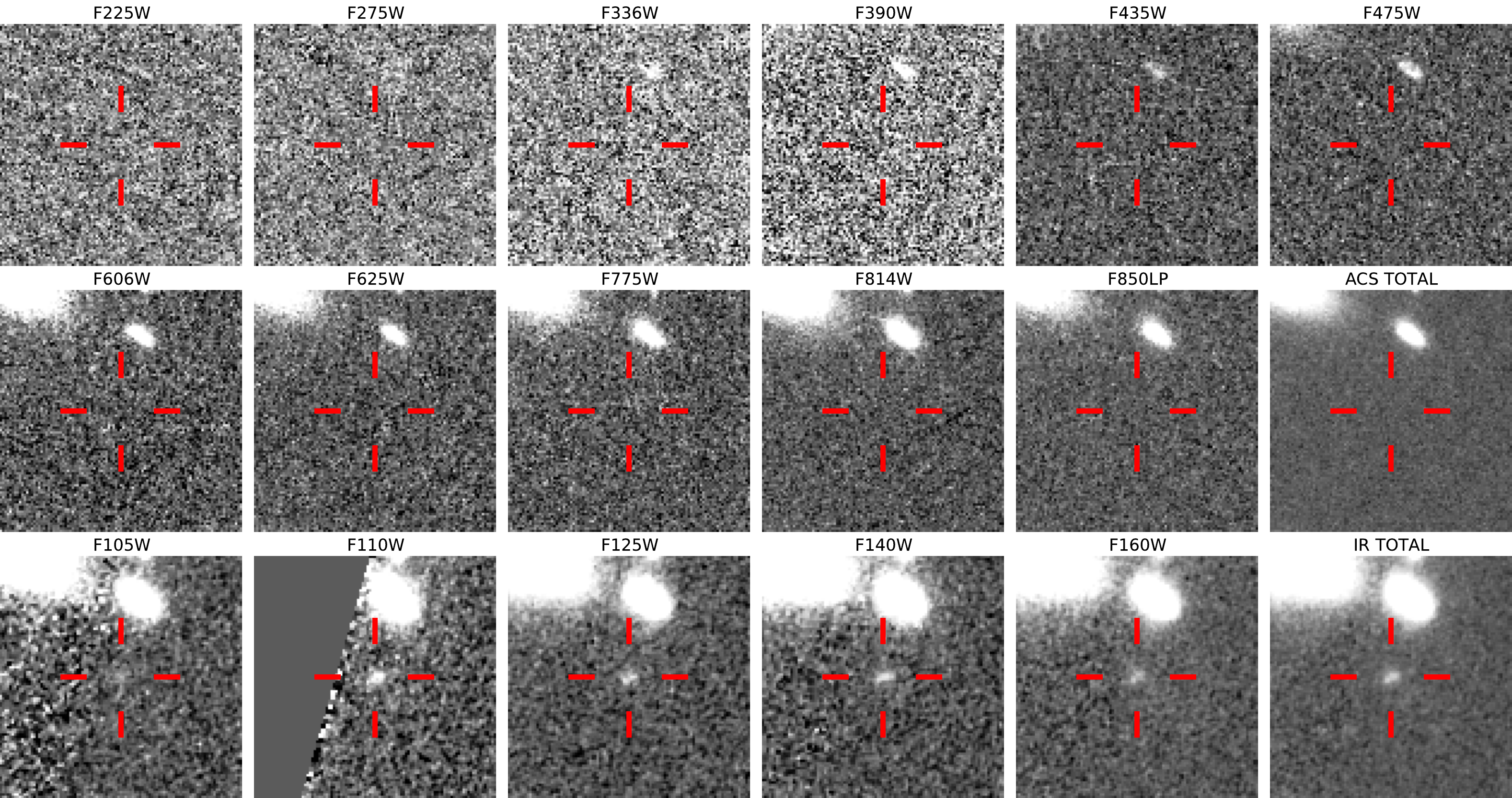}}
    \caption{Same as in Figure~\ref{fig:z6obj}, but for the $z\sim7.1$ candidate RXJ1532-0844.}
    \label{fig:z7obj}
    \end{figure*}
}

\def\figobjzeight {
    \begin{figure*}[htbp!]
    \includegraphics[width=0.5\textwidth]{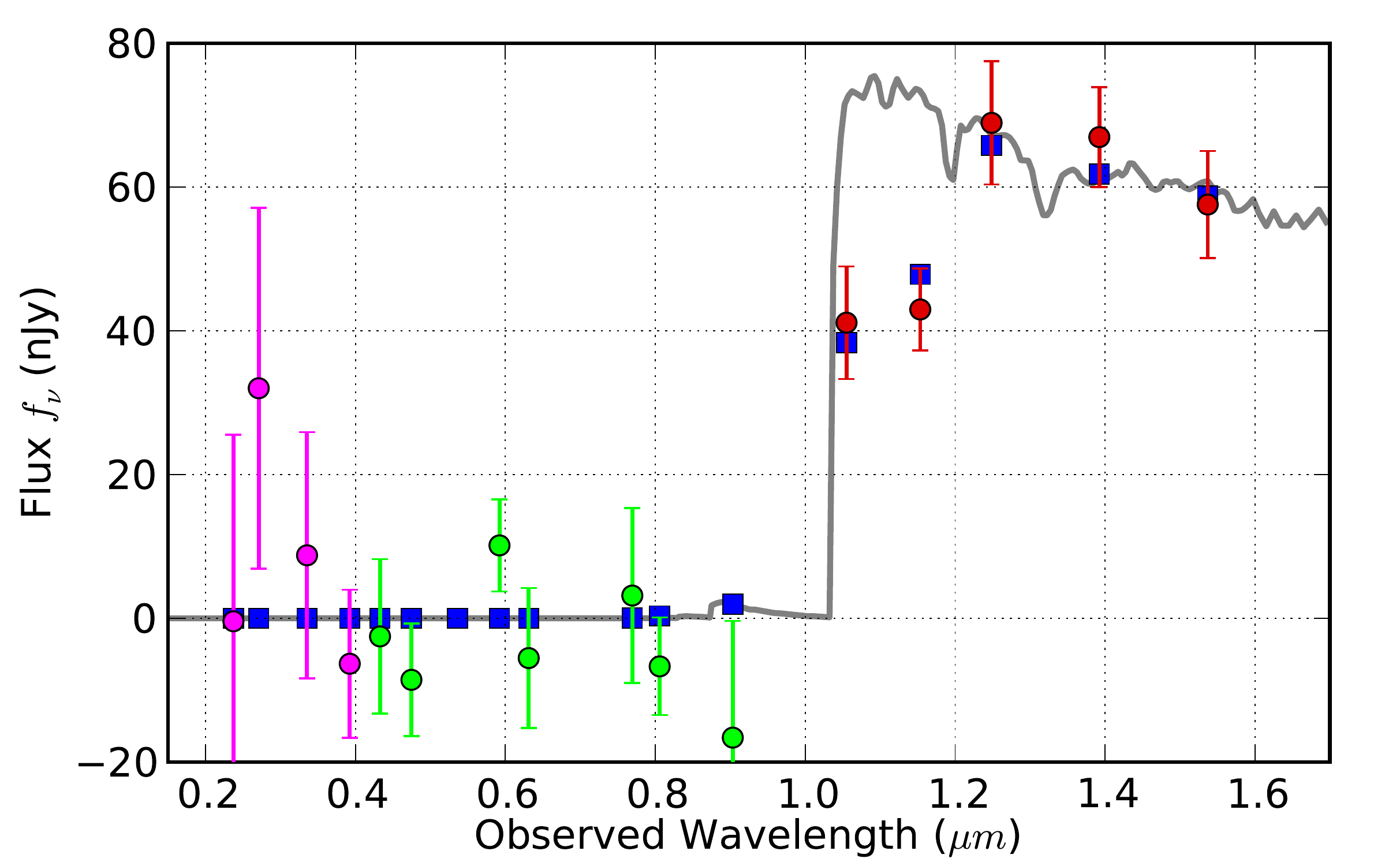}\includegraphics[width=0.5\textwidth]{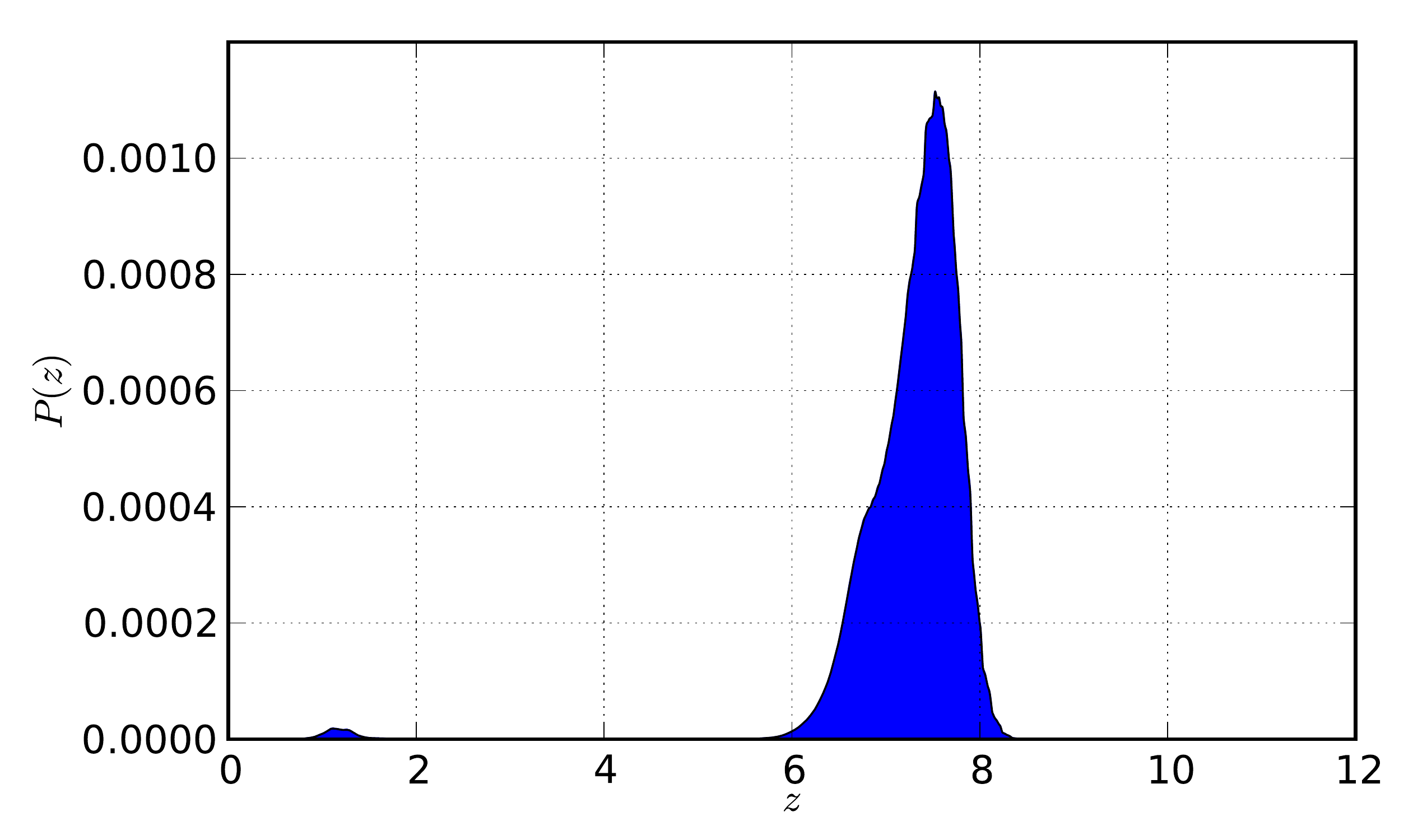}\\[3mm]
    \centerline{\includegraphics[width=0.8\textwidth]{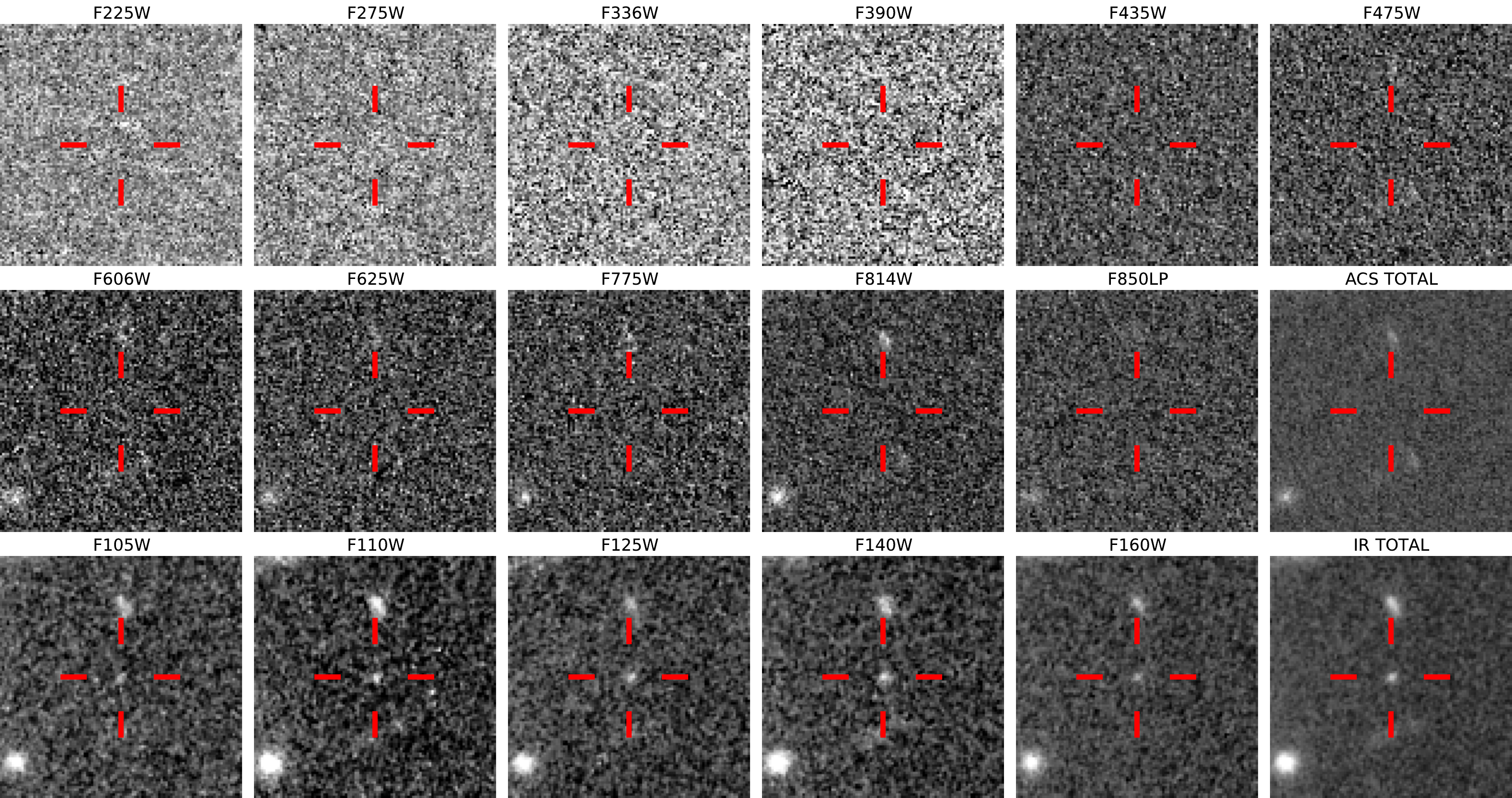}}
    \caption{Same as in Figure~\ref{fig:z6obj}, but for the $z\sim7.5$ candidate A2261-0187.}
    \label{fig:z8obj}
    \end{figure*}
}

\def\figzsixarc {
    \begin{figure*}[htbp!]
    \centerline{\includegraphics[width=0.85\textwidth]{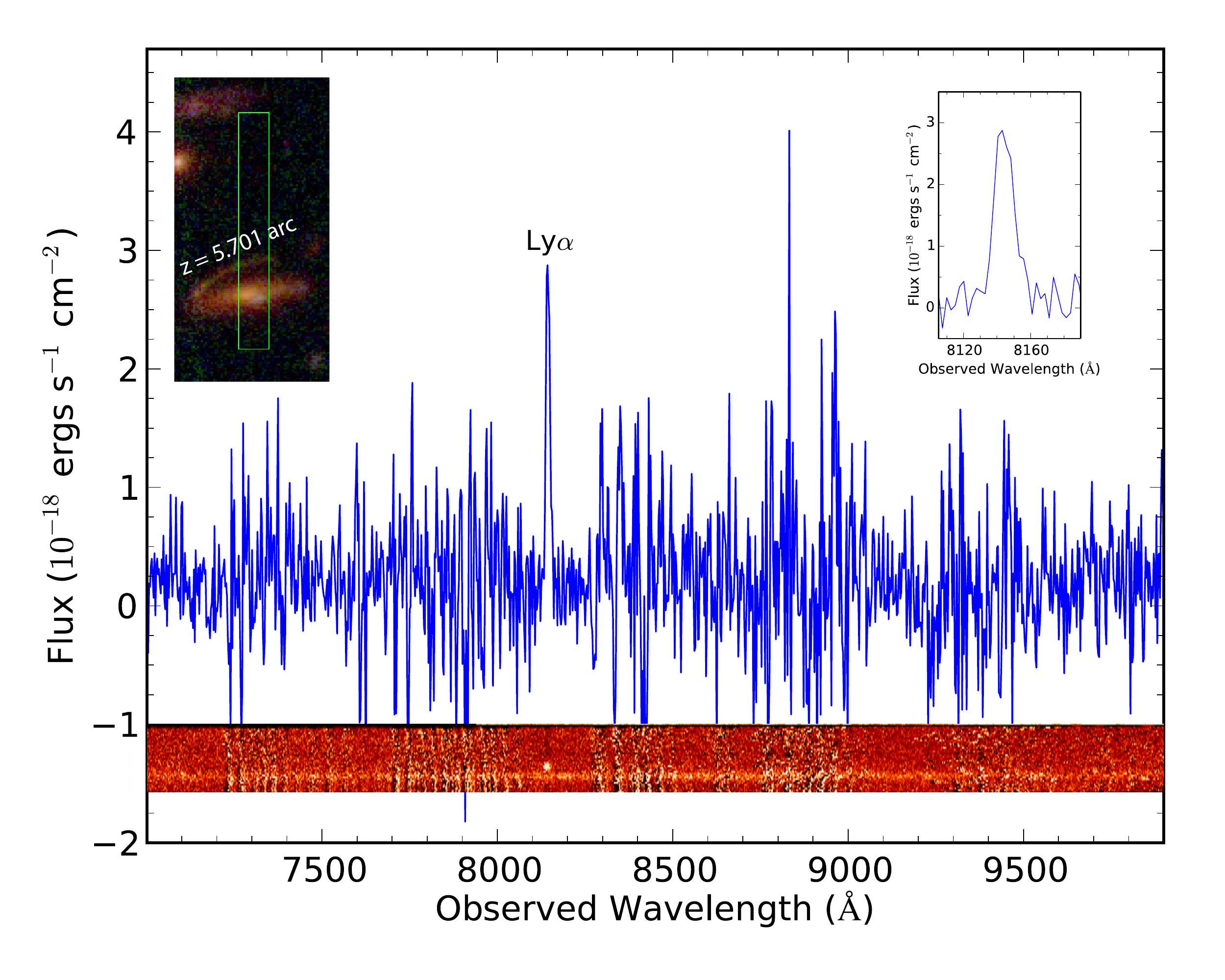}}
    \includegraphics[width=0.5\textwidth]{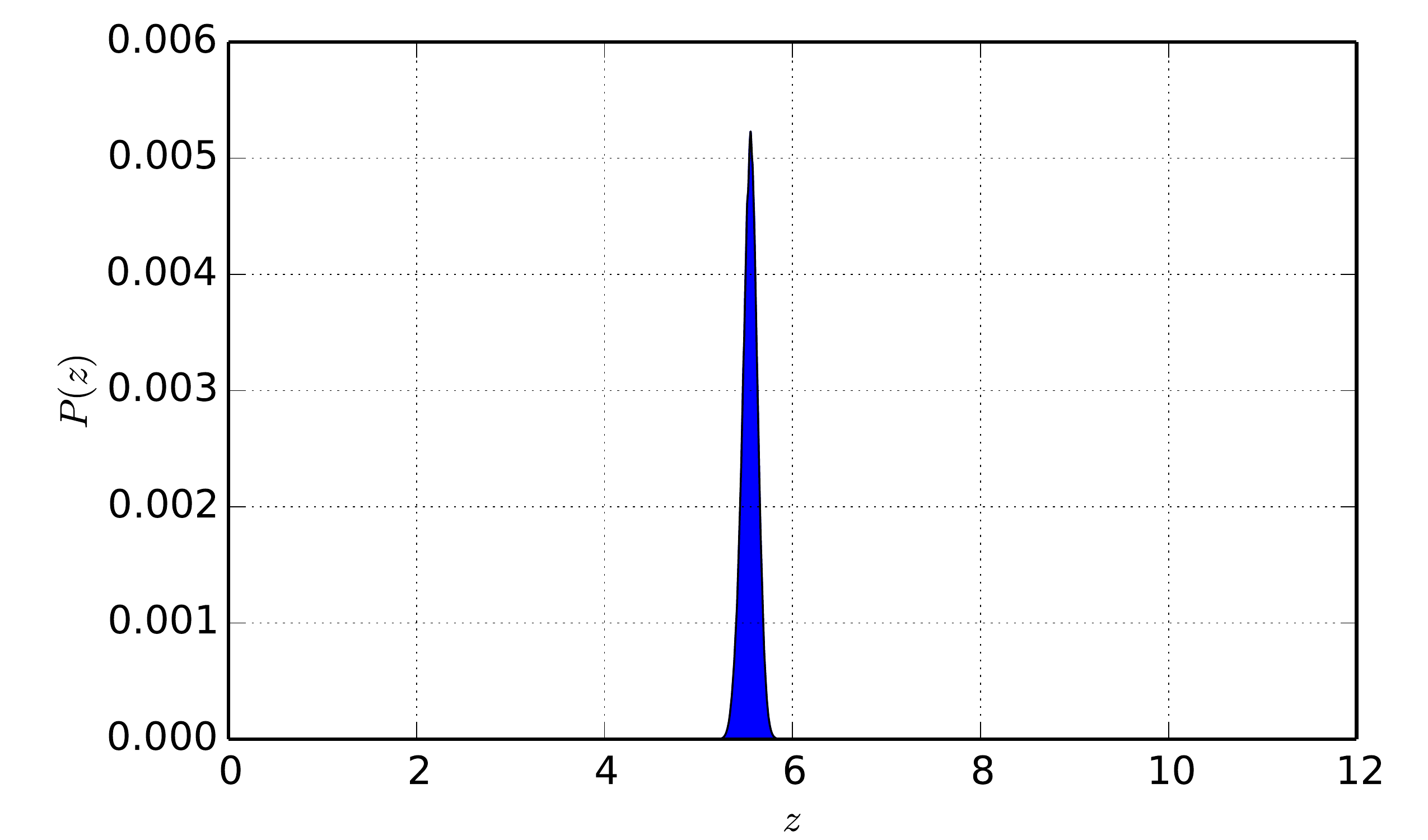}\includegraphics[width=0.5\textwidth]{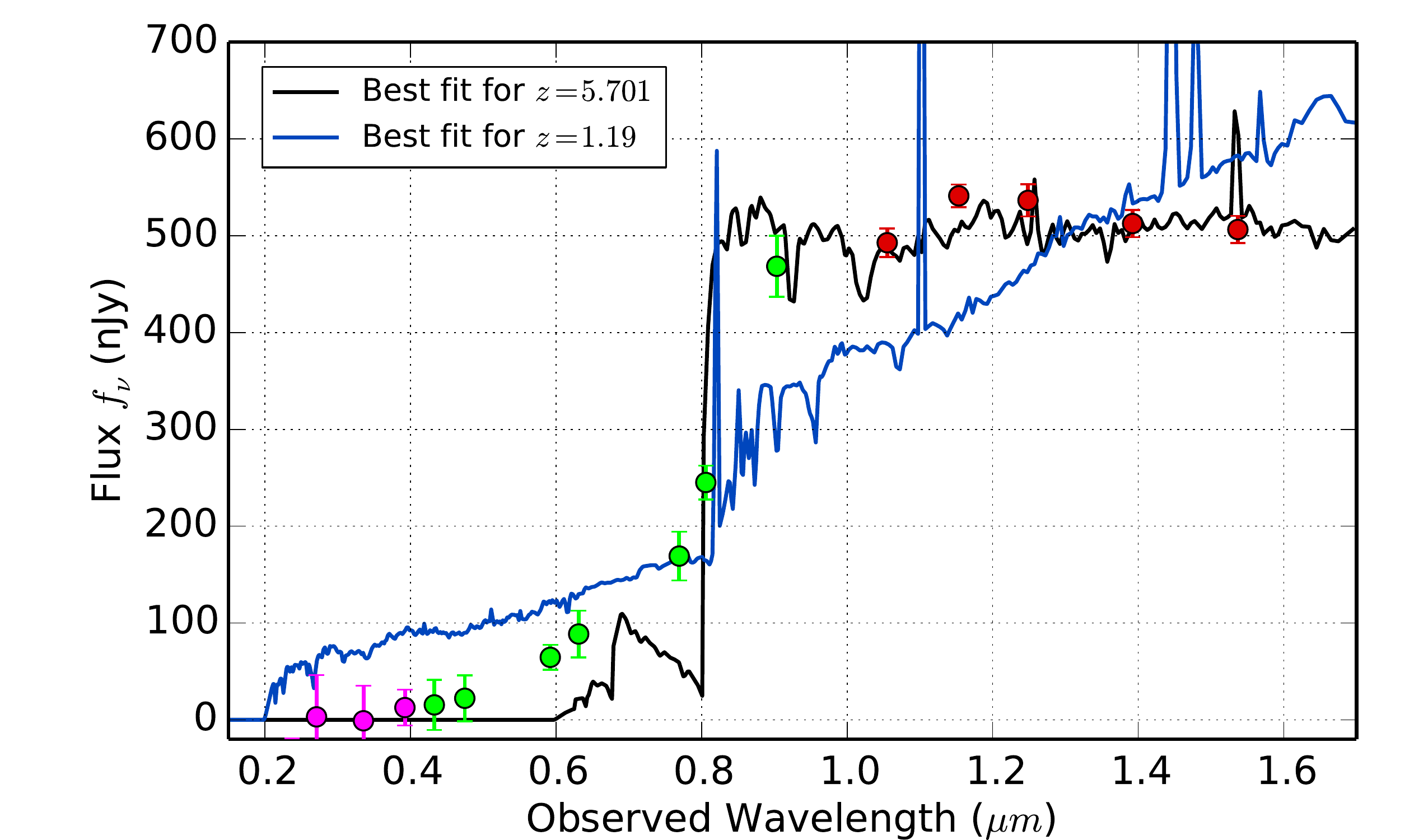}
    \caption{Observed 1D longslit spectrum of MACS1206-1796 ($m_{160} = 23.8$) obtained with {\em VLT}/VIMOS as part of the CLASH {\em VLT} program (PI: P. Rosati).  The spectrum exhibits a clear emission line at 8146 \AA, corresponding to Ly$\alpha$ at $z=5.701$.  The upper-left inset shows the slit location.  The upper-right insert shows a close up of the emission line, which shows an asymmetric profile suggestive of Ly$\alpha$.  The 2D spectrum is shown along the bottom of the plot.  Bottom left: the posterior photometric redshift distribution of this galaxy.  The peak is at $z_{\mathrm{phot}} = 5.6$, which differs from the spectroscopic measurement by only $1.8\%$.  Also note that this galaxy is part of a quadruply-lensed system behind MACS1206 and was predicted to have a redshift of $z=5.7$ based on the lens model \citep{Zitrin2012m1206}.  Bottom right:  the best-fitting BPZ SEDs at the fixed redshifts of $z=5.701$, assuming the line is Ly$\alpha$, and $z=1.19$, assuming the line is \OIIw.  The low-redshift solution provides a much poorer fit to the observed photometry.  Overall, the evidence suggests that the high-redshift solution is more likely.}
    \label{fig:z6arc}
    \end{figure*}
}

\def\figmaghist {
    \begin{figure*}[htbp!]
    \centerline{\includegraphics[width=0.5\textwidth]{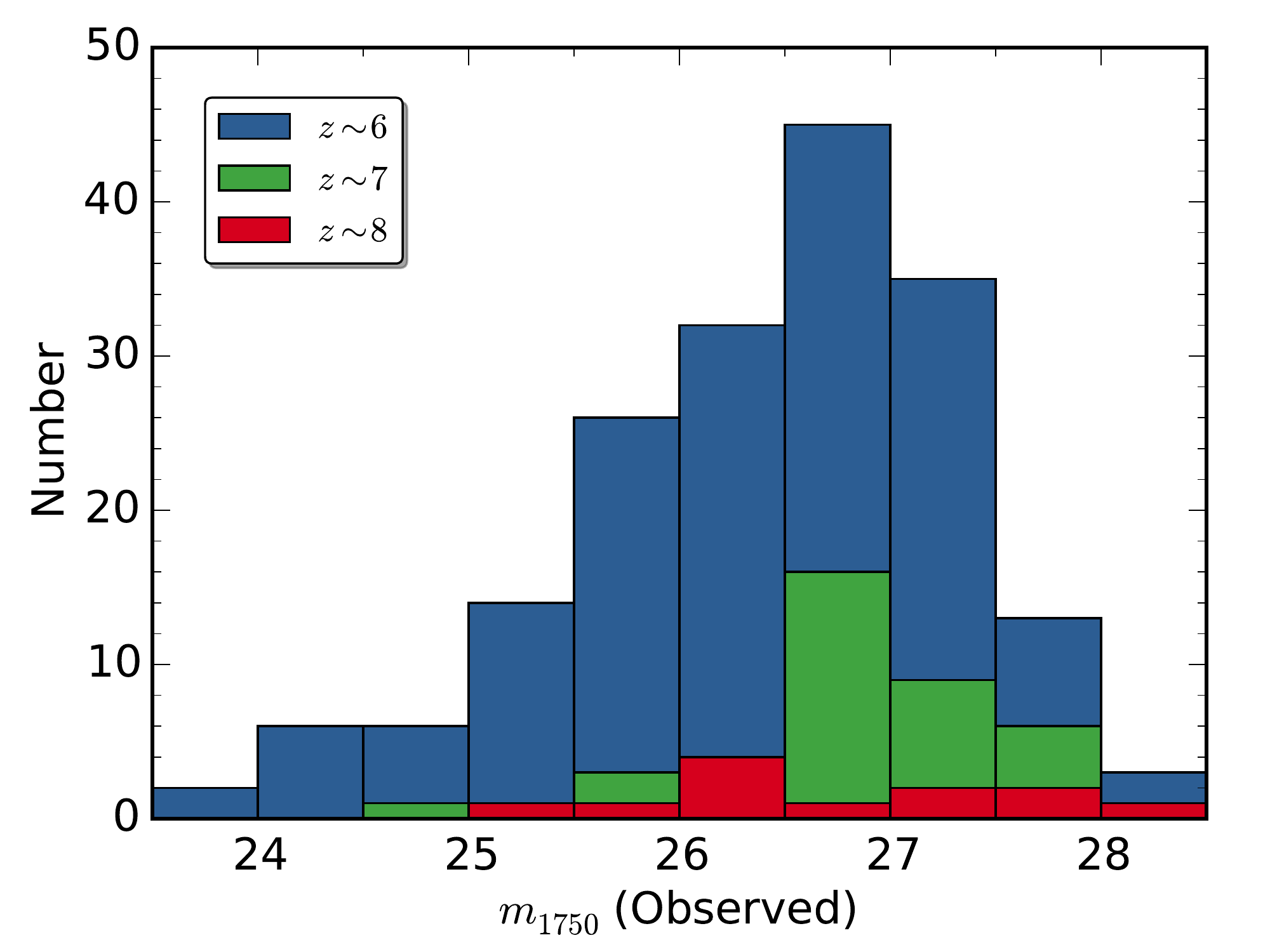}}
    \includegraphics[width=0.5\textwidth]{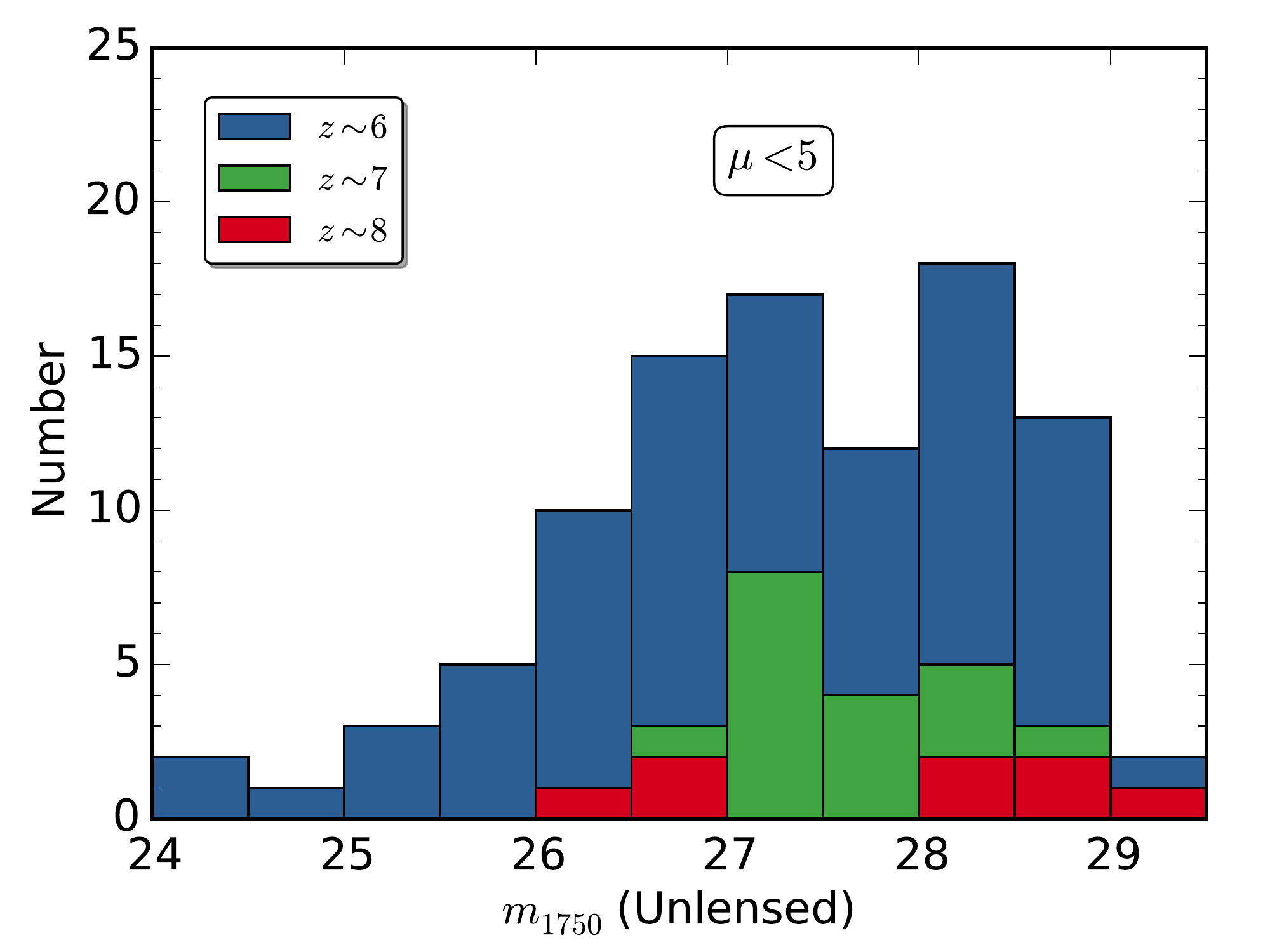}\includegraphics[width=0.5\textwidth]{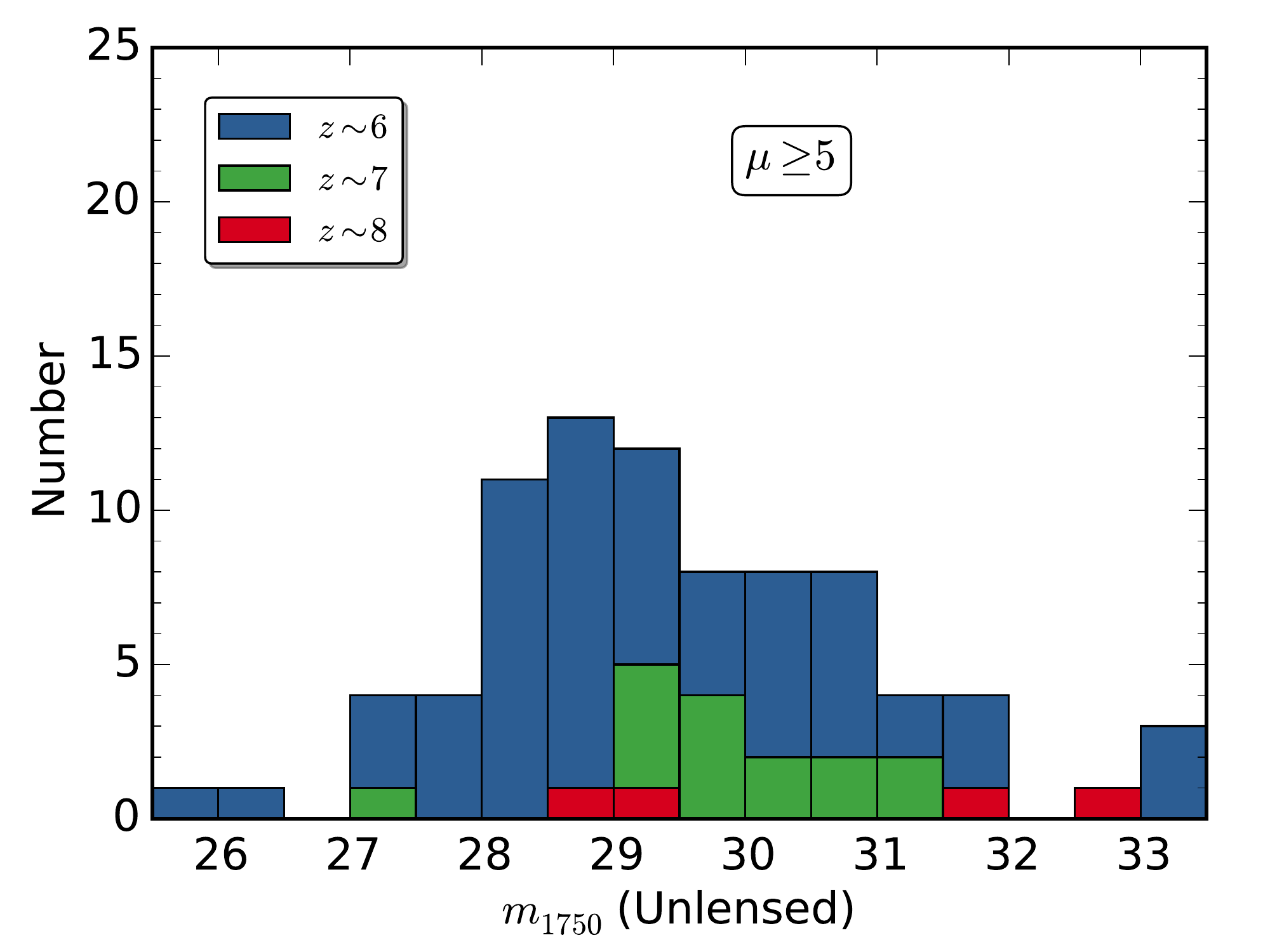}
    \caption{Histogram of the observed (top) and intrinsic (unlensed) (bottom left and right) rest-frame UV magnitudes at $\sim1750$~\AA\ for our sample of $z\sim6$ (blue), $z\sim7$ (green), and $z\sim8$ (red) galaxy candidates identified behind 17 CLASH clusters.  The highest magnifications, and hence the faintest intrinsic magnitudes, usually have the largest magnification errors due to typical uncertainties in the precise location of the critical curves (see Section~\ref{sec:muerr}).  Therefore, we separate the intrinsic magnitude histograms for candidates with magnifications $\mu < 5$ (bottom left) and $\mu \ge 5$ (bottom right).  Because of the strong lensing effect, the intrinsic magnitudes in the relatively shallow CLASH survey ($5\sigma$ limiting magnitude of $\sim27.5$ mag in the \Hband\ band) reach deeper ($> 29.5$ AB mag) than the ultra-deep HUDF12 observations.  Our observations of the 17 clusters cover a total area of $\sim22.9$~arcmin$^2$ with magnifications $\mu > 6.3$, needed to surpass the depth of the HUDF12 observations.  This is a relatively robust measure of area in spite of the model uncertainties (see Section~\ref{sec:muarea}).}
    \label{fig:maghist}
    \end{figure*}
}

\def\figmuerr {
    \begin{figure}[tbp!]
    \centerline{\includegraphics[width=1.0\columnwidth]{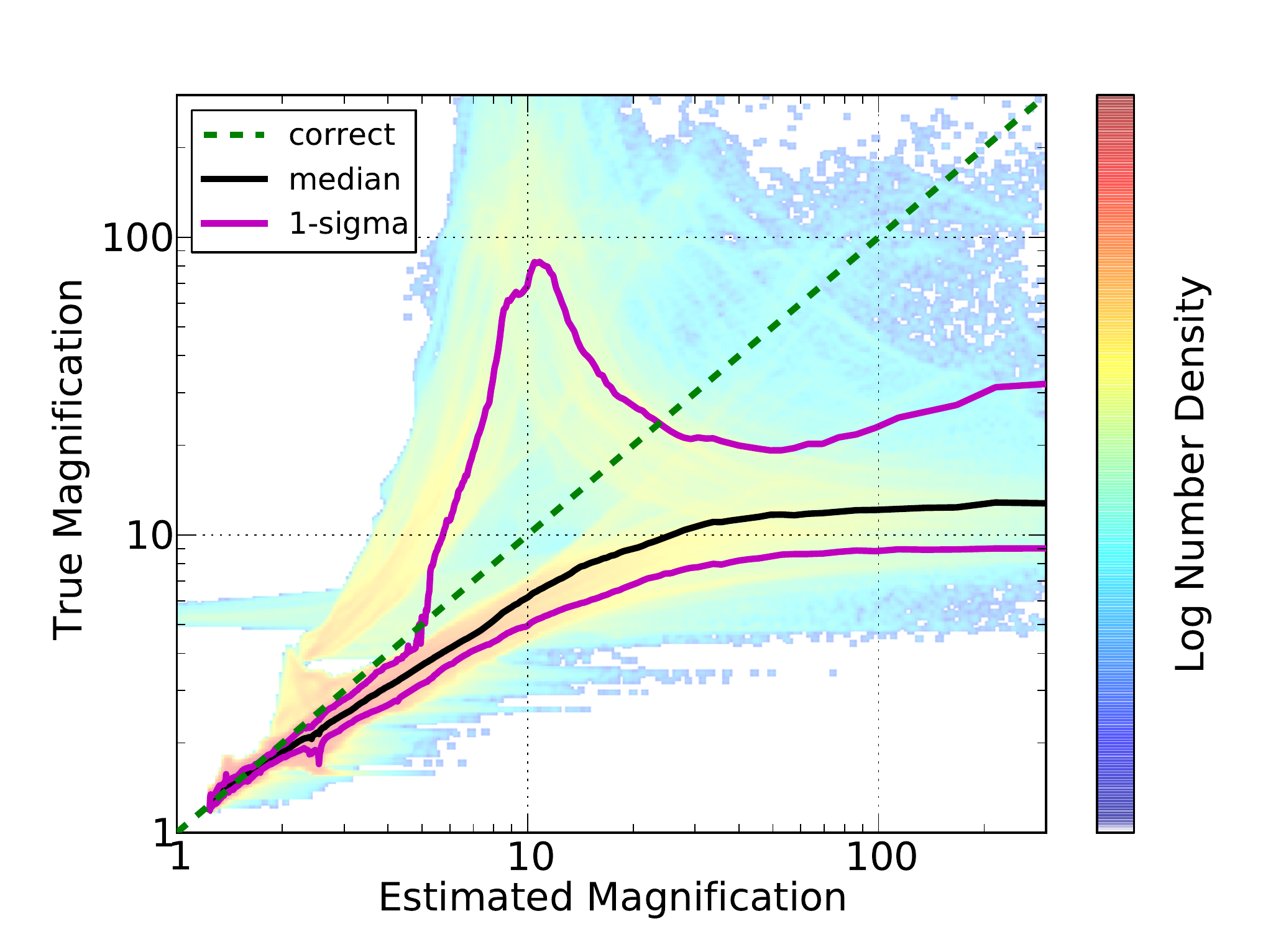}}
    \caption{Results from lens modeling of a simulated cluster:  true vs. estimated magnifications.  Values are compared at every $0\farcs05$ pixel within a \mult{2\farcm5}{2\farcm5} field of view and the density of points is plotted here.  The correct magnification values fall along the green dashed line.  The median and 68\% intervals are plotted as solid black and magenta lines, respectively.  For example, an estimated magnification of five likely corresponds to a true magnification between three and five at 68\% confidence.  Estimated model magnification values greater than 30 are often large overestimates, as the lens models do not precisely reproduce the locations of the critical curves.}
    \label{fig:muerr}
    \end{figure}
}

\def\figmuarea {
    \begin{figure}[tbp!]
    \centerline{\includegraphics[width=1.0\columnwidth]{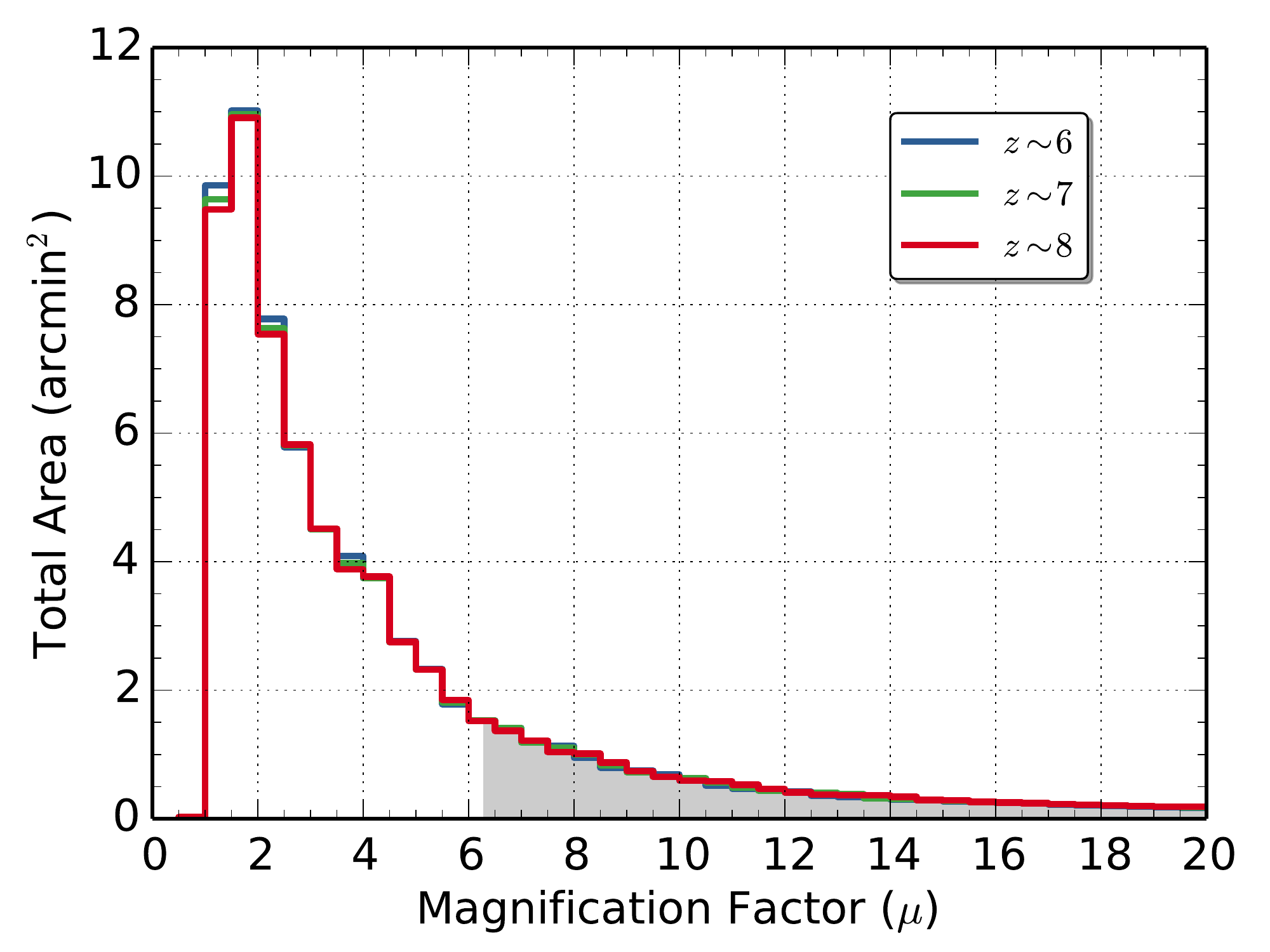}}
    \caption{Total search area over the 17 clusters as a function of magnification at $z\sim6$ (blue), $z\sim7$ (green), and $z\sim8$ (red).  Because the redshift dependence on $d_{ls}/d_{s}$ between $z\sim6$ and $z\sim8$ is very small (see Section~\ref{sec:models}), the total areas are very similar in each of the three redshift bins.  One consequence of this effect is that the search volumes behind these clusters is relatively insensitive to redshift, allowing for a differential determination of the UV LF with lower overall uncertainties \citep{Bouwens2012clash}.  While local magnifications close to the critical curves can have large uncertainties, the overall shape of this total area vs. $\mu$ curve is not significantly affected by uncertainties in lensing models.  The gray shaded region denotes the area with magnifications $\mu > 6.3$, which extends our limiting magnitude by 2 mag and corresponds to regions deeper than the HUDF12 observations.}
    \label{fig:muarea}
    \end{figure}
}

\def\figlf {
    \begin{figure*}[htbp!]
    \includegraphics[width=0.5\textwidth]{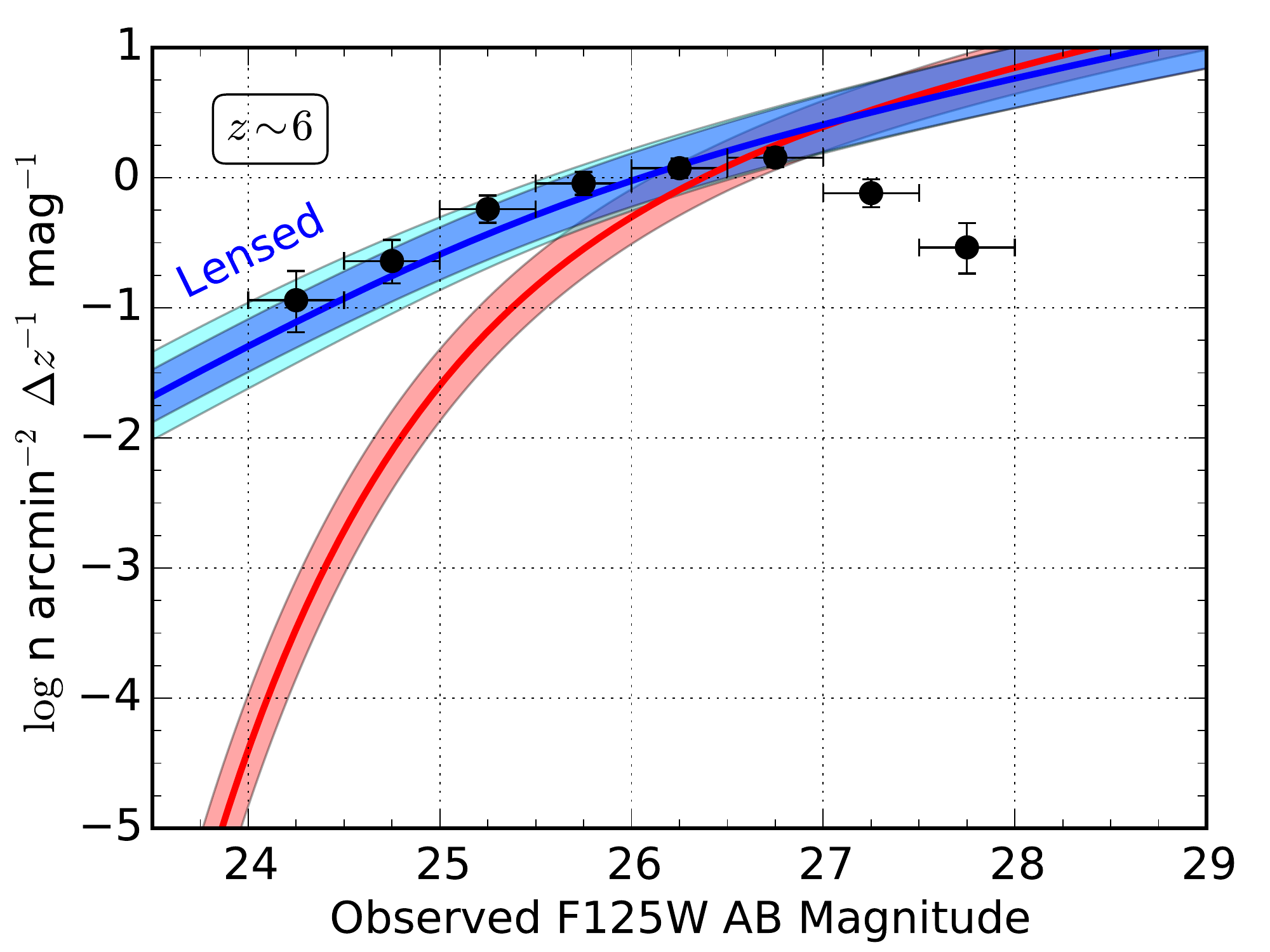}\includegraphics[width=0.5\textwidth]{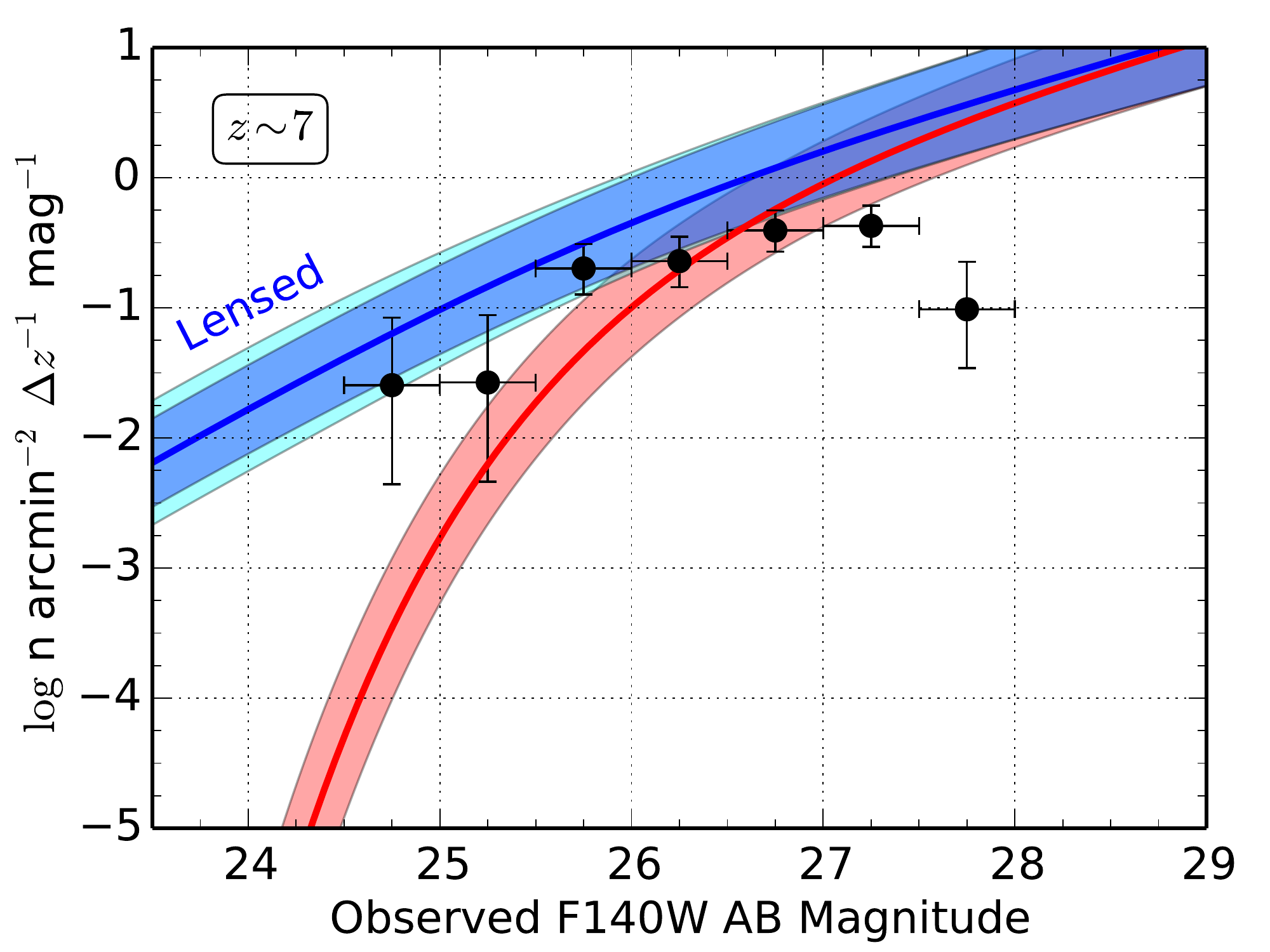}\\
    \centerline{\includegraphics[width=0.5\textwidth]{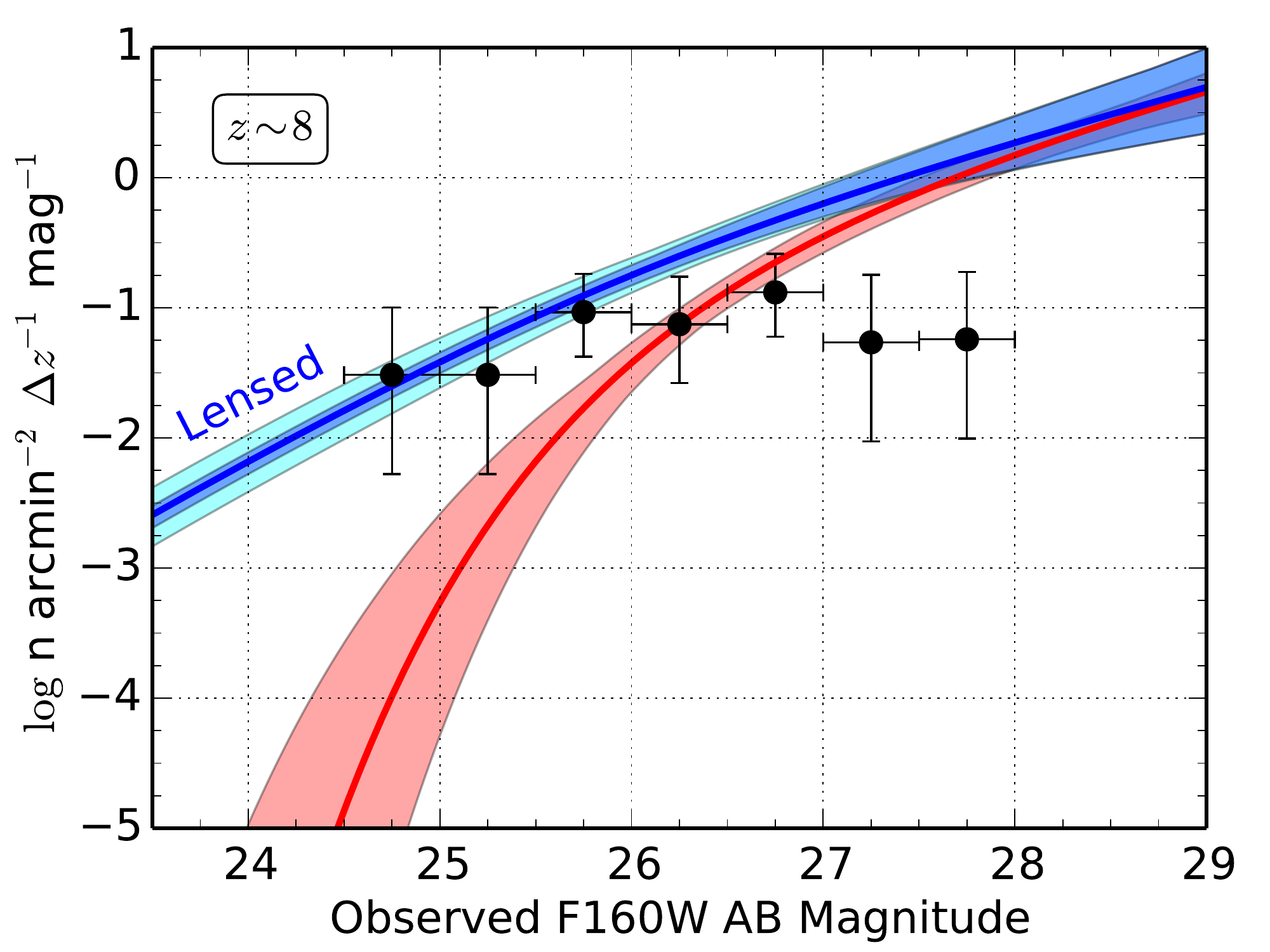}}
    \caption{Observed number counts for our lensed galaxy samples at $z\sim6$ (top left), $z\sim7$ (top right), and $z\sim8$ (bottom).  The black data points, with 68\% ($1\sigma$) confidence intervals for a Poisson distribution, represent the number densities over the 17 clusters, covering 76.9 arcmin$^2$ in total.  The red curves, with shaded $1\sigma$ regions, are the expected number densities calculated from ``blank'' field UV LFs.  The $z\sim6$ and $z\sim7$ LFs are derived from the UDF09+ERS deep fields \citep{Bouwens2011b}, while the $z\sim8$ LF was derived from a combination of wide and deep BoRG+HUDF09+ERS data \citep{Bradley2012b}.  The blue curves are the expected number densities derived by simulating the lensing effect on the field LFs using the cluster lens models.  The cyan regions include the additional errors introduced by the uncertainties in the cluster lens models.   As can be seen in the figure, the observed number counts of our lensed high-redshift sample are roughly consistent with the expected lensed number counts down to $\sim27$ mag, where we begin to suffer significant ($>50$\%) incompleteness.  This is especially true at $z\sim6$, where we have the best statistics.  We also note that our observed number densities are higher than one would expect from a ``blank'' field survey at the brightest magnitudes, where the lensing effects are most significant because of the steepness of the LF.}
    \label{fig:lf}
    \end{figure*}
}

\def\tabclobs {
\ifemulateapj
    \begin{deluxetable*}{llccccll}
\else
    \begin{deluxetable}{llccccll}
\fi
\tablecolumns{8}
\tablewidth{0pt}
\tablecaption{Observational Details for the Cluster Sample}
\tablehead{\colhead{Cluster Name} & \colhead{Cluster Alias\tnm{a}} & \colhead{$\alpha_{J2000}\tnm{b}$} & \colhead{$\delta_{J2000}\tnm{b}$} & \colhead{$z_{\mathrm{clus}}$} & \colhead{$E(B-V)$} & \colhead{Start Date} & \colhead{End Date} }
\startdata
Abell 383                   & A383     & 02:48:03.36 & $-$03:31:44.7 & 0.187 & 0.031 & 2010 Nov 18 & 2011 Mar 1 \\
MACSJ1149.6+2223\tnm{c,d,e} & MACS1149 & 11:49:35.86 & $+$22:23:55.0 & 0.544 & 0.023 & 2010 Dec 4 & 2011 Mar 9 \\
Abell 2261                  & A2261    & 17:22:27.25 & $+$32:07:58.6 & 0.224 & 0.043 & 2011 Mar 9 & 2011 Jun 13 \\
MACSJ1206.2-0847\tnm{f}     & MACS1206 & 12:06:12.28 & $-$08:48:02.4 & 0.440 & 0.063 & 2011 Apr 3 & 2011 Jul 20 \\
RXJ1347.5-1145              & RXJ1347  & 13:47:30.59 & $-$11:45:10.1 & 0.451 & 0.062 & 2011 Apr 19 & 2011 Jul 14 \\
MACSJ2129.4-0741\tnm{c,d}   & MACS2129 & 21:29:26.06\tnm{g} & $-$07:41:28.8\tnm{g} & 0.570 & 0.076 & 2011 May 15 & 2011 Aug 3 \\
MACSJ0329.7-0211\tnm{h}     & MACS0329 & 03:29:41.68 & $-$02:11:47.7 & 0.450 & 0.060 & 2011 Aug 18 & 2011 Nov 1 \\
MS2137-2353                 & MS2137   & 21:40:15.18 & $-$23:39:40.7 & 0.313 & 0.051 & 2011 Aug 21 & 2011 Nov 7 \\
MACSJ0717.5+3745\tnm{c,d,e} & MACS0717 & 07:17:31.65 & $+$37:45:18.5 & 0.548 & 0.077 & 2011 Aug 31 & 2011 Dec 9 \\
MACSJ0744.9+3927\tnm{c}     & MACS0744 & 07:44:52.80 & $+$39:27:24.4 & 0.686 & 0.058 & 2011 Sep 22 & 2011 Dec 29 \\
MACSJ0647.8+7015\tnm{c,d}   & MACS0647 & 06:47:50.03 & $+$70:14:49.7 & 0.584 & 0.111 & 2011 Oct 5 & 2011 Nov 29 \\
MACSJ1115.9+0129\tnm{f}     & MACS1115 & 11:15:52.05 & $+$01:29:56.6 & 0.352 & 0.039 & 2011 Dec 14 & 2012 Feb 24 \\
Abell 611                   & A611     & 08:00:56.83 & $+$36:03:24.1 & 0.288 & 0.057 & 2012 Jan 28 & 2012 May 17 \\
RXJ1532.9+3021              & RXJ1532  & 15:32:53.78 & $+$30:20:58.7 & 0.363 & 0.030 & 2012 Feb 3 & 2012 Apr 12 \\
MACSJ1720.3+3536\tnm{f}     & MACS1720 & 17:20:16.95 & $+$35:36:23.6 & 0.387 & 0.038 & 2012 Mar 26 & 2012 Jun 17 \\
MACSJ1931.8-2635\tnm{f}     & MACS1931 & 19:31:49.66 & $-$26:34:34.0 & 0.352 & 0.110 & 2012 Apr 10 & 2012 Jun 25 \\
MACSJ0416.1-2403\tnm{d,e,f} & MACS0416 & 04:16:09.39 & $-$24:04:03.9 & 0.396 & 0.041 & 2012 Jul 24 & 2012 Sep 27 \\
RXCJ2248.7-4431\tnm{e}      & RXCJ2248 & 22:48:44.29 & $-$44:31:48.4 & 0.348 & 0.012 & 2012 Aug 30 & 2012 Nov 19
\enddata
\tablenotetext{a}{The shortened cluster names that are generally used in this paper.}
\tablenotetext{b}{Cluster coordinates derived from X-ray data, except where noted.}
\tablenotetext{c}{\cite{Ebeling2007}.}
\tablenotetext{d}{High-magnification cluster.}
\tablenotetext{e}{Hubble Frontier Fields (HFF) cluster.}
\tablenotetext{f}{\cite{Ebeling2010}.}
\tablenotetext{g}{Cluster coordinates derived from optical data.}
\tablenotetext{h}{\cite{Mann2012}.}
\label{tbl:clobs}
\ifemulateapj
    \end{deluxetable*}
\else
    \end{deluxetable}
\fi
}

\def\tabfilts {
\ifemulateapj
    \begin{deluxetable*}{lcccc}
\else
    \begin{deluxetable}{lcccc}
\fi
\tablecolumns{5}
\tablewidth{0pt}
\tablecaption{CLASH Filter Selection, Typical Exposure Times, and Limiting Magnitudes}
\tablehead{\colhead{Detector} & \colhead{Filter} & \colhead{\HST\ Orbits} & \colhead{Exposure Time (s)\tnm{a}} & \colhead{$m_{\mathrm{lim}}$\tnm{b}}}
\startdata
WFC3/UVIS & F225W  & 1.5 & 3627 & 26.4 \\
WFC3/UVIS & F275W  & 1.5 & 3697 & 26.5 \\
WFC3/UVIS & F336W  & 1.0 & 2381 & 26.6 \\
WFC3/UVIS & F390W  & 1.0 & 2408 & 27.2 \\
ACS/WFC   & F435W  & 1.0 & 2036 & 27.2 \\
ACS/WFC   & F475W  & 1.0 & 2052 & 27.6 \\
ACS/WFC   & F606W  & 1.0 & 2028 & 27.6 \\
ACS/WFC   & F625W  & 1.0 & 2015 & 27.2 \\
ACS/WFC   & F775W  & 1.0 & 2046 & 27.0 \\
ACS/WFC   & F814W  & 2.0 & 4149 & 27.7 \\
ACS/WFC   & F850LP & 2.0 & 4162 & 26.7 \\
WFC3/IR   & F105W  & 1.0 & 2754 & 27.3 \\
WFC3/IR   & F110W  & 1.0 & 2543 & 27.8 \\
WFC3/IR   & F125W  & 1.0 & 2482 & 27.2 \\
WFC3/IR   & F140W  & 1.0 & 2323 & 27.4 \\
WFC3/IR   & F160W  & 2.0 & 5108 & 27.5
\enddata
\tablenotetext{a}{Average exposure times in each filter for the CLASH clusters presented in this paper.}
\tablenotetext{b}{Typical $5\sigma$ limiting magnitudes in an $r = 0\farcs2$ circular aperture.}
\label{tbl:filters}
\ifemulateapj
    \end{deluxetable*}
\else
    \end{deluxetable}
\fi
}

\def\tabzsix {
\ifemulateapj
    \setlength{\tabcolsep}{0.03in}
    \tabletypesize{\tiny}
    \begin{deluxetable*}{lccccccccccc}
\else
    \begin{deluxetable}{lccccccccccc}
\fi
\tablecolumns{12}
\tablewidth{0pt}
\tablecaption{Lensed $z\sim6$ Candidates Identified Behind 17 CLASH Clusters}
\tablehead{\colhead{Object ID} & \colhead{$\alpha_{J2000}$} & \colhead{$\delta_{J2000}$} & \colhead{$I_{814}$} & \colhead{$z_{850}$} & \colhead{$Y_{105}$} & \colhead{$Y_{110}$} & \colhead{$J_{125}$} & \colhead{$J_{140}$} & \colhead{$H_{160}$} & \colhead{$z_{\mathrm{phot}}$\tnm{a}} & \colhead{$\mu$\tnm{b}} }
\startdata
A2261-0309\tnm{c,d,e} & $260.6026455$ & $32.1469776$ & $28.7 \pm 0.68$ & $>28.1$ & $28.1 \pm 0.27$ & $27.8 \pm 0.15$ & $27.9 \pm 0.24$ & $27.8 \pm 0.18$ & $28.1 \pm 0.24$ & $6.2^{+1.1}_{-6.1}$ & $2.0$ \\
A2261-0478\tnm{c} & $260.6102936$ & $32.1434241$ & $>29.2$ & $28.0 \pm 0.46$ & $27.5 \pm 0.18$ & $27.2 \pm 0.11$ & $27.4 \pm 0.17$ & $27.6 \pm 0.17$ & $27.7 \pm 0.20$ & $6.4^{+0.6}_{-0.6}$ & $5.8$ \\
A2261-0632 & $260.6197982$ & $32.1408449$ & $27.5 \pm 0.23$ & $25.8 \pm 0.13$ & $26.9 \pm 0.18$ & $26.9 \pm 0.13$ & $26.8 \pm 0.18$ & $27.2 \pm 0.20$ & $26.7 \pm 0.14$ & $5.6^{+0.2}_{-5.6}$ & $7.3$ \\
A2261-0749 & $260.6247221$ & $32.1391001$ & $29.2 \pm 0.54$ & $27.8 \pm 0.38$ & $27.8 \pm 0.22$ & $28.2 \pm 0.24$ & $28.1 \pm 0.30$ & $27.7 \pm 0.19$ & $28.1 \pm 0.27$ & $5.7^{+0.8}_{-5.5}$ & $6.0$ \\
A2261-0754 & $260.6275436$ & $32.1402305$ & $28.4 \pm 0.58$ & $26.6 \pm 0.29$ & $26.0 \pm 0.11$ & $26.1 \pm 0.08$ & $26.3 \pm 0.14$ & $26.3 \pm 0.12$ & $26.2 \pm 0.11$ & $6.4^{+0.3}_{-0.3}$ & $3.8$ \\
A2261-0910 & $260.6047197$ & $32.1357219$ & $27.9 \pm 0.43$ & $25.9 \pm 0.17$ & $26.2 \pm 0.12$ & $26.0 \pm 0.08$ & $26.3 \pm 0.14$ & $26.3 \pm 0.11$ & $26.4 \pm 0.13$ & $6.4^{+0.4}_{-0.3}$ & $8.1$ \\
A2261-1206 & $260.6214225$ & $32.1310024$ & $27.1 \pm 0.22$ & $26.8 \pm 0.34$ & $25.9 \pm 0.09$ & $26.0 \pm 0.08$ & $26.0 \pm 0.11$ & $25.9 \pm 0.08$ & $26.2 \pm 0.11$ & $5.9^{+0.4}_{-0.7}$ & $13.3$ \\
A2261-1366 & $260.6077236$ & $32.1277141$ & $27.7 \pm 0.37$ & $27.6 \pm 0.68$ & $26.2 \pm 0.12$ & $26.0 \pm 0.08$ & $26.6 \pm 0.18$ & $26.3 \pm 0.13$ & $26.3 \pm 0.13$ & $6.4^{+0.3}_{-0.3}$ & $19.6$ \\
A2261-1467 & $260.618602$ & $32.1258197$ & $26.7 \pm 0.20$ & $24.8 \pm 0.12$ & $25.7 \pm 0.10$ & $25.9 \pm 0.08$ & $25.8 \pm 0.11$ & $25.9 \pm 0.10$ & $26.0 \pm 0.11$ & $5.7^{+0.3}_{-1.0}$ & $5.4$ \\
A2261-1679 & $260.6306215$ & $32.1219816$ & $26.8 \pm 0.16$ & $26.5 \pm 0.23$ & $26.6 \pm 0.16$ & $26.7 \pm 0.11$ & \nodata & $26.3 \pm 0.14$ & $26.8 \pm 0.19$ & $5.6^{+0.3}_{-0.6}$ & $1.5$ \\
A383-1320\tnm{f} & $42.0191948$ & $-3.5329581$ & $26.2 \pm 0.11$ & $25.5 \pm 0.12$ & $25.1 \pm 0.05$ & $25.3 \pm 0.05$ & $25.3 \pm 0.07$ & $25.3 \pm 0.06$ & $25.3 \pm 0.06$ & $6.1^{+0.1}_{-0.2}$ & $14.0$ \\
A383-2211 & $42.0058038$ & $-3.5495579$ & $26.4 \pm 0.10$ & $25.7 \pm 0.10$ & $25.3 \pm 0.09$ & \nodata & $25.5 \pm 0.11$ & \nodata & $25.2 \pm 0.10$ & $6.0^{+0.2}_{-0.3}$ & $2.5$ \\
A383-2569\tnm{c} & $42.0122906$ & $-3.5614552$ & $27.4 \pm 0.15$ & $26.7 \pm 0.19$ & $26.9 \pm 0.21$ & \nodata & $26.5 \pm 0.16$ & \nodata & $26.9 \pm 0.24$ & $5.6^{+0.4}_{-5.2}$ & $40.8$ \\
A383-2858\tnm{f} & $42.0136347$ & $-3.5263679$ & $25.9 \pm 0.09$ & $24.8 \pm 0.07$ & $24.9 \pm 0.05$ & $25.0 \pm 0.04$ & $24.8 \pm 0.05$ & $25.0 \pm 0.05$ & $24.9 \pm 0.05$ & $6.0^{+0.1}_{-0.1}$ & $20.5$ \\
A611-0449 & $120.2532197$ & $36.0660342$ & $>28.1$ & $26.7 \pm 0.29$ & $26.9 \pm 0.19$ & $26.7 \pm 0.13$ & $27.1 \pm 0.24$ & $26.7 \pm 0.16$ & $26.6 \pm 0.14$ & $6.0^{+0.6}_{-5.4}$ & $2.9$ \\
A611-0621 & $120.2115196$ & $36.0623423$ & $27.7 \pm 0.27$ & $27.1 \pm 0.30$ & $26.9 \pm 0.18$ & $26.9 \pm 0.12$ & $26.8 \pm 0.30$ & $26.7 \pm 0.13$ & $27.0 \pm 0.18$ & $5.9^{+0.4}_{-5.6}$ & $1.0$ \\
A611-0729 & $120.2219378$ & $36.0596831$ & $27.6 \pm 0.28$ & $>27.5$ & $26.0 \pm 0.08$ & $26.1 \pm 0.07$ & $26.1 \pm 0.09$ & $25.9 \pm 0.06$ & $25.9 \pm 0.07$ & $6.4^{+0.5}_{-0.3}$ & $1.8$ \\
A611-1121 & $120.246557$ & $36.0507464$ & $26.6 \pm 0.15$ & $25.6 \pm 0.12$ & $25.7 \pm 0.07$ & $25.7 \pm 0.06$ & $25.7 \pm 0.07$ & $25.8 \pm 0.07$ & $25.8 \pm 0.07$ & $5.8^{+0.2}_{-0.2}$ & $1.6$ \\
MACS0329-0126 & $52.4238495$ & $-2.1790833$ & $26.2 \pm 0.12$ & $25.6 \pm 0.16$ & $25.4 \pm 0.06$ & $25.5 \pm 0.06$ & $25.5 \pm 0.08$ & $25.4 \pm 0.06$ & $25.5 \pm 0.07$ & $5.8^{+0.2}_{-0.4}$ & $3.0$ \\
MACS0329-0517 & $52.4297897$ & $-2.1881596$ & $>28.5$ & $25.6 \pm 0.18$ & $25.8 \pm 0.09$ & $25.8 \pm 0.07$ & $26.0 \pm 0.11$ & $26.0 \pm 0.09$ & $25.9 \pm 0.09$ & $6.3^{+0.3}_{-0.3}$ & $3.7$ \\
MACS0329-0527\tnm{c} & $52.420542$ & $-2.1882236$ & $27.7 \pm 0.25$ & $26.5 \pm 0.18$ & $26.8 \pm 0.13$ & $26.8 \pm 0.10$ & $26.8 \pm 0.14$ & $27.0 \pm 0.13$ & $27.2 \pm 0.17$ & $5.7^{+0.3}_{-5.6}$ & $48.6$ \\
MACS0329-0760 & $52.4332719$ & $-2.1921488$ & $>29.1$ & $27.2 \pm 0.36$ & $27.1 \pm 0.19$ & $27.1 \pm 0.15$ & $26.8 \pm 0.15$ & $27.2 \pm 0.19$ & $26.9 \pm 0.14$ & $6.3^{+0.9}_{-5.2}$ & $5.4$ \\
MACS0329-0988\tnm{g} & $52.4173769$ & $-2.1959868$ & $25.6 \pm 0.11$ & $24.8 \pm 0.11$ & $24.5 \pm 0.05$ & $24.6 \pm 0.04$ & $24.5 \pm 0.05$ & $24.6 \pm 0.05$ & $24.6 \pm 0.05$ & $6.2^{+0.2}_{-0.3}$ & $11.7$ \\
MACS0329-1132\tnm{g} & $52.4169227$ & $-2.1976907$ & $25.6 \pm 0.12$ & $24.4 \pm 0.09$ & $24.1 \pm 0.04$ & $24.2 \pm 0.03$ & $24.1 \pm 0.04$ & $24.3 \pm 0.04$ & $24.3 \pm 0.04$ & $6.2^{+0.1}_{-0.3}$ & $16.4$ \\
MACS0329-1354\tnm{g} & $52.405594$ & $-2.200912$ & $26.6 \pm 0.19$ & $25.4 \pm 0.14$ & $25.3 \pm 0.08$ & $25.3 \pm 0.06$ & $25.4 \pm 0.08$ & $25.1 \pm 0.06$ & $25.2 \pm 0.06$ & $6.1^{+0.4}_{-0.5}$ & $2.4$ \\
MACS0329-1387\tnm{g} & $52.4218271$ & $-2.2012778$ & $26.4 \pm 0.18$ & $25.5 \pm 0.16$ & $24.9 \pm 0.06$ & $25.1 \pm 0.05$ & $25.0 \pm 0.06$ & $25.1 \pm 0.06$ & $25.2 \pm 0.07$ & $6.3^{+0.3}_{-0.5}$ & $3.9$ \\
MACS0329-1391 & $52.4065478$ & $-2.2013038$ & $27.6 \pm 0.33$ & $25.7 \pm 0.13$ & $26.2 \pm 0.11$ & $26.2 \pm 0.08$ & $26.3 \pm 0.13$ & $26.2 \pm 0.10$ & $26.2 \pm 0.11$ & $5.8^{+0.2}_{-0.3}$ & $2.5$ \\
MACS0329-1532 & $52.4465362$ & $-2.2037826$ & $28.3 \pm 0.45$ & $26.3 \pm 0.18$ & $26.0 \pm 0.48$ & \nodata & $26.3 \pm 0.13$ & $26.3 \pm 0.13$ & $26.0 \pm 0.11$ & $6.0^{+0.4}_{-4.9}$ & $2.7$ \\
MACS0329-1672 & $52.4094084$ & $-2.2069138$ & $27.6 \pm 0.23$ & $26.8 \pm 0.24$ & $27.8 \pm 0.33$ & $26.9 \pm 0.12$ & $26.9 \pm 0.16$ & $27.3 \pm 0.18$ & $27.5 \pm 0.23$ & $5.5^{+0.3}_{-5.5}$ & $2.5$ \\
MACS0329-1921 & $52.4218372$ & $-2.2155338$ & $>28.8$ & $27.0 \pm 0.33$ & $27.0 \pm 0.21$ & $26.9 \pm 0.15$ & $27.4 \pm 0.33$ & $27.0 \pm 0.21$ & $27.0 \pm 0.21$ & $6.2^{+0.7}_{-5.5}$ & $3.1$ \\
MACS0416-0419\tnm{h} & $64.0399924$ & $-24.0618194$ & $28.8 \pm 0.60$ & $27.6 \pm 0.50$ & $27.0 \pm 0.18$ & $27.3 \pm 0.17$ & $27.5 \pm 0.25$ & $27.2 \pm 0.18$ & $27.7 \pm 0.27$ & $6.1^{+0.5}_{-5.9}$ & $11.1$ \\
MACS0416-0427 & $64.0478483$ & $-24.0620688$ & $28.7 \pm 0.58$ & $27.6 \pm 0.51$ & $26.7 \pm 0.15$ & $26.8 \pm 0.12$ & $27.0 \pm 0.19$ & $27.3 \pm 0.21$ & $26.9 \pm 0.15$ & $6.4^{+0.4}_{-0.6}$ & $47.4$ \\
MACS0416-0546 & $64.020806$ & $-24.0641676$ & $>29.3$ & $26.9 \pm 0.23$ & $27.6 \pm 0.24$ & $28.4 \pm 0.36$ & $28.0 \pm 0.32$ & $27.9 \pm 0.26$ & $27.5 \pm 0.19$ & $5.8^{+0.9}_{-5.5}$ & $1.5$ \\
MACS0416-1821\tnm{m} & $64.039537$ & $-24.0885394$ & $27.2 \pm 0.22$ & $26.1 \pm 0.15$ & $26.5 \pm 0.15$ & $26.7 \pm 0.13$ & $26.6 \pm 0.15$ & $26.4 \pm 0.11$ & $26.6 \pm 0.13$ & $5.6^{+0.2}_{-0.6}$ & $1.6$ \\
MACS0416-1950 & $64.0365104$ & $-24.0923003$ & $27.4 \pm 0.21$ & $26.7 \pm 0.25$ & $26.2 \pm 0.11$ & $26.4 \pm 0.09$ & $26.2 \pm 0.10$ & $26.4 \pm 0.10$ & $26.5 \pm 0.11$ & $6.0^{+0.3}_{-0.6}$ & $1.5$ \\
MACS0647-0185 & $101.9421635$ & $70.2651281$ & $27.8 \pm 0.24$ & $26.3 \pm 0.27$ & $27.0 \pm 0.20$ & $26.7 \pm 0.11$ & $27.0 \pm 0.20$ & $27.2 \pm 0.19$ & $26.7 \pm 0.13$ & $5.8^{+0.3}_{-5.7}$ & $4.8$ \\
MACS0647-0273 & $101.9340017$ & $70.2633308$ & $27.6 \pm 0.17$ & $27.2 \pm 0.39$ & $26.6 \pm 0.13$ & $26.7 \pm 0.10$ & $27.1 \pm 0.19$ & $26.7 \pm 0.12$ & $26.8 \pm 0.13$ & $5.9^{+0.3}_{-0.7}$ & $6.7$ \\
MACS0647-0339 & $101.9245873$ & $70.2618682$ & $27.7 \pm 0.25$ & $26.0 \pm 0.18$ & $25.8 \pm 0.12$ & $26.0 \pm 0.06$ & \nodata & $25.8 \pm 0.11$ & $26.1 \pm 0.12$ & $6.2^{+0.2}_{-0.4}$ & $9.3$ \\
MACS0647-0345 & $101.9757709$ & $70.2617321$ & $27.5 \pm 0.19$ & $27.0 \pm 0.35$ & $26.6 \pm 0.16$ & $27.1 \pm 0.16$ & $26.6 \pm 0.15$ & $26.9 \pm 0.16$ & $26.8 \pm 0.16$ & $5.5^{+0.5}_{-5.0}$ & $6.9$ \\
MACS0647-0386 & $101.9562571$ & $70.2611963$ & $29.3 \pm 0.61$ & $27.9 \pm 0.58$ & $28.1 \pm 0.41$ & $26.7 \pm 0.09$ & $27.6 \pm 0.27$ & $27.3 \pm 0.18$ & $27.5 \pm 0.22$ & $6.3^{+0.6}_{-0.5}$ & $10.0$ \\
MACS0647-0392 & $101.9571644$ & $70.2608913$ & $27.7 \pm 0.20$ & $26.8 \pm 0.27$ & $28.0 \pm 0.42$ & $26.7 \pm 0.10$ & $27.2 \pm 0.21$ & $27.1 \pm 0.17$ & $27.3 \pm 0.20$ & $5.6^{+0.3}_{-5.4}$ & $10.6$ \\
MACS0647-0425 & $101.9544319$ & $70.260507$ & $26.7 \pm 0.12$ & $25.6 \pm 0.15$ & $26.1 \pm 0.13$ & $25.8 \pm 0.06$ & $26.2 \pm 0.13$ & $26.0 \pm 0.09$ & $26.2 \pm 0.12$ & $5.7^{+0.2}_{-0.2}$ & $12.6$ \\
MACS0647-0523 & $101.9854555$ & $70.2589206$ & $27.4 \pm 0.17$ & $>27.3$ & $26.4 \pm 0.13$ & $26.1 \pm 0.07$ & $26.4 \pm 0.12$ & $26.5 \pm 0.11$ & $26.3 \pm 0.10$ & $6.2^{+0.2}_{-0.8}$ & $9.1$ \\
MACS0647-0577 & $101.9979739$ & $70.2583512$ & $27.1 \pm 0.12$ & $26.3 \pm 0.20$ & $26.6 \pm 0.15$ & $26.3 \pm 0.09$ & $26.8 \pm 0.18$ & $26.7 \pm 0.14$ & $26.5 \pm 0.12$ & $5.5^{+0.3}_{-5.4}$ & $6.9$ \\
MACS0647-1055 & $101.916902$ & $70.2515356$ & $27.4 \pm 0.17$ & $27.5 \pm 0.49$ & $26.8 \pm 0.17$ & $26.3 \pm 0.08$ & $26.5 \pm 0.13$ & $26.9 \pm 0.16$ & $26.6 \pm 0.13$ & $5.9^{+0.2}_{-0.8}$ & $99.3$ \\
MACS0647-1138 & $101.9324453$ & $70.2501418$ & $26.3 \pm 0.10$ & $25.9 \pm 0.20$ & $25.5 \pm 0.09$ & $25.3 \pm 0.05$ & $25.5 \pm 0.08$ & $25.5 \pm 0.07$ & $25.7 \pm 0.09$ & $5.9^{+0.1}_{-0.3}$ & $13.7$ \\
MACS0647-1317 & $101.9287022$ & $70.2468151$ & $27.6 \pm 0.16$ & $26.8 \pm 0.25$ & $26.9 \pm 0.16$ & $26.9 \pm 0.11$ & $27.4 \pm 0.22$ & $27.2 \pm 0.17$ & $27.2 \pm 0.17$ & $5.6^{+0.3}_{-0.9}$ & $26.4$ \\
MACS0647-1393 & $101.9852823$ & $70.2454242$ & $27.0 \pm 0.12$ & $26.5 \pm 0.22$ & $26.6 \pm 0.16$ & $26.2 \pm 0.08$ & $26.6 \pm 0.15$ & $26.6 \pm 0.13$ & $26.7 \pm 0.14$ & $5.5^{+0.2}_{-0.6}$ & $19.1$ \\
MACS0647-1448 & $101.9216$ & $70.244592$ & $27.6 \pm 0.16$ & $27.4 \pm 0.43$ & $27.1 \pm 0.19$ & $26.5 \pm 0.08$ & $26.7 \pm 0.14$ & $26.7 \pm 0.11$ & $27.1 \pm 0.17$ & $5.9^{+0.2}_{-5.6}$ & $95.7$ \\
MACS0647-1474 & $101.9599681$ & $70.2441216$ & $28.8 \pm 0.40$ & $27.7 \pm 0.44$ & $27.9 \pm 0.36$ & $27.2 \pm 0.14$ & $27.2 \pm 0.19$ & $27.7 \pm 0.24$ & $27.3 \pm 0.18$ & $5.9^{+0.6}_{-5.6}$ & $60.2$ \\
MACS0647-1500 & $101.9699917$ & $70.2435429$ & $27.7 \pm 0.19$ & $>28.2$ & $26.7 \pm 0.15$ & $26.3 \pm 0.07$ & $27.1 \pm 0.19$ & $26.9 \pm 0.14$ & $27.0 \pm 0.15$ & $6.2^{+0.2}_{-6.2}$ & $20.8$ \\
MACS0647-1545 & $101.9566804$ & $70.2428289$ & $28.1 \pm 0.16$ & $26.8 \pm 0.16$ & $27.7 \pm 0.21$ & $27.0 \pm 0.08$ & $27.6 \pm 0.19$ & $27.6 \pm 0.17$ & $27.7 \pm 0.18$ & $5.7^{+0.2}_{-0.3}$ & $>100$ \\
MACS0647-1670\tnm{c,d,e} & $101.898829$ & $70.2407169$ & $27.4 \pm 0.17$ & $27.0 \pm 0.36$ & $26.2 \pm 0.13$ & \nodata & $26.1 \pm 0.09$ & $26.0 \pm 0.09$ & $26.2 \pm 0.12$ & $5.9^{+0.4}_{-0.6}$ & $1.0$ \\
MACS0647-1701 & $101.9084623$ & $70.2398933$ & $26.6 \pm 0.11$ & $26.0 \pm 0.20$ & $26.0 \pm 0.11$ & $25.9 \pm 0.07$ & $26.1 \pm 0.12$ & $26.1 \pm 0.10$ & $26.0 \pm 0.09$ & $5.6^{+0.2}_{-4.9}$ & $4.9$ \\
MACS0647-1706 & $101.937453$ & $70.2397838$ & $26.7 \pm 0.12$ & $26.0 \pm 0.20$ & $25.8 \pm 0.10$ & $25.8 \pm 0.08$ & $25.7 \pm 0.09$ & $26.2 \pm 0.12$ & $26.2 \pm 0.12$ & $5.9^{+0.2}_{-0.2}$ & $27.1$ \\
MACS0647-1728 & $101.9020768$ & $70.2392049$ & $27.6 \pm 0.19$ & $27.3 \pm 0.43$ & $27.0 \pm 0.22$ & \nodata & $26.9 \pm 0.16$ & $27.0 \pm 0.20$ & $27.0 \pm 0.20$ & $5.5^{+0.6}_{-5.0}$ & $1.0$ \\
MACS0647-1780 & $101.9500603$ & $70.2382908$ & $27.8 \pm 0.18$ & $26.8 \pm 0.20$ & $27.2 \pm 0.20$ & $26.8 \pm 0.09$ & $27.5 \pm 0.25$ & $27.5 \pm 0.21$ & $27.7 \pm 0.24$ & $5.5^{+0.3}_{-0.4}$ & $22.3$ \\
MACS0647-1976 & $101.9542124$ & $70.2353545$ & $27.9 \pm 0.25$ & $26.6 \pm 0.25$ & $26.1 \pm 0.10$ & $26.3 \pm 0.08$ & $26.3 \pm 0.11$ & $26.3 \pm 0.10$ & $26.7 \pm 0.14$ & $6.3^{+0.2}_{-0.2}$ & $8.2$ \\
MACS0647-2233 & $101.9610981$ & $70.2268856$ & $27.1 \pm 0.15$ & $25.8 \pm 0.18$ & $25.3 \pm 0.07$ & \nodata & $25.2 \pm 0.06$ & $25.1 \pm 0.05$ & $25.2 \pm 0.06$ & $6.2^{+0.2}_{-0.2}$ & $2.7$ \\
MACS0717-0145 & $109.3820224$ & $37.7676218$ & $27.6 \pm 0.20$ & $26.3 \pm 0.18$ & $26.9 \pm 0.20$ & $26.8 \pm 0.10$ & \nodata & $26.9 \pm 0.20$ & $27.1 \pm 0.22$ & $5.7^{+0.2}_{-0.4}$ & $4.8$ \\
MACS0717-0166 & $109.3914431$ & $37.7670479$ & $27.8 \pm 0.25$ & $26.9 \pm 0.33$ & $27.1 \pm 0.21$ & $26.8 \pm 0.12$ & $27.0 \pm 0.21$ & $27.0 \pm 0.17$ & $27.4 \pm 0.24$ & $5.9^{+0.3}_{-0.7}$ & $3.3$ \\
MACS0717-0234 & $109.3991201$ & $37.7649582$ & $28.0 \pm 0.39$ & $26.3 \pm 0.25$ & $26.5 \pm 0.15$ & $26.1 \pm 0.08$ & $26.5 \pm 0.18$ & $26.1 \pm 0.10$ & $26.3 \pm 0.13$ & $6.1^{+0.4}_{-0.5}$ & $5.2$ \\
MACS0717-0247\tnm{c} & $109.3797704$ & $37.7646903$ & $26.7 \pm 0.13$ & $25.9 \pm 0.16$ & $25.6 \pm 0.06$ & $25.6 \pm 0.05$ & $25.5 \pm 0.07$ & $25.4 \pm 0.05$ & $25.4 \pm 0.05$ & $5.6^{+0.2}_{-0.2}$ & $11.3$ \\
MACS0717-0390 & $109.3730817$ & $37.7616756$ & $28.4 \pm 0.29$ & $27.9 \pm 0.51$ & $27.3 \pm 0.17$ & $27.4 \pm 0.14$ & $27.5 \pm 0.22$ & $27.5 \pm 0.19$ & $27.4 \pm 0.17$ & $6.0^{+0.4}_{-5.6}$ & $2.1$ \\
MACS0717-0844 & $109.4045791$ & $37.754928$ & $27.2 \pm 0.15$ & $26.6 \pm 0.28$ & $26.5 \pm 0.14$ & $26.5 \pm 0.11$ & $26.6 \pm 0.17$ & $26.7 \pm 0.14$ & $26.8 \pm 0.16$ & $5.6^{+0.3}_{-5.5}$ & $>100$ \\
MACS0717-0859\tnm{n} & $109.4090649$ & $37.7546861$ & $28.0 \pm 0.26$ & $26.2 \pm 0.16$ & $26.4 \pm 0.11$ & $26.3 \pm 0.08$ & $26.4 \pm 0.12$ & $26.5 \pm 0.11$ & $26.9 \pm 0.15$ & $6.1^{+0.2}_{-0.2}$ & $15.6$ \\
MACS0717-1077 & $109.3862252$ & $37.7519284$ & $28.3 \pm 0.36$ & $27.0 \pm 0.31$ & $26.9 \pm 0.17$ & $27.0 \pm 0.14$ & $27.1 \pm 0.22$ & $27.0 \pm 0.16$ & $27.0 \pm 0.17$ & $6.0^{+0.4}_{-5.3}$ & $2.8$ \\
MACS0717-1230\tnm{c} & $109.3738235$ & $37.7494447$ & $26.7 \pm 0.12$ & $25.9 \pm 0.16$ & $25.6 \pm 0.07$ & $25.5 \pm 0.05$ & $25.7 \pm 0.08$ & $25.5 \pm 0.05$ & $25.6 \pm 0.06$ & $5.5^{+0.3}_{-0.2}$ & $>100$ \\
MACS0717-1373 & $109.3661401$ & $37.747541$ & $28.4 \pm 0.28$ & $>28.0$ & $27.9 \pm 0.30$ & $27.3 \pm 0.14$ & $27.5 \pm 0.24$ & $27.8 \pm 0.23$ & $27.4 \pm 0.18$ & $5.6^{+0.8}_{-5.3}$ & $4.6$ \\
MACS0717-1730\tnm{n} & $109.4077276$ & $37.7427406$ & $28.4 \pm 0.32$ & $25.9 \pm 0.11$ & $26.3 \pm 0.16$ & \nodata & $26.8 \pm 0.16$ & $26.7 \pm 0.16$ & $26.8 \pm 0.18$ & $6.0^{+0.2}_{-0.3}$ & $>100$ \\
MACS0717-1749 & $109.3907301$ & $37.742226$ & $26.3 \pm 0.11$ & $25.7 \pm 0.21$ & $25.3 \pm 0.08$ & $25.3 \pm 0.06$ & $25.4 \pm 0.10$ & $25.5 \pm 0.09$ & $25.4 \pm 0.08$ & $6.0^{+0.2}_{-0.5}$ & $42.2$ \\
MACS0717-1825 & $109.4087679$ & $37.7408053$ & $25.5 \pm 0.03$ & $24.7 \pm 0.04$ & $24.8 \pm 0.04$ & \nodata & $24.9 \pm 0.03$ & $24.9 \pm 0.03$ & $24.8 \pm 0.03$ & $5.6^{+0.1}_{-0.1}$ & $15.7$ \\
MACS0717-1944 & $109.3782878$ & $37.7391635$ & $27.7 \pm 0.21$ & $26.4 \pm 0.20$ & $27.5 \pm 0.28$ & $27.3 \pm 0.18$ & $27.0 \pm 0.21$ & $27.0 \pm 0.16$ & $27.2 \pm 0.19$ & $5.6^{+0.3}_{-5.5}$ & $8.1$ \\
MACS0717-1991 & $109.3923348$ & $37.7380827$ & $26.0 \pm 0.09$ & $25.6 \pm 0.24$ & $25.5 \pm 0.09$ & $25.7 \pm 0.08$ & $25.5 \pm 0.11$ & $25.5 \pm 0.08$ & $25.6 \pm 0.09$ & $5.5^{+0.2}_{-0.5}$ & $16.1$ \\
MACS0744-0323 & $116.2177736$ & $39.4712757$ & $28.3 \pm 0.32$ & $27.6 \pm 0.52$ & $26.7 \pm 0.14$ & $27.4 \pm 0.18$ & $27.2 \pm 0.21$ & $27.4 \pm 0.22$ & $27.4 \pm 0.22$ & $6.1^{+0.4}_{-4.9}$ & $5.6$ \\
MACS0744-0329 & $116.2276242$ & $39.4711928$ & $27.0 \pm 0.13$ & $26.3 \pm 0.25$ & $26.1 \pm 0.10$ & $26.5 \pm 0.10$ & $26.2 \pm 0.13$ & $26.2 \pm 0.10$ & $26.3 \pm 0.12$ & $5.7^{+0.3}_{-0.6}$ & $4.4$ \\
MACS0744-0563 & $116.2211501$ & $39.4669929$ & $26.1 \pm 0.08$ & $25.7 \pm 0.22$ & $25.3 \pm 0.07$ & $25.1 \pm 0.04$ & $25.2 \pm 0.07$ & $25.2 \pm 0.06$ & $25.1 \pm 0.06$ & $5.5^{+0.3}_{-0.2}$ & $13.2$ \\
MACS0744-0999 & $116.1905996$ & $39.4606661$ & $26.7 \pm 0.10$ & $26.0 \pm 0.14$ & $25.6 \pm 0.09$ & \nodata & $25.6 \pm 0.06$ & $25.7 \pm 0.06$ & $25.6 \pm 0.07$ & $5.7^{+0.5}_{-0.2}$ & $5.5$ \\
MACS0744-1026 & $116.2464832$ & $39.460415$ & $>29.8$ & $27.2 \pm 0.28$ & $27.9 \pm 0.29$ & $27.1 \pm 0.12$ & $27.8 \pm 0.28$ & $27.9 \pm 0.24$ & $27.6 \pm 0.21$ & $6.3^{+0.6}_{-1.0}$ & $2.9$ \\
MACS0744-1063 & $116.219486$ & $39.4598451$ & $28.6 \pm 0.42$ & $27.9 \pm 0.67$ & $27.3 \pm 0.25$ & $26.9 \pm 0.14$ & $26.9 \pm 0.19$ & $26.9 \pm 0.16$ & $26.9 \pm 0.17$ & $6.1^{+0.9}_{-5.6}$ & $7.0$ \\
MACS0744-1141 & $116.1994527$ & $39.4566391$ & $27.7 \pm 0.23$ & $26.4 \pm 0.22$ & $26.6 \pm 0.15$ & $27.0 \pm 0.15$ & $26.8 \pm 0.18$ & $26.4 \pm 0.11$ & $26.5 \pm 0.12$ & $5.6^{+0.3}_{-5.3}$ & $28.4$ \\
MACS0744-1253 & $116.2306779$ & $39.4565064$ & $26.0 \pm 0.08$ & $25.8 \pm 0.21$ & $25.5 \pm 0.08$ & $25.3 \pm 0.06$ & $25.6 \pm 0.11$ & $25.6 \pm 0.08$ & $25.6 \pm 0.09$ & $5.5^{+0.2}_{-0.6}$ & $13.7$ \\
MACS0744-1313 & $116.241686$ & $39.456442$ & $27.4 \pm 0.16$ & $26.8 \pm 0.27$ & $26.9 \pm 0.16$ & $27.1 \pm 0.16$ & $27.0 \pm 0.20$ & $26.9 \pm 0.14$ & $27.7 \pm 0.32$ & $5.5^{+0.3}_{-4.9}$ & $4.7$ \\
MACS0744-1379 & $116.2029144$ & $39.4555249$ & $27.6 \pm 0.18$ & $26.9 \pm 0.27$ & $27.0 \pm 0.16$ & $26.9 \pm 0.11$ & $27.2 \pm 0.22$ & $27.0 \pm 0.15$ & $27.3 \pm 0.20$ & $5.7^{+0.3}_{-0.9}$ & $47.4$ \\
MACS0744-1419 & $116.2477909$ & $39.4548039$ & $26.8 \pm 0.11$ & $26.1 \pm 0.17$ & $25.9 \pm 0.11$ & \nodata & $25.6 \pm 0.07$ & $25.9 \pm 0.07$ & $25.8 \pm 0.10$ & $5.6^{+0.4}_{-0.2}$ & $2.7$ \\
MACS0744-1476 & $116.1958217$ & $39.4538057$ & $28.2 \pm 0.39$ & $27.6 \pm 0.61$ & $26.8 \pm 0.20$ & $26.7 \pm 0.13$ & $26.9 \pm 0.23$ & $26.5 \pm 0.14$ & $26.5 \pm 0.14$ & $6.2^{+0.5}_{-5.6}$ & $>100$ \\
MACS0744-1600 & $116.2324074$ & $39.4528428$ & $27.9 \pm 0.23$ & $27.0 \pm 0.31$ & $27.5 \pm 0.25$ & $27.2 \pm 0.17$ & $26.7 \pm 0.16$ & $26.9 \pm 0.13$ & $27.1 \pm 0.18$ & $5.6^{+0.3}_{-5.4}$ & $27.0$ \\
MACS0744-1704 & $116.2249599$ & $39.4514537$ & $27.4 \pm 0.15$ & $26.9 \pm 0.32$ & $26.7 \pm 0.14$ & $26.7 \pm 0.11$ & $26.6 \pm 0.15$ & $26.7 \pm 0.12$ & $26.8 \pm 0.15$ & $5.8^{+0.2}_{-0.7}$ & $11.6$ \\
MACS0744-1744 & $116.2147513$ & $39.4520614$ & $28.8 \pm 0.50$ & $>27.6$ & $27.5 \pm 0.28$ & $26.8 \pm 0.13$ & $27.1 \pm 0.23$ & $27.1 \pm 0.16$ & $27.0 \pm 0.18$ & $6.3^{+1.0}_{-5.5}$ & $33.0$ \\
MACS0744-2374 & $116.2210822$ & $39.4411485$ & $27.2 \pm 0.13$ & $26.0 \pm 0.13$ & $26.4 \pm 0.11$ & $26.5 \pm 0.10$ & $26.6 \pm 0.15$ & $26.8 \pm 0.12$ & $26.7 \pm 0.14$ & $5.8^{+0.2}_{-0.2}$ & $3.7$ \\
MACS0744-2580 & $116.2169117$ & $39.4374838$ & $27.8 \pm 0.17$ & $26.0 \pm 0.15$ & $27.2 \pm 0.20$ & $27.4 \pm 0.17$ & $27.3 \pm 0.23$ & $27.1 \pm 0.16$ & $27.5 \pm 0.22$ & $5.6^{+0.2}_{-5.5}$ & $2.6$ \\
MACS1115-0091 & $168.975381$ & $1.516936$ & $27.5 \pm 0.29$ & $26.9 \pm 0.42$ & $25.6 \pm 0.09$ & $25.7 \pm 0.06$ & \nodata & $26.1 \pm 0.12$ & $25.9 \pm 0.11$ & $6.4^{+0.2}_{-0.2}$ & $1.9$ \\
MACS1115-0240 & $168.977624$ & $1.5122685$ & $>28.6$ & $>27.7$ & $27.8 \pm 0.28$ & $27.3 \pm 0.15$ & $27.4 \pm 0.23$ & $27.6 \pm 0.25$ & $27.5 \pm 0.21$ & $6.3^{+0.9}_{-6.0}$ & $2.2$ \\
MACS1115-0352\tnm{i} & $168.9719677$ & $1.5100216$ & $28.7 \pm 0.65$ & $27.2 \pm 0.43$ & $26.7 \pm 0.15$ & $26.9 \pm 0.15$ & $27.2 \pm 0.26$ & $27.1 \pm 0.21$ & $27.2 \pm 0.22$ & $6.2^{+0.5}_{-6.0}$ & $2.8$ \\
MACS1115-0907 & $168.9552697$ & $1.4992604$ & $>29.4$ & $27.9 \pm 0.46$ & $27.7 \pm 0.24$ & $27.5 \pm 0.16$ & $27.9 \pm 0.33$ & $27.8 \pm 0.27$ & $27.5 \pm 0.18$ & $6.3^{+0.7}_{-5.9}$ & $5.3$ \\
MACS1115-1618 & $168.9572896$ & $1.484229$ & $28.1 \pm 0.50$ & $27.1 \pm 0.43$ & $26.4 \pm 0.16$ & $26.2 \pm 0.10$ & $26.3 \pm 0.15$ & $26.6 \pm 0.19$ & $26.4 \pm 0.14$ & $6.3^{+0.4}_{-5.5}$ & $2.3$ \\
MACS1115-1658 & $168.9511485$ & $1.482588$ & $27.6 \pm 0.33$ & $26.1 \pm 0.18$ & $26.1 \pm 0.16$ & \nodata & $25.9 \pm 0.10$ & $26.0 \pm 0.15$ & $26.0 \pm 0.12$ & $5.8^{+0.4}_{-5.3}$ & $1.9$ \\
MACS1149-0094 & $177.4004414$ & $22.4180794$ & $27.2 \pm 0.18$ & $25.8 \pm 0.15$ & $25.9 \pm 0.14$ & $25.8 \pm 0.07$ & \nodata & $25.6 \pm 0.11$ & $25.6 \pm 0.10$ & $5.9^{+0.2}_{-0.3}$ & $4.0$ \\
MACS1149-0144\tnm{c} & $177.3907057$ & $22.4169154$ & $27.2 \pm 0.14$ & $>28.2$ & $26.1 \pm 0.07$ & $26.1 \pm 0.06$ & $26.1 \pm 0.13$ & $26.1 \pm 0.08$ & $26.0 \pm 0.08$ & $6.0^{+0.2}_{-0.4}$ & $6.0$ \\
MACS1149-0226 & $177.4120212$ & $22.415777$ & $26.2 \pm 0.09$ & $24.9 \pm 0.08$ & $25.2 \pm 0.08$ & \nodata & $24.9 \pm 0.05$ & $24.8 \pm 0.05$ & $24.9 \pm 0.06$ & $5.8^{+0.1}_{-0.1}$ & $2.3$ \\
MACS1149-0432 & $177.4129973$ & $22.4117036$ & $27.9 \pm 0.19$ & $26.8 \pm 0.21$ & $27.2 \pm 0.17$ & $27.3 \pm 0.14$ & $27.1 \pm 0.18$ & $27.4 \pm 0.19$ & $27.6 \pm 0.24$ & $5.7^{+0.3}_{-0.7}$ & $2.1$ \\
MACS1149-0524 & $177.4030246$ & $22.409501$ & $27.8 \pm 0.20$ & $26.4 \pm 0.16$ & $27.0 \pm 0.15$ & $27.2 \pm 0.14$ & $26.9 \pm 0.17$ & $27.4 \pm 0.20$ & $27.5 \pm 0.24$ & $5.5^{+0.3}_{-0.4}$ & $9.2$ \\
MACS1149-0626 & $177.4070876$ & $22.4078431$ & $27.3 \pm 0.20$ & $25.7 \pm 0.14$ & $26.5 \pm 0.15$ & $26.4 \pm 0.11$ & $26.5 \pm 0.17$ & $26.7 \pm 0.17$ & $26.4 \pm 0.13$ & $5.6^{+0.2}_{-5.5}$ & $4.6$ \\
MACS1149-0992 & $177.3789714$ & $22.4024227$ & $26.0 \pm 0.05$ & $25.5 \pm 0.09$ & \nodata & \nodata & $25.3 \pm 0.06$ & \nodata & $25.2 \pm 0.09$ & $5.5^{+0.4}_{-0.1}$ & $3.3$ \\
MACS1149-1210 & $177.3888435$ & $22.4028565$ & $27.9 \pm 0.22$ & $26.6 \pm 0.19$ & $27.1 \pm 0.17$ & $27.4 \pm 0.17$ & $27.3 \pm 0.23$ & $27.5 \pm 0.23$ & $26.8 \pm 0.13$ & $5.6^{+0.2}_{-5.5}$ & $>100$ \\
MACS1149-1401 & $177.3825494$ & $22.3954137$ & $27.5 \pm 0.21$ & $26.8 \pm 0.28$ & $26.4 \pm 0.14$ & $26.5 \pm 0.09$ & $26.3 \pm 0.15$ & $26.8 \pm 0.22$ & $26.2 \pm 0.11$ & $5.6^{+0.5}_{-5.0}$ & $4.8$ \\
MACS1149-1632 & $177.3925946$ & $22.3917302$ & $28.4 \pm 0.28$ & $26.7 \pm 0.16$ & $27.2 \pm 0.15$ & $27.6 \pm 0.17$ & $27.3 \pm 0.19$ & $27.5 \pm 0.19$ & $27.4 \pm 0.18$ & $5.8^{+0.3}_{-0.4}$ & $21.9$ \\
MACS1149-2104 & $177.3883908$ & $22.3823891$ & $28.5 \pm 0.38$ & $26.7 \pm 0.20$ & $26.5 \pm 0.11$ & $26.7 \pm 0.10$ & $26.6 \pm 0.14$ & $27.2 \pm 0.19$ & $26.9 \pm 0.16$ & $6.2^{+0.3}_{-0.2}$ & $2.1$ \\
MACS1149-2198 & $177.3960473$ & $22.3797645$ & $28.6 \pm 0.32$ & $27.5 \pm 0.29$ & $27.3 \pm 0.19$ & $27.2 \pm 0.13$ & $27.0 \pm 0.17$ & $27.2 \pm 0.18$ & $27.2 \pm 0.17$ & $6.0^{+0.4}_{-0.5}$ & $3.2$ \\
MACS1206-0457\tnm{j} & $181.5513858$ & $-8.7912941$ & $25.8 \pm 0.09$ & $25.2 \pm 0.12$ & $25.3 \pm 0.07$ & $25.2 \pm 0.05$ & $25.3 \pm 0.08$ & $25.2 \pm 0.06$ & $25.3 \pm 0.06$ & $5.5^{+0.2}_{-0.3}$ & $3.6$ \\
MACS1206-0704 & $181.5649579$ & $-8.7943784$ & $28.5 \pm 0.51$ & $>28.2$ & $27.1 \pm 0.19$ & $27.1 \pm 0.14$ & $27.2 \pm 0.23$ & $26.9 \pm 0.16$ & $27.4 \pm 0.24$ & $6.4^{+0.8}_{-5.8}$ & $4.4$ \\
MACS1206-0861\tnm{j} & $181.5440852$ & $-8.7970801$ & $26.1 \pm 0.10$ & $25.3 \pm 0.10$ & $25.4 \pm 0.06$ & $25.4 \pm 0.05$ & $25.4 \pm 0.07$ & $25.4 \pm 0.06$ & $25.6 \pm 0.07$ & $5.6^{+0.2}_{-0.2}$ & $3.1$ \\
MACS1206-0894 & $181.5361803$ & $-8.7973156$ & $27.3 \pm 0.27$ & $26.6 \pm 0.29$ & $26.3 \pm 0.14$ & $26.3 \pm 0.11$ & $26.2 \pm 0.14$ & $26.4 \pm 0.15$ & $26.2 \pm 0.12$ & $5.6^{+0.6}_{-5.1}$ & $3.2$ \\
MACS1206-0895 & $181.5505714$ & $-8.7972768$ & $27.9 \pm 0.28$ & $26.8 \pm 0.22$ & $27.0 \pm 0.16$ & $27.0 \pm 0.12$ & $27.1 \pm 0.19$ & $27.5 \pm 0.23$ & $27.4 \pm 0.22$ & $5.5^{+0.3}_{-0.5}$ & $83.2$ \\
MACS1206-0908 & $181.5505101$ & $-8.797488$ & $28.0 \pm 0.39$ & $26.8 \pm 0.30$ & $26.4 \pm 0.12$ & $26.5 \pm 0.10$ & $26.6 \pm 0.15$ & $26.8 \pm 0.16$ & $26.6 \pm 0.14$ & $6.2^{+0.3}_{-0.5}$ & $13.8$ \\
MACS1206-1135\tnm{j} & $181.5552114$ & $-8.8010583$ & $26.1 \pm 0.09$ & $25.3 \pm 0.10$ & $25.5 \pm 0.07$ & $25.4 \pm 0.05$ & $25.6 \pm 0.07$ & $25.6 \pm 0.07$ & $25.5 \pm 0.06$ & $5.5^{+0.1}_{-0.3}$ & $1.2$ \\
MACS1206-1311 & $181.571211$ & $-8.8031521$ & $>29.3$ & $27.6 \pm 0.37$ & $27.1 \pm 0.18$ & $27.6 \pm 0.16$ & $26.9 \pm 0.24$ & $27.2 \pm 0.21$ & $27.5 \pm 0.21$ & $6.4^{+0.8}_{-0.9}$ & $3.3$ \\
MACS1206-1670 & $181.5419297$ & $-8.8107101$ & $27.9 \pm 0.43$ & $27.0 \pm 0.40$ & $26.8 \pm 0.19$ & $26.6 \pm 0.13$ & $26.9 \pm 0.22$ & $26.4 \pm 0.12$ & $26.6 \pm 0.15$ & $5.7^{+0.7}_{-5.3}$ & $2.7$ \\
MACS1206-1796\tnm{j,k} & $181.5470009$ & $-8.8122557$ & $24.5 \pm 0.05$ & $24.0 \pm 0.07$ & $23.9 \pm 0.03$ & $23.8 \pm 0.02$ & $23.8 \pm 0.03$ & $23.9 \pm 0.03$ & $23.8 \pm 0.03$ & $5.6^{+0.1}_{-0.2}$ & $2.1$ \\
MACS1206-1811 & $181.5417209$ & $-8.8122549$ & $>27.9$ & $26.6 \pm 0.32$ & $26.0 \pm 0.10$ & $26.0 \pm 0.07$ & $26.0 \pm 0.11$ & $26.0 \pm 0.09$ & $26.1 \pm 0.10$ & $6.4^{+0.5}_{-0.4}$ & $2.0$ \\
MACS1206-1917 & $181.5383621$ & $-8.8144235$ & $26.7 \pm 0.18$ & $25.5 \pm 0.16$ & $25.9 \pm 0.09$ & $25.8 \pm 0.06$ & $25.8 \pm 0.09$ & $25.8 \pm 0.08$ & $26.1 \pm 0.11$ & $5.8^{+0.2}_{-0.2}$ & $1.7$ \\
MACS1206-2101 & $181.5404978$ & $-8.8196689$ & $29.3 \pm 0.68$ & $27.0 \pm 0.22$ & $27.9 \pm 0.33$ & \nodata & $28.1 \pm 0.29$ & $27.6 \pm 0.21$ & $27.6 \pm 0.24$ & $5.8^{+0.5}_{-5.5}$ & $1.5$ \\
MACS1206-2160 & $181.5474602$ & $-8.8222067$ & $26.3 \pm 0.14$ & $26.2 \pm 0.28$ & $25.6 \pm 0.11$ & $25.7 \pm 0.06$ & \nodata & $25.6 \pm 0.11$ & $25.7 \pm 0.12$ & $5.5^{+0.5}_{-0.3}$ & $1.5$ \\
MACS1720-0134\tnm{c} & $260.0693781$ & $35.6264574$ & $26.3 \pm 0.10$ & $25.6 \pm 0.13$ & $24.9 \pm 0.04$ & $24.9 \pm 0.03$ & $24.8 \pm 0.04$ & $24.9 \pm 0.03$ & $24.8 \pm 0.03$ & $6.1^{+0.1}_{-0.4}$ & $2.0$ \\
MACS1720-0477 & $260.0740341$ & $35.6187819$ & $27.6 \pm 0.24$ & $26.6 \pm 0.21$ & $26.9 \pm 0.17$ & $26.8 \pm 0.10$ & $26.9 \pm 0.16$ & $27.1 \pm 0.17$ & $27.0 \pm 0.17$ & $5.6^{+0.3}_{-0.6}$ & $5.9$ \\
MACS1720-0645\tnm{c} & $260.0734907$ & $35.6155909$ & $27.9 \pm 0.27$ & $26.6 \pm 0.21$ & $26.4 \pm 0.10$ & $26.5 \pm 0.07$ & $26.3 \pm 0.09$ & $26.4 \pm 0.09$ & $26.6 \pm 0.11$ & $6.1^{+0.3}_{-0.4}$ & $75.8$ \\
MACS1720-0768 & $260.0874493$ & $35.6128696$ & $29.0 \pm 0.70$ & $27.6 \pm 0.49$ & $26.5 \pm 0.14$ & $26.5 \pm 0.09$ & $26.7 \pm 0.16$ & $26.3 \pm 0.10$ & $26.7 \pm 0.14$ & $6.5^{+0.5}_{-0.4}$ & $1.6$ \\
MACS1720-0803 & $260.0843099$ & $35.6125046$ & $>28.7$ & $27.3 \pm 0.35$ & $27.0 \pm 0.18$ & $26.9 \pm 0.11$ & $26.8 \pm 0.16$ & $27.0 \pm 0.15$ & $26.8 \pm 0.13$ & $6.2^{+0.7}_{-0.8}$ & $1.8$ \\
MACS1720-1114\tnm{i} & $260.0847192$ & $35.6068842$ & $28.3 \pm 0.37$ & $26.6 \pm 0.19$ & $26.8 \pm 0.14$ & $27.1 \pm 0.12$ & $27.2 \pm 0.19$ & $27.1 \pm 0.14$ & $27.9 \pm 0.29$ & $5.9^{+0.3}_{-0.3}$ & $1.8$ \\
MACS1720-1199\tnm{c} & $260.079187$ & $35.6053751$ & $26.6 \pm 0.11$ & $27.8 \pm 0.64$ & $25.8 \pm 0.07$ & $25.8 \pm 0.05$ & $25.7 \pm 0.07$ & $25.8 \pm 0.06$ & $25.8 \pm 0.06$ & $5.5^{+0.5}_{-0.2}$ & $3.1$ \\
MACS1720-1304 & $260.0740896$ & $35.6033961$ & $27.4 \pm 0.20$ & $26.5 \pm 0.19$ & $27.3 \pm 0.25$ & $27.1 \pm 0.14$ & $26.6 \pm 0.14$ & $27.2 \pm 0.18$ & $27.5 \pm 0.25$ & $5.6^{+0.3}_{-0.5}$ & $45.0$ \\
MACS1720-1860 & $260.0524853$ & $35.5921507$ & $27.7 \pm 0.22$ & $27.1 \pm 0.28$ & $26.4 \pm 0.16$ & \nodata & $27.2 \pm 0.20$ & $27.1 \pm 0.17$ & $26.8 \pm 0.17$ & $5.6^{+0.5}_{-5.3}$ & $1.6$ \\
MACS1931-0094 & $292.9503289$ & $-26.5554582$ & $27.4 \pm 0.43$ & $>27.4$ & $25.9 \pm 0.15$ & $25.7 \pm 0.07$ & \nodata & \nodata & $25.7 \pm 0.10$ & $6.5^{+1.0}_{-0.5}$ & $2.3$ \\
MACS1931-0201\tnm{c} & $292.9547422$ & $-26.5587983$ & $25.0 \pm 0.06$ & $24.5 \pm 0.07$ & $24.2 \pm 0.03$ & $24.1 \pm 0.02$ & $24.0 \pm 0.03$ & $24.1 \pm 0.02$ & $24.0 \pm 0.02$ & $5.5^{+0.1}_{-0.1}$ & $5.5$ \\
MACS1931-0307\tnm{c,d} & $292.9496458$ & $-26.5613531$ & $>28.0$ & $26.5 \pm 0.26$ & $26.4 \pm 0.11$ & $26.3 \pm 0.08$ & $26.2 \pm 0.11$ & $26.3 \pm 0.09$ & $26.2 \pm 0.09$ & $6.1^{+0.6}_{-0.4}$ & $3.2$ \\
MACS1931-0332\tnm{c} & $292.9510777$ & $-26.5621776$ & $26.4 \pm 0.15$ & $25.6 \pm 0.15$ & $25.1 \pm 0.05$ & $25.1 \pm 0.04$ & $25.0 \pm 0.05$ & $25.0 \pm 0.04$ & $25.0 \pm 0.04$ & $5.8^{+0.3}_{-0.2}$ & $3.6$ \\
MACS1931-0390\tnm{c,d} & $292.9541233$ & $-26.5629601$ & $26.3 \pm 0.15$ & $26.1 \pm 0.26$ & $24.9 \pm 0.04$ & $24.8 \pm 0.03$ & $24.6 \pm 0.04$ & $24.8 \pm 0.04$ & $24.8 \pm 0.04$ & $6.3^{+0.2}_{-0.2}$ & $12.6$ \\
MACS1931-0414\tnm{c} & $292.9519357$ & $-26.5629018$ & $27.5 \pm 0.28$ & $26.4 \pm 0.21$ & $26.8 \pm 0.15$ & $26.7 \pm 0.11$ & $26.5 \pm 0.13$ & $26.5 \pm 0.11$ & $26.6 \pm 0.11$ & $5.6^{+0.4}_{-5.4}$ & $4.1$ \\
MACS1931-0437\tnm{c} & $292.978389$ & $-26.5636071$ & $26.1 \pm 0.09$ & $25.3 \pm 0.09$ & $25.5 \pm 0.08$ & $25.5 \pm 0.04$ & \nodata & \nodata & $25.5 \pm 0.06$ & $5.6^{+0.1}_{-0.2}$ & $1.2$ \\
MACS1931-0444 & $292.9624842$ & $-26.5637544$ & $27.7 \pm 0.38$ & $>27.9$ & $27.2 \pm 0.18$ & $27.0 \pm 0.12$ & $27.3 \pm 0.23$ & $27.2 \pm 0.17$ & $27.2 \pm 0.17$ & $6.1^{+0.6}_{-6.0}$ & $3.7$ \\
MACS1931-0464\tnm{c} & $292.9777005$ & $-26.5642029$ & $25.6 \pm 0.08$ & $25.2 \pm 0.10$ & $24.7 \pm 0.05$ & $24.7 \pm 0.03$ & \nodata & \nodata & $24.6 \pm 0.04$ & $5.5^{+0.2}_{-0.1}$ & $1.2$ \\
MACS1931-0507\tnm{c} & $292.9547566$ & $-26.5654048$ & $27.5 \pm 0.33$ & $26.0 \pm 0.18$ & $25.8 \pm 0.07$ & $25.9 \pm 0.07$ & $25.5 \pm 0.07$ & $25.8 \pm 0.07$ & $25.7 \pm 0.06$ & $5.9^{+0.3}_{-0.3}$ & $16.5$ \\
MACS1931-0543\tnm{c} & $292.9730338$ & $-26.5656782$ & $26.5 \pm 0.16$ & $25.9 \pm 0.17$ & $25.2 \pm 0.05$ & $25.1 \pm 0.04$ & $25.1 \pm 0.05$ & $25.0 \pm 0.04$ & $25.1 \pm 0.04$ & $6.1^{+0.2}_{-0.2}$ & $1.4$ \\
MACS1931-0584\tnm{c} & $292.9471261$ & $-26.5661392$ & $28.0 \pm 0.40$ & $27.4 \pm 0.45$ & $26.7 \pm 0.13$ & $26.5 \pm 0.09$ & $26.4 \pm 0.11$ & $26.7 \pm 0.12$ & $26.9 \pm 0.14$ & $6.3^{+0.5}_{-0.5}$ & $6.9$ \\
MACS1931-0596\tnm{c} & $292.9469143$ & $-26.5664889$ & $27.3 \pm 0.31$ & $26.1 \pm 0.20$ & $25.5 \pm 0.06$ & $25.4 \pm 0.05$ & $25.2 \pm 0.05$ & $25.3 \pm 0.05$ & $25.3 \pm 0.05$ & $6.2^{+0.3}_{-0.2}$ & $6.4$ \\
MACS1931-0633 & $292.9592822$ & $-26.5670855$ & $27.7 \pm 0.39$ & $26.9 \pm 0.37$ & $26.7 \pm 0.16$ & $26.1 \pm 0.08$ & $26.7 \pm 0.19$ & $26.6 \pm 0.15$ & $26.6 \pm 0.14$ & $6.2^{+0.3}_{-6.1}$ & $22.6$ \\
MACS1931-0776\tnm{c} & $292.9737171$ & $-26.5703672$ & $26.1 \pm 0.11$ & $25.5 \pm 0.13$ & $25.2 \pm 0.06$ & $25.2 \pm 0.04$ & $25.1 \pm 0.06$ & $25.2 \pm 0.05$ & $25.0 \pm 0.04$ & $5.6^{+0.2}_{-0.2}$ & $1.4$ \\
MACS1931-0808 & $292.9746042$ & $-26.5708672$ & $>28.2$ & $27.2 \pm 0.47$ & $26.4 \pm 0.14$ & $26.5 \pm 0.13$ & $26.8 \pm 0.21$ & $26.7 \pm 0.16$ & $26.4 \pm 0.13$ & $6.4^{+0.5}_{-0.9}$ & $1.5$ \\
MACS1931-0938\tnm{c,d,e} & $292.9797587$ & $-26.5731188$ & $28.1 \pm 0.36$ & $27.7 \pm 0.45$ & $27.1 \pm 0.23$ & \nodata & $26.9 \pm 0.15$ & $27.0 \pm 0.14$ & $27.1 \pm 0.27$ & $6.0^{+0.7}_{-5.5}$ & $1.3$ \\
MACS1931-0945\tnm{c} & $292.9782034$ & $-26.5732344$ & $28.4 \pm 0.50$ & $26.7 \pm 0.24$ & $26.8 \pm 0.16$ & $26.4 \pm 0.09$ & $26.4 \pm 0.13$ & $26.7 \pm 0.13$ & $26.3 \pm 0.09$ & $6.0^{+0.4}_{-5.2}$ & $1.3$ \\
MACS1931-0953\tnm{c,d} & $292.9651244$ & $-26.5733583$ & $26.9 \pm 0.16$ & $26.5 \pm 0.23$ & $26.3 \pm 0.10$ & $26.3 \pm 0.09$ & $26.1 \pm 0.10$ & $26.2 \pm 0.08$ & $26.3 \pm 0.09$ & $5.5^{+0.4}_{-0.3}$ & $2.6$ \\
MACS1931-1070\tnm{c,d} & $292.9492074$ & $-26.5750469$ & $27.5 \pm 0.33$ & $26.9 \pm 0.35$ & $25.8 \pm 0.08$ & $25.9 \pm 0.06$ & $25.7 \pm 0.08$ & $25.8 \pm 0.07$ & $25.7 \pm 0.07$ & $6.3^{+0.5}_{-0.4}$ & $3.1$ \\
MACS1931-1171 & $292.9611301$ & $-26.5760123$ & $28.1 \pm 0.51$ & $26.3 \pm 0.26$ & $26.6 \pm 0.16$ & $26.5 \pm 0.12$ & $26.9 \pm 0.23$ & $26.7 \pm 0.15$ & $26.5 \pm 0.14$ & $6.0^{+0.4}_{-0.6}$ & $11.4$ \\
MACS1931-1477\tnm{c} & $292.9716046$ & $-26.5805159$ & $26.2 \pm 0.12$ & $25.2 \pm 0.10$ & $25.4 \pm 0.07$ & $25.6 \pm 0.07$ & $25.5 \pm 0.07$ & $25.3 \pm 0.05$ & $25.4 \pm 0.06$ & $5.7^{+0.2}_{-0.3}$ & $1.6$ \\
MACS1931-1552 & $292.9382122$ & $-26.5816018$ & $28.7 \pm 0.66$ & $>27.5$ & $26.8 \pm 0.14$ & $27.5 \pm 0.19$ & $27.1 \pm 0.20$ & $27.3 \pm 0.20$ & $27.2 \pm 0.18$ & $6.4^{+0.6}_{-1.2}$ & $1.4$ \\
MACS1931-1597 & $292.9662127$ & $-26.5821962$ & $27.7 \pm 0.34$ & $26.1 \pm 0.17$ & $26.3 \pm 0.10$ & $26.9 \pm 0.15$ & $26.4 \pm 0.13$ & $27.1 \pm 0.19$ & $26.5 \pm 0.12$ & $5.7^{+0.3}_{-0.4}$ & $2.9$ \\
MACS1931-1665\tnm{c} & $292.9682582$ & $-26.5835702$ & $26.3 \pm 0.11$ & $25.4 \pm 0.10$ & $25.6 \pm 0.07$ & $25.6 \pm 0.05$ & $25.5 \pm 0.07$ & $25.5 \pm 0.06$ & $25.6 \pm 0.06$ & $5.6^{+0.2}_{-0.3}$ & $2.3$ \\
MACS1931-1736 & $292.9353502$ & $-26.584787$ & $27.9 \pm 0.28$ & $27.0 \pm 0.24$ & $27.4 \pm 0.23$ & $27.1 \pm 0.11$ & \nodata & \nodata & $27.4 \pm 0.19$ & $5.6^{+0.4}_{-5.3}$ & $1.3$ \\
MACS1931-1771 & $292.9371846$ & $-26.5856266$ & $27.2 \pm 0.20$ & $26.8 \pm 0.27$ & $26.6 \pm 0.15$ & $26.5 \pm 0.08$ & \nodata & \nodata & $26.6 \pm 0.13$ & $5.9^{+0.3}_{-0.9}$ & $1.3$ \\
MACS1931-1870 & $292.9650269$ & $-26.5874801$ & $>28.2$ & $27.6 \pm 0.58$ & $27.1 \pm 0.20$ & $27.0 \pm 0.15$ & $27.3 \pm 0.24$ & $26.7 \pm 0.13$ & $27.0 \pm 0.16$ & $6.2^{+0.8}_{-6.1}$ & $2.2$ \\
MACS1931-1970 & $292.9657214$ & $-26.5895885$ & $27.1 \pm 0.22$ & $26.0 \pm 0.19$ & $26.6 \pm 0.17$ & $26.5 \pm 0.12$ & $26.8 \pm 0.22$ & $26.5 \pm 0.14$ & $26.6 \pm 0.15$ & $5.6^{+0.3}_{-5.2}$ & $1.9$ \\
MACS1931-2039\tnm{c} & $292.9600484$ & $-26.5912831$ & $27.9 \pm 0.44$ & $26.7 \pm 0.31$ & $25.7 \pm 0.07$ & $25.6 \pm 0.06$ & $25.5 \pm 0.07$ & $25.5 \pm 0.06$ & $25.6 \pm 0.06$ & $6.4^{+0.6}_{-0.3}$ & $12.4$ \\
MACS1931-2170\tnm{d} & $292.9490186$ & $-26.5948644$ & $>27.8$ & $26.6 \pm 0.32$ & $26.1 \pm 0.17$ & \nodata & $26.0 \pm 0.13$ & $25.9 \pm 0.09$ & $25.9 \pm 0.17$ & $6.3^{+0.6}_{-5.3}$ & $2.2$ \\
MACS1931-2226\tnm{c} & $292.9599516$ & $-26.5966804$ & $27.6 \pm 0.26$ & $27.1 \pm 0.32$ & $25.9 \pm 0.09$ & $26.1 \pm 0.07$ & \nodata & \nodata & $26.2 \pm 0.09$ & $6.4^{+0.2}_{-0.2}$ & $2.6$ \\
MACS2129-0464\tnm{c} & $322.3484132$ & $-7.6802292$ & $29.0 \pm 0.44$ & $>27.9$ & $27.6 \pm 0.24$ & $27.5 \pm 0.16$ & $27.7 \pm 0.24$ & $27.2 \pm 0.14$ & $27.4 \pm 0.17$ & $6.2^{+0.9}_{-5.5}$ & $1.8$ \\
MACS2129-0575\tnm{c} & $322.3546912$ & $-7.6826197$ & $26.4 \pm 0.12$ & $25.8 \pm 0.13$ & $25.2 \pm 0.05$ & $25.2 \pm 0.04$ & $25.2 \pm 0.05$ & $25.2 \pm 0.04$ & $25.2 \pm 0.04$ & $5.5^{+0.2}_{-0.2}$ & $3.0$ \\
MACS2129-0833 & $322.3439866$ & $-7.6869395$ & $27.1 \pm 0.14$ & $26.5 \pm 0.25$ & $26.4 \pm 0.12$ & $26.7 \pm 0.12$ & $26.9 \pm 0.19$ & $26.9 \pm 0.16$ & $26.5 \pm 0.11$ & $5.6^{+0.3}_{-0.8}$ & $6.7$ \\
MACS2129-1121 & $322.3424847$ & $-7.6926449$ & $26.6 \pm 0.17$ & $25.4 \pm 0.18$ & $24.5 \pm 0.05$ & $24.4 \pm 0.03$ & $24.4 \pm 0.04$ & $24.4 \pm 0.03$ & $24.4 \pm 0.04$ & $6.4^{+0.2}_{-0.2}$ & $5.4$ \\
MACS2129-1134 & $322.3482721$ & $-7.692628$ & $27.0 \pm 0.13$ & $26.4 \pm 0.23$ & $27.0 \pm 0.20$ & $26.4 \pm 0.09$ & $26.8 \pm 0.18$ & $26.8 \pm 0.14$ & $26.6 \pm 0.12$ & $5.5^{+0.2}_{-1.3}$ & $17.8$ \\
MACS2129-1256 & $322.3534109$ & $-7.6948568$ & $27.7 \pm 0.19$ & $26.8 \pm 0.23$ & $27.4 \pm 0.24$ & $27.0 \pm 0.12$ & $27.1 \pm 0.19$ & $27.3 \pm 0.19$ & $27.7 \pm 0.25$ & $5.6^{+0.3}_{-1.3}$ & $4.5$ \\
MACS2129-1307 & $322.3509386$ & $-7.6933313$ & $27.9 \pm 0.35$ & $26.5 \pm 0.26$ & $26.2 \pm 0.14$ & $26.0 \pm 0.08$ & $26.4 \pm 0.16$ & $26.3 \pm 0.12$ & $26.5 \pm 0.14$ & $6.3^{+0.2}_{-0.3}$ & $4.3$ \\
MACS2129-1322 & $322.3511495$ & $-7.6957508$ & $27.2 \pm 0.17$ & $25.8 \pm 0.13$ & $26.0 \pm 0.10$ & $25.8 \pm 0.06$ & $25.9 \pm 0.08$ & $25.8 \pm 0.07$ & $26.1 \pm 0.08$ & $5.9^{+0.2}_{-0.4}$ & $55.4$ \\
MACS2129-1498\tnm{c} & $322.3581356$ & $-7.6993401$ & $28.0 \pm 0.25$ & $27.1 \pm 0.27$ & $26.9 \pm 0.15$ & $26.8 \pm 0.11$ & $27.1 \pm 0.18$ & $26.9 \pm 0.13$ & $27.1 \pm 0.16$ & $6.0^{+0.3}_{-0.7}$ & $5.0$ \\
MACS2129-1939 & $322.3676293$ & $-7.7080924$ & $28.0 \pm 0.28$ & $26.9 \pm 0.27$ & $27.0 \pm 0.20$ & $27.2 \pm 0.17$ & $26.9 \pm 0.18$ & $27.2 \pm 0.19$ & $26.9 \pm 0.15$ & $5.6^{+0.5}_{-5.2}$ & $1.3$ \\
MS2137-0300 & $325.0524519$ & $-23.6442799$ & $26.6 \pm 0.13$ & $25.6 \pm 0.13$ & $26.0 \pm 0.10$ & $26.3 \pm 0.09$ & $26.2 \pm 0.12$ & $26.2 \pm 0.11$ & $26.2 \pm 0.11$ & $5.7^{+0.2}_{-0.2}$ & $1.2$ \\
MS2137-0450\tnm{c} & $325.0834263$ & $-23.6521488$ & $25.4 \pm 0.05$ & $24.6 \pm 0.08$ & $24.2 \pm 0.03$ & \nodata & $24.0 \pm 0.02$ & $24.0 \pm 0.02$ & $24.1 \pm 0.03$ & $5.8^{+0.1}_{-0.2}$ & $1.4$ \\
MS2137-0593 & $325.0748296$ & $-23.6445635$ & $25.2 \pm 0.05$ & $24.6 \pm 0.09$ & $24.2 \pm 0.03$ & $24.2 \pm 0.02$ & \nodata & $24.2 \pm 0.03$ & $24.1 \pm 0.03$ & $5.7^{+0.2}_{-0.1}$ & $1.3$ \\
MS2137-0830\tnm{c} & $325.0504338$ & $-23.6609238$ & $26.3 \pm 0.13$ & $25.3 \pm 0.11$ & $25.0 \pm 0.05$ & $25.0 \pm 0.04$ & $24.8 \pm 0.05$ & $24.9 \pm 0.04$ & $24.8 \pm 0.04$ & $5.9^{+0.2}_{-0.2}$ & $1.8$ \\
MS2137-1555\tnm{c} & $325.0753169$ & $-23.6772512$ & $25.3 \pm 0.09$ & $24.6 \pm 0.08$ & $23.9 \pm 0.03$ & $23.9 \pm 0.01$ & \nodata & $23.8 \pm 0.03$ & $23.7 \pm 0.02$ & $6.1^{+0.1}_{-0.4}$ & $1.2$ \\
RXJ1347-0471 & $206.8871171$ & $-11.7450133$ & $27.6 \pm 0.28$ & $27.5 \pm 0.55$ & $25.9 \pm 0.11$ & $25.9 \pm 0.05$ & $25.9 \pm 0.11$ & $25.6 \pm 0.07$ & $25.7 \pm 0.06$ & $6.4^{+0.3}_{-0.3}$ & $5.5$ \\
RXJ1347-0649 & $206.874861$ & $-11.748161$ & $27.1 \pm 0.14$ & $26.3 \pm 0.18$ & $26.4 \pm 0.14$ & $26.5 \pm 0.07$ & $26.5 \pm 0.14$ & $26.7 \pm 0.14$ & $26.6 \pm 0.10$ & $5.5^{+0.2}_{-0.5}$ & $5.4$ \\
RXJ1347-0828 & $206.9030145$ & $-11.7503692$ & $29.1 \pm 0.66$ & $>28.3$ & $27.2 \pm 0.23$ & $27.2 \pm 0.11$ & $27.3 \pm 0.23$ & $27.1 \pm 0.17$ & $27.1 \pm 0.14$ & $6.5^{+0.9}_{-5.0}$ & $3.0$ \\
RXJ1347-0989 & $206.8953767$ & $-11.7494898$ & $27.8 \pm 0.53$ & $27.2 \pm 0.73$ & $25.3 \pm 0.11$ & $25.6 \pm 0.06$ & $25.6 \pm 0.14$ & $25.4 \pm 0.09$ & $25.2 \pm 0.06$ & $6.5^{+0.5}_{-5.5}$ & $10.2$ \\
RXJ1347-1202 & $206.8886406$ & $-11.7542822$ & $28.0 \pm 0.42$ & $26.1 \pm 0.22$ & $25.7 \pm 0.10$ & $25.8 \pm 0.05$ & $25.8 \pm 0.11$ & $25.8 \pm 0.08$ & $26.2 \pm 0.10$ & $6.3^{+0.3}_{-0.3}$ & $35.9$ \\
RXJ1347-1760 & $206.8904761$ & $-11.7646984$ & $28.4 \pm 0.38$ & $26.8 \pm 0.24$ & $27.0 \pm 0.19$ & $27.3 \pm 0.11$ & $27.6 \pm 0.29$ & $27.6 \pm 0.25$ & $27.5 \pm 0.18$ & $5.6^{+0.4}_{-5.6}$ & $4.1$ \\
RXCJ2248-0401\tnm{l} & $342.1712956$ & $-44.519812$ & $27.4 \pm 0.20$ & $25.9 \pm 0.13$ & $25.9 \pm 0.08$ & $26.2 \pm 0.07$ & $25.9 \pm 0.08$ & $26.2 \pm 0.09$ & $26.3 \pm 0.10$ & $6.0^{+0.2}_{-0.2}$ & $1.5$ \\
RXCJ2248-0619 & $342.1763786$ & $-44.5235886$ & $>29.3$ & $27.4 \pm 0.28$ & $28.5 \pm 0.42$ & $27.7 \pm 0.17$ & $28.0 \pm 0.29$ & $27.7 \pm 0.20$ & $28.0 \pm 0.26$ & $5.9^{+0.9}_{-5.7}$ & $2.5$ \\
RXCJ2248-0743\tnm{c} & $342.1733274$ & $-44.5259336$ & $29.6 \pm 0.61$ & $28.1 \pm 0.47$ & $27.8 \pm 0.21$ & $27.8 \pm 0.16$ & $27.8 \pm 0.22$ & $27.8 \pm 0.20$ & $27.7 \pm 0.19$ & $6.1^{+0.8}_{-5.6}$ & $2.6$ \\
RXCJ2248-0847 & $342.1890479$ & $-44.5300249$ & $25.2 \pm 0.06$ & $24.5 \pm 0.08$ & $24.3 \pm 0.03$ & $24.2 \pm 0.02$ & $24.3 \pm 0.04$ & $24.2 \pm 0.03$ & $24.1 \pm 0.03$ & $5.5^{+0.1}_{-0.1}$ & $3.0$ \\
RXCJ2248-0979 & $342.1893685$ & $-44.5309358$ & $28.3 \pm 0.37$ & $27.1 \pm 0.34$ & $27.1 \pm 0.17$ & $27.2 \pm 0.15$ & $27.0 \pm 0.17$ & $27.0 \pm 0.15$ & $27.3 \pm 0.21$ & $5.9^{+0.4}_{-5.8}$ & $5.5$ \\
RXCJ2248-1291\tnm{l} & $342.1908893$ & $-44.5374615$ & $25.9 \pm 0.07$ & $25.1 \pm 0.10$ & $24.8 \pm 0.04$ & $24.9 \pm 0.03$ & $24.9 \pm 0.04$ & $25.1 \pm 0.04$ & $25.2 \pm 0.04$ & $6.0^{+0.1}_{-0.1}$ & $2.6$ \\
RXCJ2248-1551 & $342.1679855$ & $-44.5425066$ & $27.0 \pm 0.14$ & $26.4 \pm 0.19$ & $26.3 \pm 0.10$ & $26.4 \pm 0.08$ & $26.3 \pm 0.11$ & $26.5 \pm 0.12$ & $26.5 \pm 0.11$ & $5.7^{+0.2}_{-0.4}$ & $58.2$ \\
RXCJ2248-1563 & $342.176127$ & $-44.5426723$ & $28.2 \pm 0.27$ & $26.9 \pm 0.23$ & $27.4 \pm 0.20$ & $27.5 \pm 0.16$ & $27.5 \pm 0.23$ & $27.5 \pm 0.19$ & $27.7 \pm 0.23$ & $5.6^{+0.3}_{-4.8}$ & $3.2$ \\
RXCJ2248-1704 & $342.1974342$ & $-44.546269$ & $28.1 \pm 0.29$ & $26.7 \pm 0.24$ & $27.3 \pm 0.19$ & $27.4 \pm 0.15$ & $27.2 \pm 0.19$ & $27.2 \pm 0.15$ & $27.6 \pm 0.22$ & $5.7^{+0.4}_{-5.6}$ & $1.4$ \\
RXCJ2248-1722 & $342.1777187$ & $-44.5469379$ & $28.3 \pm 0.41$ & $27.0 \pm 0.34$ & $26.6 \pm 0.14$ & $26.9 \pm 0.13$ & $26.9 \pm 0.19$ & $26.8 \pm 0.15$ & $26.8 \pm 0.14$ & $5.8^{+0.7}_{-0.8}$ & $1.7$ \\
RXCJ2248-1773 & $342.1696535$ & $-44.5485453$ & $28.2 \pm 0.32$ & $26.7 \pm 0.22$ & $26.7 \pm 0.13$ & $26.9 \pm 0.13$ & $26.6 \pm 0.13$ & $26.8 \pm 0.14$ & $26.8 \pm 0.14$ & $6.0^{+0.4}_{-0.4}$ & $1.9$
\enddata
\tablecomments{Magnitudes are expressed as observed (lensed) isophotal magnitudes (ISOMAG).}
\tablenotetext{a}{Photometric redshift estimate with $2\sigma$ (95\%) confidence intervals (see Section~\ref{sec:bpz}).  Objects with large lower bounds have a secondary peak at lower redshift ($z\sim1-2$) that contains at least $5\%$ of the posterior probability.}
\tablenotetext{b}{Magnification estimate from the lens models (see Section~\ref{sec:models}).  Because of uncertainties in the precise location of the critical curves, objects with magnifications $>100$ are simply quoted as such.  Objects outside the region constrained by the strong lensing models have been assigned a magnification of $1.1$.}
\tablenotetext{c}{Unresolved object with FWHM $<0.22$ arcsec in the image plane.  There is a small chance that brighter unresolved candidates could be low-mass stars even though we explored such possibilities (see Section~\ref{sec:contaminants}).}
\tablenotetext{d}{Using LePhare, we find that these unresolved candidates have a good fit with stellar templates, with $\chi^{2}_{\mathrm{star}}$ comparable or less than $\chi^{2}_{\mathrm{galaxy}}$.}
\tablenotetext{e}{While BPZ prefers a high-redshift solution, we note that LePhare slightly prefers a low-redshift solution ($z\sim1$) over the high-redshift solution for these three galaxies.  Given this and the possible fit with stellar templates, these candidates should be considered less confident than the others.}
\tablenotetext{f}{Spectroscopically confirmed multiply-imaged galaxy at $z=6.027$ \citep{Richard2011}.}
\tablenotetext{g}{Quadruply lensed galaxy at $z\sim6.2$ \citep{Zitrin2012m0329}.}
\tablenotetext{h}{\cite{Zitrin2013m0416} report that MACS0419-0419 is part of a visually identified double system at $z_{\mathrm{phot}} \sim 6.1$.  The second candidate (at R.A. = 04:16:09.946, Dec = $-$24:03:45.31) fell below our S/N threshold and thus does not appear in this catalog.}
\tablenotetext{i}{MACS1115-0352 and MACS1720-1114 have very blue SEDs.  While our best-fit photometric redshifts suggest that these are high-redshift candidates, a possible alternative solution is that they could be low-redshift extreme emission-line galaxies with rest-frame equivalent widths of $\sim2000$~\AA\ (X. Huang et al. 2014, submitted).}
\tablenotetext{j}{Quadruply lensed galaxy at $z_{\mathrm{phot}}\sim5.6$ ($z_{spec}=5.701$) \citep{Zitrin2012macs1206}.}
\tablenotetext{k}{VLT/VIMOS spectroscopy confirms this galaxy at $z=5.701$ (see Section~\ref{sec:selection}).}
\tablenotetext{l}{\cite{Monna2014} found that these two candidates, along with three others, are part of a quintuply-lensed system with $z_{\mathrm{phot}} \sim5.9$.  Based on their lens model, the magnifications for RXCJ2248-0401 and RXCJ2248-1291 (ID4 and ID3, respectively in \cite{Monna2014} are $2.4\pm 0.2$ and $6.0\pm1.5$, respectively.}
\tablenotetext{m}{In the recent Hubble Frontier Fields data, this candidate is detected in the ultradeep optical data and therefore it is very unlikely to be a high-redshift galaxy.}
\tablenotetext{n}{MACS0717-0859 and MACS0717-1730 are spectroscopically confirmed at $z=6.387$ \citep{Vanzella2014}.}
\label{tbl:z6}
\ifemulateapj
    \end{deluxetable*}
    \setlength{\tabcolsep}{0.07in}
\else
    \end{deluxetable}
\fi
}

\def\tabzseven {
\ifemulateapj
    \tabletypesize{\tiny}
    \begin{deluxetable*}{lccccccccccc}
\else
    \begin{deluxetable}{lccccccccccc}
\fi
\tablecolumns{12}
\tablewidth{0pt}
\tablecaption{Lensed $z\sim7$ Candidates Identified Behind 17 CLASH Clusters}
\tablehead{\colhead{Object ID} & \colhead{$\alpha_{J2000}$} & \colhead{$\delta_{J2000}$} & \colhead{$I_{814}$} & \colhead{$z_{850}$} & \colhead{$Y_{105}$} & \colhead{$Y_{110}$} & \colhead{$J_{125}$} & \colhead{$J_{140}$} & \colhead{$H_{160}$} & \colhead{$z_{\mathrm{phot}}$\tnm{a}} & \colhead{$\mu$\tnm{b}} }
\startdata
A2261-0450 & $260.6124593$ & $32.1438429$ & $>28.8$ & $26.6 \pm 0.31$ & $25.6 \pm 0.06$ & $25.5 \pm 0.04$ & $25.5 \pm 0.06$ & $25.5 \pm 0.05$ & $25.5 \pm 0.06$ & $6.8^{+0.2}_{-0.3}$ & $5.6$ \\
A2261-0731 & $260.6232556$ & $32.1393984$ & $>29.7$ & $>28.9$ & $28.1 \pm 0.27$ & $28.2 \pm 0.24$ & $28.1 \pm 0.28$ & $27.7 \pm 0.18$ & $27.9 \pm 0.22$ & $6.9^{+1.0}_{-5.9}$ & $7.7$ \\
A2261-0772 & $260.6059024$ & $32.1388049$ & $>29.2$ & $>28.4$ & $27.3 \pm 0.18$ & $27.6 \pm 0.18$ & $27.7 \pm 0.26$ & $27.6 \pm 0.21$ & $27.4 \pm 0.19$ & $6.5^{+0.8}_{-5.4}$ & $6.3$ \\
A2261-1559\tnm{c} & $260.6006401$ & $32.1243407$ & $>28.9$ & $27.8 \pm 0.45$ & $26.7 \pm 0.11$ & $27.1 \pm 0.11$ & $26.8 \pm 0.13$ & $27.0 \pm 0.13$ & $27.0 \pm 0.14$ & $6.5^{+0.5}_{-0.4}$ & $2.7$ \\
A383-1734 & $42.0224155$ & $-3.5407973$ & $>29.4$ & $>28.3$ & $27.4 \pm 0.16$ & $27.8 \pm 0.19$ & $27.4 \pm 0.17$ & $27.3 \pm 0.16$ & $27.9 \pm 0.24$ & $6.8^{+0.7}_{-0.7}$ & $3.2$ \\
A383-2079\tnm{c,d} & $42.0132618$ & $-3.5490106$ & $>29.1$ & $>27.9$ & $26.4 \pm 0.13$ & \nodata & $25.9 \pm 0.07$ & $25.9 \pm 0.08$ & $26.4 \pm 0.14$ & $7.4^{+0.4}_{-0.5}$ & $3.1$ \\
A611-0890\tnm{c} & $120.2528024$ & $36.055796$ & $>29.4$ & $>28.6$ & $27.8 \pm 0.24$ & $27.7 \pm 0.18$ & $28.1 \pm 0.33$ & $27.4 \pm 0.16$ & $27.6 \pm 0.20$ & $6.7^{+0.9}_{-5.8}$ & $1.8$ \\
A611-1213 & $120.2156808$ & $36.0487504$ & $>28.3$ & $>28.1$ & $27.6 \pm 0.25$ & $27.0 \pm 0.13$ & $27.2 \pm 0.19$ & $26.8 \pm 0.13$ & $27.2 \pm 0.18$ & $6.6^{+1.0}_{-6.0}$ & $1.4$ \\
MACS0329-0388 & $52.4311631$ & $-2.185244$ & $>27.8$ & $>27.4$ & $26.4 \pm 0.16$ & $26.2 \pm 0.09$ & $26.3 \pm 0.15$ & $26.2 \pm 0.11$ & $26.3 \pm 0.12$ & $6.6^{+0.7}_{-5.1}$ & $4.2$ \\
MACS0329-0714 & $52.4406209$ & $-2.1911504$ & $>29.2$ & $>28.4$ & $27.7 \pm 0.28$ & $27.3 \pm 0.16$ & $27.9 \pm 0.34$ & $27.6 \pm 0.22$ & $27.4 \pm 0.21$ & $6.6^{+0.8}_{-6.2}$ & $3.2$ \\
MACS0329-1853 & $52.4180851$ & $-2.2124641$ & $>29.0$ & $>28.3$ & $27.3 \pm 0.24$ & $26.9 \pm 0.14$ & $26.9 \pm 0.17$ & $27.2 \pm 0.20$ & $27.2 \pm 0.19$ & $6.8^{+0.7}_{-0.7}$ & $5.5$ \\
MACS0416-0036 & $64.0260447$ & $-24.0509958$ & $>29.2$ & $>27.7$ & $27.3 \pm 0.34$ & $27.3 \pm 0.20$ & \nodata & \nodata & $26.8 \pm 0.16$ & $7.0^{+1.2}_{-6.0}$ & $1.3$ \\
MACS0416-2028\tnm{c,e} & $64.0335661$ & $-24.0968499$ & $26.6 \pm 0.16$ & $25.6 \pm 0.15$ & $23.9 \pm 0.03$ & \nodata & $23.4 \pm 0.01$ & $23.5 \pm 0.01$ & $23.6 \pm 0.02$ & $7.2^{+0.1}_{-0.1}$ & $1.5$ \\
MACS0647-0169 & $101.9940284$ & $70.264168$ & $28.6 \pm 0.48$ & $26.2 \pm 0.20$ & $25.4 \pm 0.09$ & $25.2 \pm 0.03$ & \nodata & $25.0 \pm 0.04$ & $25.1 \pm 0.05$ & $6.6^{+0.5}_{-0.2}$ & $4.7$ \\
MACS0647-0424 & $101.9364959$ & $70.2605516$ & $>29.2$ & $>28.5$ & $28.1 \pm 0.33$ & $27.0 \pm 0.10$ & $27.4 \pm 0.19$ & $27.5 \pm 0.18$ & $27.5 \pm 0.18$ & $6.6^{+0.7}_{-0.6}$ & $19.2$ \\
MACS0647-1051\tnm{f} & $101.9209905$ & $70.2515462$ & $28.9 \pm 0.56$ & $27.8 \pm 0.68$ & $26.5 \pm 0.14$ & $26.2 \pm 0.08$ & $26.4 \pm 0.13$ & $26.7 \pm 0.13$ & $26.7 \pm 0.14$ & $6.5^{+0.4}_{-0.3}$ & $32.6$ \\
MACS0647-1205 & $101.9194993$ & $70.2490521$ & $>29.3$ & $27.5 \pm 0.59$ & $26.7 \pm 0.16$ & $26.1 \pm 0.07$ & $26.4 \pm 0.13$ & $26.2 \pm 0.09$ & $26.1 \pm 0.09$ & $6.7^{+0.6}_{-0.4}$ & $75.5$ \\
MACS0647-1541\tnm{f} & $101.9215164$ & $70.2429165$ & $>29.7$ & $28.0 \pm 0.57$ & $27.2 \pm 0.18$ & $27.5 \pm 0.16$ & $27.2 \pm 0.18$ & $27.5 \pm 0.19$ & $27.7 \pm 0.22$ & $6.6^{+0.6}_{-0.6}$ & $48.8$ \\
MACS0647-1825 & $101.9964309$ & $70.237579$ & $>29.5$ & $>27.5$ & $28.2 \pm 0.52$ & $27.1 \pm 0.16$ & $27.1 \pm 0.21$ & $27.0 \pm 0.17$ & $27.0 \pm 0.18$ & $7.2^{+1.0}_{-6.0}$ & $14.3$ \\
MACS0647-2175 & $101.983916$ & $70.2294832$ & $>28.9$ & $>28.1$ & $27.6 \pm 0.25$ & $27.0 \pm 0.12$ & $27.4 \pm 0.24$ & $27.6 \pm 0.23$ & $27.5 \pm 0.20$ & $6.5^{+0.7}_{-0.6}$ & $2.1$ \\
MACS0647-2263 & $101.965785$ & $70.2265848$ & $28.7 \pm 0.35$ & $>27.8$ & $26.8 \pm 0.16$ & \nodata & $26.5 \pm 0.11$ & $27.0 \pm 0.17$ & $27.3 \pm 0.24$ & $6.5^{+0.6}_{-0.4}$ & $2.8$ \\
MACS0744-0225\tnm{g} & $116.2208608$ & $39.4736635$ & $>29.2$ & $>28.4$ & $27.9 \pm 0.31$ & $27.2 \pm 0.14$ & $27.4 \pm 0.23$ & $27.6 \pm 0.21$ & $27.9 \pm 0.30$ & $6.6^{+0.8}_{-5.2}$ & $5.0$ \\
MACS0744-1182 & $116.2064105$ & $39.4583017$ & $>29.3$ & $>27.5$ & $27.7 \pm 0.32$ & $27.8 \pm 0.26$ & $27.4 \pm 0.26$ & $27.5 \pm 0.23$ & $27.4 \pm 0.22$ & $6.7^{+1.3}_{-6.0}$ & $10.8$ \\
MACS0744-1590 & $116.2504125$ & $39.4530105$ & $>29.0$ & $27.4 \pm 0.48$ & $26.1 \pm 0.13$ & \nodata & $26.0 \pm 0.10$ & $26.0 \pm 0.08$ & $25.7 \pm 0.10$ & $6.7^{+0.5}_{-0.4}$ & $3.2$ \\
MACS0744-1695\tnm{g} & $116.21974$ & $39.4533508$ & $28.7 \pm 0.50$ & $27.5 \pm 0.54$ & $26.1 \pm 0.09$ & $26.5 \pm 0.11$ & $26.1 \pm 0.11$ & $26.2 \pm 0.08$ & $26.5 \pm 0.13$ & $6.6^{+0.4}_{-0.3}$ & $16.5$ \\
MACS1115-0850 & $168.9677412$ & $1.5003116$ & $>28.3$ & $27.4 \pm 0.57$ & $27.7 \pm 0.45$ & $26.1 \pm 0.09$ & $26.2 \pm 0.14$ & $26.9 \pm 0.23$ & $26.7 \pm 0.18$ & $7.2^{+0.5}_{-5.8}$ & $57.4$ \\
MACS1206-0777 & $181.537366$ & $-8.7956188$ & $>29.0$ & $>28.1$ & $27.9 \pm 0.40$ & $27.2 \pm 0.18$ & $27.2 \pm 0.23$ & $27.3 \pm 0.23$ & $27.1 \pm 0.21$ & $7.2^{+0.9}_{-6.3}$ & $2.8$ \\
MACS1720-1003 & $260.0507479$ & $35.6090866$ & $>29.0$ & $>27.4$ & $26.9 \pm 0.16$ & $27.3 \pm 0.16$ & $27.0 \pm 0.19$ & $26.6 \pm 0.11$ & $27.1 \pm 0.19$ & $6.6^{+0.9}_{-0.7}$ & $1.6$ \\
MACS1720-1047 & $260.081558$ & $35.6081686$ & $28.8 \pm 0.54$ & $27.6 \pm 0.46$ & $26.8 \pm 0.15$ & $26.7 \pm 0.09$ & $26.4 \pm 0.11$ & $26.5 \pm 0.10$ & $26.9 \pm 0.14$ & $6.6^{+0.7}_{-0.7}$ & $2.4$ \\
MACS1931-0217\tnm{c,d} & $292.9524953$ & $-26.5592239$ & $>28.5$ & $27.1 \pm 0.42$ & $26.3 \pm 0.10$ & $26.1 \pm 0.07$ & $25.8 \pm 0.08$ & $25.9 \pm 0.07$ & $25.7 \pm 0.06$ & $6.8^{+0.6}_{-0.7}$ & $3.3$ \\
MACS1931-0513 & $292.9768277$ & $-26.5653146$ & $>28.7$ & $>27.4$ & $26.8 \pm 0.24$ & $26.7 \pm 0.13$ & \nodata & \nodata & $26.6 \pm 0.16$ & $6.7^{+1.0}_{-0.9}$ & $1.3$ \\
MACS1931-0649 & $292.9528795$ & $-26.5675066$ & $>29.1$ & $>28.2$ & $27.0 \pm 0.16$ & $27.6 \pm 0.20$ & $26.6 \pm 0.13$ & $26.9 \pm 0.13$ & $27.2 \pm 0.17$ & $7.3^{+0.5}_{-1.1}$ & $25.5$ \\
MACS1931-1359 & $292.9353237$ & $-26.5790513$ & $>28.6$ & $>27.8$ & $26.4 \pm 0.13$ & $26.4 \pm 0.10$ & $26.8 \pm 0.21$ & $26.6 \pm 0.15$ & $26.5 \pm 0.14$ & $6.6^{+0.5}_{-0.4}$ & $1.3$ \\
MACS2129-0218 & $322.3508469$ & $-7.6752448$ & $>29.3$ & $>28.0$ & $27.2 \pm 0.23$ & $26.7 \pm 0.11$ & $27.5 \pm 0.29$ & $27.2 \pm 0.19$ & $26.7 \pm 0.13$ & $6.5^{+0.6}_{-0.6}$ & $1.3$ \\
MACS2129-1064 & $322.3716472$ & $-7.6910304$ & $>29.6$ & $>27.9$ & $27.6 \pm 0.27$ & $28.2 \pm 0.34$ & $27.7 \pm 0.29$ & $27.5 \pm 0.20$ & $27.8 \pm 0.26$ & $6.7^{+1.3}_{-5.9}$ & $5.2$ \\
MACS2129-1408 & $322.3532384$ & $-7.6974442$ & $>28.7$ & $>28.1$ & $26.9 \pm 0.17$ & $26.9 \pm 0.12$ & $27.0 \pm 0.18$ & $27.2 \pm 0.17$ & $27.1 \pm 0.17$ & $6.6^{+0.5}_{-0.4}$ & $7.0$ \\
MS2137-0207 & $325.0558387$ & $-23.6455503$ & $29.1 \pm 0.74$ & $>27.6$ & $26.5 \pm 0.11$ & $26.5 \pm 0.08$ & $26.5 \pm 0.12$ & $26.6 \pm 0.12$ & $26.5 \pm 0.11$ & $6.7^{+0.4}_{-0.4}$ & $1.3$ \\
RXJ1347-0186 & $206.8964092$ & $-11.7377958$ & $>29.2$ & $>27.8$ & $26.9 \pm 0.31$ & \nodata & $26.6 \pm 0.16$ & $26.9 \pm 0.21$ & $26.8 \pm 0.19$ & $6.9^{+1.2}_{-5.6}$ & $4.6$ \\
RXJ1347-0310 & $206.8760007$ & $-11.7411934$ & $>29.3$ & $>28.2$ & $26.8 \pm 0.19$ & $27.1 \pm 0.11$ & $27.1 \pm 0.23$ & $27.0 \pm 0.17$ & $27.2 \pm 0.16$ & $6.7^{+0.6}_{-0.6}$ & $8.2$ \\
RXJ1347-0346 & $206.8823155$ & $-11.7421821$ & $>29.2$ & $28.0 \pm 0.66$ & $26.4 \pm 0.14$ & $26.6 \pm 0.07$ & $26.3 \pm 0.12$ & $26.6 \pm 0.13$ & $27.2 \pm 0.17$ & $6.7^{+0.4}_{-0.3}$ & $71.3$ \\
RXJ1347-1046 & $206.9008589$ & $-11.7542095$ & $>28.8$ & $26.8 \pm 0.31$ & $26.2 \pm 0.12$ & $26.1 \pm 0.05$ & $25.9 \pm 0.09$ & $26.1 \pm 0.09$ & $26.1 \pm 0.07$ & $6.8^{+0.4}_{-0.5}$ & $2.7$ \\
RXJ1347-1316 & $206.8769757$ & $-11.7576773$ & $>29.3$ & $27.7 \pm 0.49$ & $27.3 \pm 0.27$ & $27.1 \pm 0.10$ & $26.6 \pm 0.15$ & $27.0 \pm 0.16$ & $26.9 \pm 0.12$ & $6.6^{+1.1}_{-0.9}$ & $2.6$ \\
RXJ1347-2025 & $206.8760786$ & $-11.7729955$ & $>29.7$ & $28.1 \pm 0.58$ & $27.9 \pm 0.42$ & $27.5 \pm 0.14$ & $27.4 \pm 0.21$ & $27.7 \pm 0.26$ & $27.5 \pm 0.17$ & $6.6^{+1.1}_{-5.2}$ & $5.7$ \\
\enddata
\tablecomments{Magnitudes are expressed as observed (lensed) isophotal magnitudes (ISOMAG).}
\tablenotetext{a}{Photometric redshift estimate with $2\sigma$ (95\%) confidence intervals (see Section~\ref{sec:bpz}).  Objects with large lower bounds have a secondary peak at lower redshift ($z\sim1-2$) that contains at least $5\%$ of the posterior probability.}
\tablenotetext{b}{Magnification estimate from the lens models (see Section~\ref{sec:models}).  Because of uncertainties in the precise location of the critical curves, objects with magnifications $>100$ are simply quoted as such.  Objects outside the region constrained by the strong lensing models have been assigned a magnification of $1.1$.}
\tablenotetext{c}{Unresolved object with FWHM $<0.22$ arcsec in the image plane.  There is a small chance that brighter unresolved candidates could be low-mass stars even though we explored such possibilities (see Section~\ref{sec:contaminants}).}
\tablenotetext{d}{Using LePhare, we find that these unresolved candidates have a good fit with stellar templates, with $\chi^{2}_{\mathrm{star}}$ comparable or less than $\chi^{2}_{\mathrm{galaxy}}$.}
\tablenotetext{e}{Despite having a best-fit photometric redshift of $z_{\mathrm{phot}} = 7.2$, we suspect this unresolved object is most likely a star.  Based on our current understanding of the $z\sim7$ LF \citep[e.g.,][]{Bouwens2011b}, the probability of detecting a slightly-magnified $z\sim7$ galaxy with $\Hband = 23.6$ in the small area covered by the clusters in this paper is exceedingly small.  We list this candidate for completeness, but do not use it for subsequent analysis given its suspect nature.}
\tablenotetext{f}{The lens models suggest that this is likely a multiple system at $z\sim6.5$.}
\tablenotetext{g}{The lens models suggest that this is likely a multiple system at $z\sim6.6$.}
\label{tbl:z7}
\ifemulateapj
    \end{deluxetable*}
\else
    \end{deluxetable}
\fi
}

\def\tabzeight {
\ifemulateapj
    \tabletypesize{\tiny}
    \begin{deluxetable*}{lccccccccccc}
\else
    \begin{deluxetable}{lccccccccccc}
\fi
\tablecolumns{12}
\tablewidth{0pt}
\tablecaption{Lensed $z\sim8$ Candidates Identified Behind 17 CLASH Clusters}
\tablehead{\colhead{Object ID} & \colhead{$\alpha_{J2000}$} & \colhead{$\delta_{J2000}$} & \colhead{$I_{814}$} & \colhead{$z_{850}$} & \colhead{$Y_{105}$} & \colhead{$Y_{110}$} & \colhead{$J_{125}$} & \colhead{$J_{140}$} & \colhead{$H_{160}$} & \colhead{$z_{\mathrm{phot}}$\tnm{a}} & \colhead{$\mu$\tnm{b}} }
\startdata
A2261-0187 & $260.6073833$ & $32.1495175$ & $>29.3$ & $>28.4$ & $27.4 \pm 0.19$ & $27.3 \pm 0.13$ & $26.8 \pm 0.13$ & $26.8 \pm 0.11$ & $27.0 \pm 0.13$ & $7.5^{+0.4}_{-1.2}$ & $2.9$ \\
A383-2311 & $42.0149347$ & $-3.5526143$ & $>29.3$ & $28.0 \pm 0.60$ & $27.7 \pm 0.35$ & \nodata & $26.9 \pm 0.15$ & $26.7 \pm 0.16$ & $27.2 \pm 0.23$ & $7.5^{+0.9}_{-6.4}$ & $4.5$ \\
A611-0193 & $120.2513506$ & $36.0734262$ & $28.5 \pm 0.54$ & $28.1 \pm 0.74$ & $27.0 \pm 0.19$ & \nodata & $26.0 \pm 0.09$ & $26.0 \pm 0.10$ & $26.0 \pm 0.17$ & $7.9^{+0.3}_{-6.5}$ & $1.6$ \\
MACS0416-1830\tnm{c} & $64.0379772$ & $-24.0888285$ & $>29.1$ & $>28.7$ & $29.0 \pm 0.58$ & $28.6 \pm 0.32$ & $28.0 \pm 0.25$ & $27.8 \pm 0.19$ & $28.2 \pm 0.26$ & $8.1^{+0.7}_{-7.7}$ & $1.6$ \\
MACS0647-1411 & $101.9204414$ & $70.2448881$ & $28.9 \pm 0.58$ & $>27.9$ & $26.7 \pm 0.18$ & $26.6 \pm 0.11$ & $26.1 \pm 0.10$ & $26.0 \pm 0.08$ & $26.1 \pm 0.08$ & $7.6^{+0.5}_{-6.3}$ & $>100$ \\
MACS1115-0356 & $168.962901$ & $1.5097067$ & $28.4 \pm 0.62$ & $>27.9$ & $27.2 \pm 0.25$ & $27.0 \pm 0.18$ & $26.3 \pm 0.13$ & $26.1 \pm 0.10$ & $26.1 \pm 0.10$ & $8.0^{+0.4}_{-6.7}$ & $12.0$ \\
MACS1206-1581 & $181.5512693$ & $-8.8084446$ & $>29.4$ & $>28.7$ & $28.8 \pm 0.53$ & $28.6 \pm 0.31$ & $27.8 \pm 0.24$ & $28.0 \pm 0.24$ & $27.8 \pm 0.20$ & $7.9^{+0.8}_{-7.0}$ & $4.7$ \\
MACS1720-0696\tnm{d,e} & $260.0466407$ & $35.6144436$ & $>28.2$ & $>27.9$ & $26.5 \pm 0.15$ & $25.9 \pm 0.05$ & \nodata & $25.6 \pm 0.07$ & $25.8 \pm 0.09$ & $7.7^{+0.2}_{-1.0}$ & $1.3$ \\
MACS1720-1231 & $260.0530519$ & $35.6047895$ & $>29.6$ & $>28.7$ & $29.1 \pm 0.58$ & $28.3 \pm 0.23$ & $27.9 \pm 0.24$ & $27.6 \pm 0.16$ & $27.7 \pm 0.19$ & $8.3^{+0.6}_{-7.1}$ & $1.7$ \\
MACS1931-1990 & $292.9578406$ & $-26.5913165$ & $>28.0$ & $>27.2$ & $25.9 \pm 0.11$ & $26.0 \pm 0.09$ & $25.5 \pm 0.08$ & $25.4 \pm 0.07$ & $25.5 \pm 0.07$ & $7.5^{+0.3}_{-0.9}$ & $27.8$ \\
RXJ1347-0943 & $206.8912473$ & $-11.7526045$ & $>29.1$ & $>28.1$ & $27.3 \pm 0.31$ & $26.6 \pm 0.08$ & $26.4 \pm 0.14$ & $26.4 \pm 0.12$ & $26.5 \pm 0.10$ & $7.5^{+0.5}_{-1.0}$ & $>100$ \\
RXCJ2248-1301 & $342.208379$ & $-44.5375171$ & $>28.7$ & $>28.0$ & $27.4 \pm 0.32$ & $27.1 \pm 0.13$ & \nodata & $26.5 \pm 0.14$ & $26.4 \pm 0.13$ & $7.7^{+0.6}_{-6.5}$ & $1.4$
\enddata
\tablecomments{Magnitudes are expressed as observed (lensed) isophotal magnitudes (ISOMAG).}
\tablenotetext{a}{Photometric redshift estimate with $2\sigma$ (95\%) confidence intervals (see Section~\ref{sec:bpz}).  Objects with large lower bounds have a secondary peak at lower redshift ($z\sim1-2$) that contains at least $5\%$ of the posterior probability.}
\tablenotetext{b}{Magnification estimate from the lens models (see Section~\ref{sec:models}).  Because of uncertainties in the precise location of the critical curves, objects with magnifications $>100$ are simply quoted as such.}
\tablenotetext{c}{In the recent Hubble Frontier Fields data, this candidate is detected in the ultradeep optical data and therefore it is very unlikely to be a high-redshift galaxy.}
\tablenotetext{d}{Unresolved object with FWHM $<0.22$ arcsec in the image plane.  There is a small chance that brighter unresolved candidates could be low-mass stars even though we explored such possibilities (see Section~\ref{sec:contaminants}).}
\tablenotetext{e}{Using LePhare, we find that this unresolved candidate has a good fit with stellar templates, with $\chi^{2}_{\mathrm{star}}$ comparable or less than $\chi^{2}_{\mathrm{galaxy}}$.}
\label{tbl:z8}
\ifemulateapj
    \end{deluxetable*}
\else
    \end{deluxetable}
\fi
}

\def\tabnobj {
\ifemulateapj
    \tabletypesize{\scriptsize}
    \begin{deluxetable*}{lcccccccc}
\else
    \tabletypesize{\scriptsize}
    \begin{deluxetable}{lccccccccc}
\fi
\tablecolumns{9}
\tablewidth{0pt}
\tablecaption{Number of High-Redshift Candidates at $z\sim6$, $z\sim7$, and $z\sim8$ and Area Surveyed at $z\sim6$ at Low, Intermediate, and High Magnification}
\tablehead{\colhead{Cluster} &
\multicolumn{3}{c}{Number of Candidates\tnm{a}} &
\multicolumn{4}{c}{Area (arcmin$^2$)\tnm{b}} &
\colhead{$\mu_{\mathrm{med}}$\tnm{c}} \\
\cline{5-8}
\colhead{} &
\colhead{$z\sim6$} &
\colhead{$z\sim7$} &
\colhead{$z\sim8$} &
\colhead{$\mu \le 2.3$} &
\colhead{$2.3 < \mu < 5.4$} &
\colhead{$\mu \ge 5.4$} &
\colhead{Total} &
\colhead{}
}
\startdata
Abell 383                 &   3 &  2 &  1 & 0.0   &  3.2 &  1.2 &  4.4 &  4.4 \\
MACSJ1149.6+2223\tnm{d}   &  12 &  0 &  0 & 0.0   &  1.2 &  3.3 &  4.5 &  9.8 \\
Abell 2261                &  10 &  4 &  1 & 1.8   &  1.8 &  0.8 &  4.4 &  2.5 \\
MACSJ1206.2-0847          &  11 &  1 &  1 & 1.5   &  2.0 &  1.1 &  4.6 &  2.9 \\
RXJ1347.5-1145            &   6 &  6 &  1 & 0.3   &  1.9 &  2.1 &  4.3 &  5.2 \\
MACSJ2129.4-0741\tnm{d,e} &  10 &  3 &  0 & 2.3   &  1.6 &  0.8 &  4.7 &  2.3 \\
MACSJ0329.7-0211          &   9 &  3 &  0 & 0.4   &  2.6 &  1.5 &  4.5 &  3.9 \\
MS2137-2353               &   5 &  1 &  0 & 3.9   &  0.6 &  0.3 &  4.8 &  1.5 \\
MACSJ0717.5+3745\tnm{d}   &  15 &  0 &  0 & 0.1   &  1.7 &  2.5 &  4.3 &  6.5 \\
MACSJ0744.9+3927          &  17 &  3 &  0 & 0.0   &  0.7 &  4.0 &  4.6 &  9.4 \\
MACSJ0647.8+7015\tnm{d}   &  24 &  7 &  1 & 0.02  &  0.8 &  3.6 &  4.4 & 15.4 \\
MACSJ1115.9+0129          &   6 &  1 &  1 & 0.9   &  2.5 &  1.3 &  4.6 &  3.3 \\
Abell 611                 &   4 &  2 &  1 & 3.6   &  0.9 &  0.3 &  4.7 &  1.6 \\
RXJ1532.9+3021\tnm{f}     &  16 &  4 &  1 & \nodata  & \nodata & \nodata &  \nodata &  \nodata \\
MACSJ1720.3+3536          &   9 &  2 &  2 & 2.7   &  1.1 &  0.6 &  4.4 &  1.9 \\
MACSJ1931.8-2635          &  32 &  4 &  1 & 2.6   &  1.3 &  0.6 &  4.5 &  2.1 \\
MACSJ0416.1-2403          &   5 &  2 &  1 & 2.7   &  1.1 &  0.8 &  4.6 &  2.0 \\
RXCJ2248.7-4431           &  10 &  0 &  1 & 2.8   &  1.0 &  0.8 &  4.6 &  1.9 \\
\hline
Total                     & 204 & 45 & 13 & 25.7  & 25.9 & 25.6 & 76.9 &  3.4 \\
\enddata
\tablenotetext{a}{The number of unique high-redshift candidates.  Candidates that our cluster models find are likely multiple images are counted only once.}
\tablenotetext{b}{WFC3/IR coverage area at $z\sim6$ at low ($\mu \le 2.3$), intermediate ($2.3 < \mu < 5.4$), and high ($\mu \ge 5.4$) magnification, defined such that each magnification bin has approximately the same total area over the 18 clusters.  We exclude the search area behind the clusters lost as a result of intervening foreground sources.  Because of the weak redshift dependence on $d_{ls}/d_{s}$ between $z\sim6$ and $z\sim8$, the areas surveyed at $z\sim7$ and $z\sim8$ are very similar to those at $z\sim6$.}
\tablenotetext{c}{Median cluster magnification for $z\sim6$ over the WFC3/IR coverage area excluding foreground sources.}
\tablenotetext{d}{High magnification cluster.}
\tablenotetext{e}{This cluster was selected as one of the five CLASH high magnification clusters, but our lens modeling indicates the overall magnification strength of this cluster is more similar to the X-ray selected CLASH clusters.}
\tablenotetext{f}{For completeness, we report the high-redshift candidates identified behind this cluster (see Appendix~\ref{sec:rxj1532}), but we do not include them in the analysis of this paper given the uncertain nature of the strong lensing model for this cluster (see Section~\ref{sec:models}).}
\label{tbl:nobj}
\ifemulateapj
    \end{deluxetable*}
\else
    \end{deluxetable}
\fi
}

\def\tabrxjonefive {
\ifemulateapj
    \tabletypesize{\tiny}
    \begin{deluxetable*}{lccccccccccc}
\else
    \begin{deluxetable}{lccccccccccc}
\fi
\tablecolumns{12}
\tablewidth{0pt}
\tablecaption{Lensed Candidates Identified Behind RXJ1532.9+3021}
\tablehead{\colhead{Object ID} & \colhead{$\alpha_{J2000}$} & \colhead{$\delta_{J2000}$} & \colhead{$I_{814}$} & \colhead{$z_{850}$} & \colhead{$Y_{105}$} & \colhead{$Y_{110}$} & \colhead{$J_{125}$} & \colhead{$J_{140}$} & \colhead{$H_{160}$} & \colhead{$z_{\mathrm{phot}}$\tnm{a}} & \colhead{$\mu$\tnm{b}} }
\startdata
RXJ1532-0080 & $233.2324699$ & $30.3692524$ & $28.7 \pm 0.54$ & $26.9 \pm 0.30$ & $26.5 \pm 0.17$ & \nodata & $26.6 \pm 0.13$ & $26.5 \pm 0.13$ & $26.6 \pm 0.17$ & $6.3^{+0.4}_{-5.3}$ & $5.0$ \\
RXJ1532-0324 & $233.2474354$ & $30.3620154$ & $26.7 \pm 0.19$ & $26.3 \pm 0.28$ & $26.0 \pm 0.16$ & $25.3 \pm 0.05$ & \nodata & $25.9 \pm 0.17$ & $25.7 \pm 0.13$ & $6.2^{+0.2}_{-5.9}$ & $2.5$ \\
RXJ1532-0390 & $233.2245355$ & $30.3604005$ & $>28.6$ & $28.2 \pm 0.73$ & $27.5 \pm 0.25$ & $27.1 \pm 0.13$ & $27.6 \pm 0.28$ & $27.4 \pm 0.19$ & $27.6 \pm 0.24$ & $6.3^{+0.8}_{-5.5}$ & $4.0$ \\
RXJ1532-0471 & $233.2370996$ & $30.3589943$ & $28.1 \pm 0.34$ & $>27.7$ & $26.8 \pm 0.15$ & $27.1 \pm 0.15$ & $26.7 \pm 0.15$ & $26.9 \pm 0.15$ & $26.8 \pm 0.14$ & $6.1^{+0.8}_{-5.7}$ & $4.5$ \\
RXJ1532-0488 & $233.2463449$ & $30.358271$ & $27.5 \pm 0.17$ & $26.8 \pm 0.19$ & $26.9 \pm 0.17$ & $27.3 \pm 0.14$ & \nodata & $26.9 \pm 0.14$ & $27.4 \pm 0.24$ & $5.5^{+0.3}_{-0.7}$ & $3.8$ \\
RXJ1532-0594 & $233.2336512$ & $30.3554857$ & $28.5 \pm 0.43$ & $26.9 \pm 0.22$ & $26.9 \pm 0.14$ & $27.0 \pm 0.11$ & $27.7 \pm 0.27$ & $27.8 \pm 0.26$ & $27.3 \pm 0.17$ & $5.9^{+0.4}_{-0.3}$ & $4.3$ \\
RXJ1532-0682 & $233.2312664$ & $30.3542801$ & $27.4 \pm 0.29$ & $26.7 \pm 0.34$ & $26.2 \pm 0.12$ & $26.5 \pm 0.12$ & $26.7 \pm 0.19$ & $26.6 \pm 0.15$ & $26.6 \pm 0.15$ & $6.0^{+0.4}_{-0.6}$ & $14.7$ \\
RXJ1532-0683 & $233.2141861$ & $30.3545029$ & $>29.0$ & $28.3 \pm 0.71$ & $27.3 \pm 0.19$ & $27.4 \pm 0.15$ & $27.1 \pm 0.17$ & $27.9 \pm 0.26$ & $27.4 \pm 0.19$ & $6.4^{+0.6}_{-6.2}$ & $6.6$ \\
RXJ1532-0712 & $233.2502805$ & $30.3527546$ & $27.7 \pm 0.24$ & $26.8 \pm 0.24$ & $26.5 \pm 0.17$ & \nodata & $27.2 \pm 0.19$ & $26.9 \pm 0.16$ & $26.7 \pm 0.16$ & $5.6^{+0.5}_{-5.2}$ & $3.8$ \\
RXJ1532-0716 & $233.2312519$ & $30.3529432$ & $>28.6$ & $26.3 \pm 0.20$ & $27.1 \pm 0.22$ & $26.7 \pm 0.11$ & $27.0 \pm 0.20$ & $27.3 \pm 0.22$ & $26.8 \pm 0.16$ & $5.7^{+0.5}_{-5.7}$ & $>100$ \\
RXJ1532-0771 & $233.2130381$ & $30.3519917$ & $29.2 \pm 0.70$ & $27.0 \pm 0.26$ & $27.2 \pm 0.18$ & $27.5 \pm 0.16$ & $27.6 \pm 0.23$ & $27.4 \pm 0.17$ & $27.5 \pm 0.20$ & $5.8^{+0.6}_{-5.3}$ & $4.3$ \\
RXJ1532-0916 & $233.2457329$ & $30.3488797$ & $28.8 \pm 0.66$ & $27.2 \pm 0.37$ & $25.8 \pm 0.12$ & \nodata & $26.6 \pm 0.15$ & $26.8 \pm 0.20$ & $26.7 \pm 0.19$ & $6.4^{+0.3}_{-0.3}$ & $5.1$ \\
RXJ1532-1100 & $233.2138441$ & $30.3442524$ & $27.8 \pm 0.32$ & $27.1 \pm 0.36$ & $26.5 \pm 0.13$ & $26.6 \pm 0.10$ & $26.8 \pm 0.18$ & $26.4 \pm 0.11$ & $26.6 \pm 0.13$ & $5.7^{+0.7}_{-5.3}$ & $3.5$ \\
RXJ1532-1127 & $233.2263652$ & $30.344196$ & $27.5 \pm 0.20$ & $26.2 \pm 0.16$ & $26.7 \pm 0.13$ & $26.8 \pm 0.11$ & $26.9 \pm 0.15$ & $26.9 \pm 0.13$ & $27.3 \pm 0.18$ & $5.6^{+0.2}_{-4.7}$ & $15.6$ \\
RXJ1532-1241 & $233.2192718$ & $30.3407408$ & $28.8 \pm 0.59$ & $26.7 \pm 0.22$ & $26.8 \pm 0.15$ & $26.8 \pm 0.11$ & $26.9 \pm 0.16$ & $26.7 \pm 0.12$ & $27.0 \pm 0.16$ & $6.1^{+0.4}_{-0.5}$ & $5.4$ \\
RXJ1532-1538\tnm{c} & $233.2227919$ & $30.3306261$ & $27.9 \pm 0.32$ & $26.7 \pm 0.25$ & $25.9 \pm 0.09$ & $26.0 \pm 0.07$ & $25.7 \pm 0.07$ & $26.0 \pm 0.08$ & $26.0 \pm 0.08$ & $6.3^{+0.4}_{-0.5}$ & $1.5$ \\
RXJ1532-0844 & $233.2450679$ & $30.3506265$ & $>28.8$ & $>27.3$ & $26.5 \pm 0.17$ & $25.9 \pm 0.07$ & $26.0 \pm 0.10$ & $26.1 \pm 0.09$ & $26.0 \pm 0.09$ & $7.1^{+0.3}_{-0.8}$ & $3.4$ \\
RXJ1532-0878 & $233.2152996$ & $30.3530354$ & $28.9 \pm 0.65$ & $>28.2$ & $27.2 \pm 0.20$ & $26.7 \pm 0.10$ & $27.2 \pm 0.20$ & $26.9 \pm 0.14$ & $27.1 \pm 0.16$ & $6.5^{+0.7}_{-0.6}$ & $14.7$ \\
RXJ1532-0918 & $233.2150798$ & $30.3529243$ & $>28.9$ & $>28.0$ & $26.2 \pm 0.11$ & $26.5 \pm 0.10$ & $26.5 \pm 0.14$ & $26.8 \pm 0.16$ & $26.7 \pm 0.14$ & $6.6^{+0.4}_{-0.2}$ & $12.8$ \\
RXJ1532-1063 & $233.199579$ & $30.3455558$ & $27.9 \pm 0.48$ & $>27.2$ & $25.7 \pm 0.13$ & \nodata & $25.7 \pm 0.09$ & $25.6 \pm 0.09$ & $25.4 \pm 0.09$ & $6.5^{+0.6}_{-5.4}$ & $1.0$ \\
RXJ1532-0030 & $233.2273252$ & $30.3721732$ & $>28.7$ & $>28.1$ & $27.3 \pm 0.30$ & \nodata & $26.4 \pm 0.11$ & $26.8 \pm 0.16$ & $27.1 \pm 0.25$ & $7.7^{+0.5}_{-1.6}$ & $4.9$
\enddata
\tablecomments{Magnitudes are expressed as observed (lensed) isophotal magnitudes (ISOMAG).}
\tablenotetext{a}{Photometric redshift estimate with $2\sigma$ (95\%) confidence intervals (see Section~\ref{sec:bpz}).  Objects with large lower bounds have a secondary peak at lower redshift ($z\sim1-2$) that contains at least $5\%$ of the posterior probability.}
\tablenotetext{b}{Magnification estimate from the lens models (see Section~\ref{sec:models}).  Because of uncertainties in the precise location of the critical curves, objects with magnifications $>100$ are simply quoted as such.}
\tablenotetext{c}{Unresolved object with FWHM $<0.22$ arcsec in the image plane.  There is a small chance that brighter unresolved candidates could be low-mass stars even though we explored such possibilities (see Section~\ref{sec:contaminants}).}
\label{tbl:rxj1532}
\ifemulateapj
    \end{deluxetable*}
\else
    \end{deluxetable}
\fi
}

\ifemulateapj
    \journalinfo{Astrophysical Journal, Accepted 2014 June 17}
    \slugcomment{Astrophysical Journal, Accepted 2014 June 17}
    \shorttitle{MAGNIFIED HIGH-REDSHIFT GALAXIES AT $z \sim 6-8$}
    \shortauthors{\sc{BRADLEY ET AL.}}
\fi

\begin{document}

\title{CLASH: A Census of Magnified Star-Forming Galaxies at $\lowercase{z} \sim 6-8$\footnotemark[*]}

\footnotetext[*]{Based on observations made with the NASA/ESA {\em Hubble Space Telescope}, obtained at the Space Telescope Science Institute, which is operated by the Association of Universities for Research in Astronomy under NASA contract NAS5-26555.  These observations are associated with programs 12065-12069, 12100-12104, and 12451-12460.}

\author{L.D.~Bradley\altaffilmark{1},
A.~Zitrin\altaffilmark{2,27},
D.~Coe\altaffilmark{1},
R.~Bouwens\altaffilmark{3},
M.~Postman\altaffilmark{1},
I.~Balestra\altaffilmark{4,5},
C.~Grillo\altaffilmark{6},
A.~Monna\altaffilmark{7,8},
P.~Rosati\altaffilmark{9},
S.~Seitz\altaffilmark{7,8},
O.~Host\altaffilmark{6},
D.~Lemze\altaffilmark{10},
J.~Moustakas\altaffilmark{11},
L.A.~Moustakas\altaffilmark{12},
X.~Shu\altaffilmark{13},
W.~Zheng\altaffilmark{10},
T.~Broadhurst\altaffilmark{14,15},
M.~Carrasco\altaffilmark{2,16},
S.~Jouvel\altaffilmark{17,18},
A.~Koekemoer\altaffilmark{1},
E..~Medezinski\altaffilmark{10},
M.~Meneghetti\altaffilmark{12,19,20},
M.~Nonino\altaffilmark{4},
R.~Smit\altaffilmark{3},
K.~Umetsu\altaffilmark{21},
M.~Bartelmann\altaffilmark{2},
N.~Ben\'itez\altaffilmark{22},
M.~Donahue\altaffilmark{23},
H.~Ford\altaffilmark{10},
L.~Infante\altaffilmark{16},
Y.~Jimenez-Teja\altaffilmark{22},
D.~Kelson\altaffilmark{24},
O.~Lahav\altaffilmark{18},
D.~Maoz\altaffilmark{25},
P.~Melchior\altaffilmark{26},
J.~Merten\altaffilmark{12},
and
A.~Molino\altaffilmark{22}}
\altaffiltext{1}{Space Telescope Science Institute, 3700 San Martin Drive, Baltimore, MD 21218, USA}
\altaffiltext{2}{Institut f\"ur Theoretische Astrophysik, Zentrum f\"ur Astronomie, Institut f\"ur Theoretische Astrophysik, Albert-Ueberle-Str.~2, D-29120 Heidelberg, Germany}
\altaffiltext{3}{Leiden Observatory, Leiden University, Leiden, The Netherlands}
\altaffiltext{4}{INAF/Osservatorio Astronomico di Trieste, via G.B. Tiepolo 11, I-34143 Trieste, Italy}
\altaffiltext{5}{INAF/Osservatorio Astronomico di Capodimonte, via Moiariello 16, I-80131 Napoli, Italy}
\altaffiltext{6}{Dark Cosmology Centre, Niels Bohr Institute, University of Copenhagen, Juliane Mariesvej 30, DK-2100 Copenhagen, Denmark}
\altaffiltext{7}{University Observatory Munich, Scheinerstrasse 1, D-81679 Munich, Germany}
\altaffiltext{8}{Max Planck Institute for Extraterrestrial Physics, Giessenbachstrasse, D-85748 Garching, Germany}
\altaffiltext{9}{ESO-European Southern Observatory, D-85748 Garching bei M\"unchen, Germany}
\altaffiltext{10}{Department of Physics and Astronomy, The Johns Hopkins University, Baltimore, MD 21218, USA}
\altaffiltext{11}{Department of Physics \& Astronomy, Siena College, 515 Loudon Road, Loudonville, NY 12211, USA}
\altaffiltext{12}{Jet Propulsion Laboratory, California Institute of Technology, Pasadena, CA 91109, USA}
\altaffiltext{13}{Department of Astronomy, University of Science and Technology of China, Hefei, Anhui 230026, China}
\altaffiltext{14}{Department of Theoretical Physics and History of Science, University of the Basque Country UPV/EHU, PO Box 644, E-48080 Bilbao, Spain}
\altaffiltext{15}{Ikerbasque, Basque Foundation for Science, Alameda Urquijo, 36-5 Plaza Bizkaia E-48011, Bilbao, Spain}
\altaffiltext{16}{Centro de Astro-Ingenier\'ia, Departamento de Astronom\'ia y Astrof\'isica, Pontificia Universidad Cat\'olica de Chile,  V. Mackenna 4860, Santiago, Chile}
\altaffiltext{17}{Institut de Ci\'encies de l'Espai (IEEC-CSIC), E-08193 Bellaterra (Barcelona), Spain}
\altaffiltext{18}{Department of Physics and Astronomy, University College London, London, WC1E 6BT, UK}
\altaffiltext{19}{INAF/Osservatorio Astronomico di Bologna, via Ranzani 1, I-40127 Bologna, Italy}
\altaffiltext{20}{INFN, Sezione di Bologna, Via Ranzani 1, I-40127 Bologna, Italy}
\altaffiltext{21}{Institute of Astronomy and Astrophysics, Academia Sinica, PO Box 23-141, Taipei 10617, Taiwan}
\altaffiltext{22}{Instituto de Astrof\'isica de Andaluc\'ia (IAA-CSIC), E-18008 Granada, Spain}
\altaffiltext{23}{Department of Physics and Astronomy, Michigan State University, East Lansing, MI 48824, USA}
\altaffiltext{24}{Observatories of the Carnegie Institution of Washington, Pasadena, CA, USA}
\altaffiltext{25}{School of Physics and Astronomy, Tel Aviv University, Tel-Aviv I-69978, Israel}
\altaffiltext{26}{Center for Cosmology and Astro-Particle Physics \& Department of Physics, The Ohio State University, Columbus, OH 43210, USA}
\altaffiltext{27}{Hubble Fellow}

\begin{abstract}
    We utilize 16 band {\em Hubble Space Telescope} (\HST)
    observations of 18 lensing clusters obtained as part of the
    Cluster Lensing And Supernova survey with Hubble (CLASH)
    Multi-Cycle Treasury program to search for $z\sim6-8$ galaxies.
    We report the discovery of 204, 45, and 13 Lyman-break galaxy
    candidates at $z\sim6$, $z\sim7$, and $z\sim8$, respectively,
    identified from purely photometric redshift selections.  This
    large sample, representing nearly an order of magnitude increase
    in the number of magnified star-forming galaxies at $z\sim 6-8$
    presented to date, is unique in that we have observations in four
    WFC3/UVIS UV, seven ACS/WFC optical, and all five WFC3/IR
    broadband filters, which enable very accurate photometric redshift
    selections.  We construct detailed lensing models for 17 of the 18
    clusters to estimate object magnifications and to identify two new
    multiply lensed $z\ga6$ candidates.  The median magnifications
    over the 17 clusters are 4, 4, and 5 for the $z\sim6$, $z\sim7$,
    and $z\sim8$ samples, respectively, over an average area of 4.5
    arcmin$^2$ per cluster.  We compare our observed number counts
    with expectations based on convolving ``blank'' field UV
    luminosity functions through our cluster lens models and find
    rough agreement down to $\sim27$ mag, where we begin to suffer
    significant incompleteness.  In all three redshift bins, we find a
    higher number density at brighter observed magnitudes than the
    field predictions, empirically demonstrating for the first time
    the enhanced efficiency of lensing clusters over field surveys.
    Our number counts also are in general agreement with the lensed
    expectations from the cluster models, especially at $z\sim6$,
    where we have the best statistics.
\end{abstract}


\section{Introduction}

The improved {\em Hubble Space Telescope} has revolutionized our
ability to study galaxies in the early universe at redshifts $z\ga 6$.
The ultra-deep WFC3/IR observations of the Hubble Ultra-Deep Field
from the HUDF09 \citep{Bouwens2011b} and HUDF12 \citep{Ellis2013}
campaigns, its two ultra-deep parallel fields, and the deep wide-area
WFC3/IR Early Release Science (ERS) observations \citep{Windhorst2011}
have revealed a large sample of $\sim200$ $z\sim7-8$ Lyman-break
galaxy (LBG) candidates \citep{Bouwens2011b, Lorenzoni2011,
McLure2011, Schenker2013, McLure2013}.

Complementary WFC3/IR surveys have further increased the sample of
$z\sim7-8$ galaxies, including those obtained as part of the Cosmic
Assembly Near-Infrared Deep Extragalactic Legacy Survey
\citep[CANDELS;][]{Grogin2011, Koekemoer2011, Oesch2012, Yan2012}
Multi-Cycle Treasury (MCT) program, the Brightest of Reionizing
Galaxies \citep[BoRG;][]{Trenti2011, Trenti2012a, Bradley2012b} and
the Hubble Infrared Pure Parallel Imaging Extragalactic Survey
\citep[HIPPIES;][]{Yan2011a}.  Together, these data sets have allowed
for the first detailed studies of galaxies firmly in the reionization
epoch at $z\sim7-8$, including their physical properties
\citep[e.g.,][]{Oesch2010b, Labbe2010b}, rest-frame UV-continuum
slopes \citep[e.g.,][]{Wilkins2011, Bouwens2012beta, Dunlop2013,
Bouwens2013beta}, clustering \citep{Trenti2012b}, nebular line
emission \citep{Labbe2013, Smit2014}, and luminosity function
\citep[e.g.,][]{Bouwens2011b, Oesch2012, Bradley2012b, Schenker2013,
McLure2013}.

Gravitational lensing by massive galaxy clusters has also been
highlighted as a powerful tool in the discovery and study of the
properties of faint high-redshift galaxies \citep[e.g.,][]{Kneib2004,
Egami2005, Bradley2008, Richard2008, Zheng2009, Bradac2009,
Bouwens2009a, Hall2012, Bradac2012, Bradley2012a, Zitrin2012m0329}.
Of particular note, this includes the recent discoveries of two
$z\sim9$ candidates behind MACSJ1115.9+0129 and MACSJ1720.3+3536
\citep{Bouwens2012clash}, a $z\sim9.6$ candidate behind
MACSJ1149.6+2223 \citep{Zheng2012}, and a triply lensed candidate at
$z\sim10.7$ behind MACSJ0647.8+7015 \citep{Coe2013}, all identified by
the Cluster Lensing And Supernova survey with Hubble (CLASH;
\citealt{Postman2012}) MCT program.

Massive galaxy clusters can act as gravitational ``cosmic
telescopes'', considerably magnifying both the apparent luminosity and
size of background sources.  The flux amplification provides a deeper
effective limiting magnitude of the observations, allowing for the
identification of high-redshift galaxies that otherwise would have
remained undetected.  Likewise, the brighter apparent magnitude of
magnified high-redshift sources can place them within reach of
ground-based spectroscopy, as recently demonstrated by the
spectroscopic confirmation of a lensed LBG in Abell~383 at $z=6.027$
identified by \cite{Richard2011} and A1703-zD6 \citep{Bradley2012a} at
$z=7.045$ \citep{Schenker2012}, which is the highest-redshift lensed
galaxy with a spectroscopic confirmation.  Magnification also provides
an effective increase in spatial resolution, enabling detailed studies
of the sizes and morphologies of high-redshift galaxies that otherwise
would not be possible \citep[e.g.,][]{Franx1997, Kneib2004,
Bradley2008, Zheng2009, Swinbank2009, Bradley2012a, Zitrin2012m0329,
Zheng2012, Sharon2012}.

Here we utilize the 16 band \HST\ WFC3/UVIS, ACS/WFC, and WFC3/IR
observations of 18 lensing clusters obtained as part of the CLASH MCT
program to search for $z\sim 6-8$ galaxies.  This cluster sample
includes all five CLASH clusters selected based on their lensing
strength (the other 20 were X-ray selected) and four of the six
clusters chosen to be part of the Hubble Frontier Fields (HFF)
program.\footnote{For details, see
http://www.stsci.edu/hst/campaigns/frontier-fields/}  We identify the
high-redshift galaxy candidates from their photometric redshifts,
taking advantage of the presence of the Lyman-break feature in their
spectral energy distributions (SEDs) \citep{Steidel1996a}.  Our
resulting sample of LBG candidates represents the largest sample of
magnified star-forming galaxies at $z\sim 6-8$ presented to date.
This lensed galaxy sample is unique in that we have observations in
seven ACS optical and all five WFC3/IR broadband filters, which enable
very accurate photometric redshift selections.  Using strong lensing
models for 17 of the 18 clusters (RXJ1532 is excluded because of the
uncertainty in its strong lensing model; see
Section~\ref{sec:models}), we derive the expected number densities of
high-redshift candidates behind these clusters.

\tabclobs
\tabfilts
\figobjzsix
\figobjzseven
\figobjzeight
\figzsixarc

This paper is organized as follows.  We begin with a description of
the observations in Section~\ref{sec:obs} and discuss our photometry
and catalog construction in Section~\ref{sec:photometry}.  We discuss
the photometric redshifts in Section~\ref{sec:bpz} and our
high-redshift galaxy sample selection in Section~\ref{sec:selection}.
In Section~\ref{sec:models}, we describe our detailed cluster lens
models.  In Section~\ref{sec:nden}, we compare the number densities of
our high-redshift galaxy sample with those found in ``blank" field
surveys.  Finally, we summarize our results in
Section~\ref{sec:summary}.  Throughout this paper we adopt a cosmology
with $\Omega_{m} = 0.3$, $\Omega_{\Lambda} = 0.7$, and $H_{0} =
70$~\kmsMpc.  This provides an angular scale of 5.7~kpc~\peras,
5.2~kpc~\peras, and 4.8~kpc~\peras\ (proper) at $z = 6.0$, $7.0$, and
$8.0$, respectively.  We refer to the \HST\ F814W, F850LP, F105W,
F110W, F125W, F140W, and F160W bands as \Iband, \zband, \Yband,
\YJband, \Jband, \JHband, and \Hband, respectively.  All magnitudes
are expressed in the AB photometric system \citep{Oke1974}.

\section{Observations}
\label{sec:obs}

CLASH is a 524 orbit multi-cycle treasury program to observe 25 galaxy
clusters to a total depth of 20 orbits each, incorporating archival
\HST\ data for our cluster sample whenever possible
\citep{Postman2012}.  Each cluster is observed using WFC3/UVIS,
ACS/WFC, and WFC3/IR to obtain imaging in 16 broadband
filters\footnote{Some clusters have additional archival data with the
ACS F555W filter, which is not a standard filter in our CLASH program.
Taking advantage of the archival F555W data, four of our clusters have
observations in 17 \HST\ bands.} spanning from $0.2$ to $1.7$~$\mu$m
(for the throughput curves of each filter, see \citealt{Postman2012}
or \citealt{Jouvel2014}).  For this paper, we include the observations
of 18 clusters in the CLASH sample.  The cluster observations are
provided in Table~\ref{tbl:clobs} and the exposure details, including
filters, exposure times, and limiting magnitudes, for a typical
cluster are presented in Table~\ref{tbl:filters}.  The galaxy cluster
sample in this paper includes 11 clusters from the Massive Cluster
Survey (MACS; \citealt{Ebeling2001}), including six at $z < 0.5$
\citep{Ebeling2010, Mann2012} and five at $z > 0.5$
\citep{Ebeling2007}.  Henceforth, we will refer to the clusters by
their shortened names listed in Table~\ref{tbl:clobs}.

We calibrate the raw \HST\ data using standard techniques to remove
the instrumental bias, dark, and flat-field signatures from the data.
The ACS data are further processed to remove the bias striping and
charge-transfer inefficiency effects.  For the WFC3/IR data, we take
advantage of ``guard darks'' taken immediately preceding the first
visit of most CLASH observations.  Our calibration pipeline subtracts
the standard dark from the guard dark to create ``delta darks'', which
contain information about new hot/warm pixels and persistence of
charge from data taken in the orbits immediately prior to CLASH
observations.  Additionally, we identify bright sources in our WFC3/IR
observations to create persistence masks for data taken within CLASH
visits.  The external and internal persistence masks are flagged in
the WFC3/IR data quality arrays and used downstream to exclude
persistence regions when drizzling the data.

The data in each filter were combined with the {\tt MosaicDrizzle}
pipeline \citep{Koekemoer2003} described in detail in our overview
paper \citep{Postman2012}.  The pipeline produces cosmic-ray rejected
and aligned images for each filter using a combination of
cross-correlation and catalog matching.  The final images are drizzled
to a common pixel grid with a scale of $0\farcs065$ pixel$^{-1}$.

\section{Photometry and Source Catalogs}
\label{sec:photometry}

We used SExtractor version 2.5.0 \citep{Bertin1996} in dual-image mode
to perform object detection and photometry.  For each of our 18
clusters, we constructed a detection image by performing an
inverse-variance weighted sum of the images in all five WFC3/IR bands:
\Yband, \YJband, \Jband, \JHband, and \Hband.  The local background
was measured within a rectangular annulus (default width 24 pixels)
and sources were required to be detected at $>1\sigma$ significance
over a minimum area of nine contiguous pixels.  We measured object
colors using the flux enclosed within the isophotal apertures.  The
flux uncertainties are derived by SExtractor using an rms image (input
to SExtractor as a weight map) including all sources of noise except
for the Poisson noise of the objects.  The flux uncertainty derived by
SExtractor adds the Poisson source noise in quadrature to the noise
determined from the rms image, which is primarily background noise.

Sources that are undetected ($<1\sigma$) in a particular band are
given their $1\sigma$ upper detection limit to calculate limits for
object colors.  Total magnitudes were measured in scalable
\cite{Kron1980} apertures with a Kron factor of 2.5 and a minimum
radius of 3.5 pixels.  Our photometry is also corrected for the
foreground Galactic extinction along the line of sight to each cluster
using the \cite{Schlegel1998} IR dust emission maps.  The $E(B-V)$
color excess values for each cluster are presented in
Table~\ref{tbl:clobs}.

No correction for the varying size of the PSF across the bandpasses
has been applied to the CLASH photometric catalogs.  For the typical
sizes of our high-redshift candidates, applying such a correction
would redden the $\Yband\ - \Hband$ colors by only $\sim 0.1$ mag.
All of the CLASH \HST\ data and source photometric catalogs are
available online.\footnote{http://archive.stsci.edu/prepds/clash/}

It should be noted that the CLASH photometric catalogs do not correct
for the presence of correlated noise introduced in the drizzling
procedure \citep[e.g.,][]{Casertano2000}.  However, after we
constructed our catalogs and samples of high-redshift candidates, we
later investigated the effects of correlated noise on the
high-redshift sample selection (see Section~\ref{sec:selection}).

\section{Photometric Redshifts}
\label{sec:bpz}

In total, our catalogs contain over 38,000 sources over all the 18
clusters.  For each source, we derive photometric redshifts using the
complete 16 band (or 17 band) observed photometry spanning from $0.2$
to $1.7$~$\mu$m.  To estimate the redshifts of our candidates and to
derive the posterior redshift probability distribution functions,
$P(z)$, we used the Bayesian photometric redshift (BPZ) code
\citep{Benitez2000, Benitez2004, Coe2006}.

The photometric redshifts are based on a $\chi^{2}$ fitting procedure
to the observed measured fluxes (even if negative), and flux
uncertainties.  We utilized a combination of empirical galaxy
templates and template SEDs from PEGASE \citep{Fioc1997} that have
been recalibrated with known spectroscopic redshifts from the
FIREWORKS survey \citep{Wuyts2008}.  BPZ does not redden any of its
SED templates, but it includes a large range of templates designed and
calibrated to fit almost all galaxies.  Comparing BPZ's template set
to large data sets with high-quality spectra reveals that there are
only $\la 1\%$ outliers not covered by BPZ.  This demonstrates that
the BPZ templates encompass the range of metallicities, reddenings,
and star formation histories observed for the vast majority of real
galaxies \citep{Coe2013}.  Lyman series line-blanketing and
photoelectric absorption produced by intervening hydrogen along the
line of sight are applied to the BPZ templates following the
prescription of \cite{Madau1995}.

At present, the Bayesian prior, $P(z, m_{0})$, is not well calibrated
at faint magnitudes ($m\ga26$) or at the high redshifts $z\sim6-8$
investigated here.  Therefore, we utilized a flat prior in BPZ to
construct our catalog of high-redshift galaxy candidates.  Note that
because of the flat prior, the photometric redshifts derived here for
the specific purpose of identifying high-redshift galaxy candidates
are different from the best-fitting $z_{\mathrm{phot}}$ in the online
CLASH catalogs.  However, for comparison, the online CLASH catalogs
include a maximum likelihood redshift, $z_{\mathrm{ml}}$, which is
equivalent to using a flat prior.  For details about the BPZ priors
used in the online CLASH catalogs, please see \cite{Jouvel2014}.

The primary contaminants to the $z\sim5.5-8.5$ sample are faint red
galaxies at $z\sim1.0-1.9$.  This class of low-redshift galaxies with
very prominent Balmer breaks represents the main galaxy population
that can mimic high-redshift LBGs.  We can directly compare the
expected relative numbers of faint red $z\sim1-2$ galaxies and blue
star-forming galaxies at $z\sim5.5-8.5$ using published galaxy
luminosity functions (LF) for these two galaxy populations
\citep{Giallongo2005, Bouwens2011b, Bradley2012b}.  We base our
low-redshift expectations on \cite{Giallongo2005}, who derived LFs for
red galaxies using deep NIR observations over the HDF-North and
HDF-South fields and the K20 spectroscopic sample \citep{Cimatti2002}.
For high-redshift expectations, we use the LFs derived by
\cite{Bouwens2011b} and \cite{Bradley2012b}.

At $z\sim1.5$, the \cite{Giallongo2005} $< m^{∗}/m(\mathrm{bimodal})$
LF results correspond to $M^{*}_{B,0} = -21.62$ mag, $\phi^{*}∗ = 3.8
\times 10^{−4}$~Mpc$^{-3}$, and $\alpha = −0.53$.  Assuming the
magnitude range between $26-27$ mag (where majority of the
high-redshift galaxies are to be found) and a selection window with
$\Delta z = 1$, the LFs predict 0.18 faint red galaxies per arcmin$^2$
at $z\sim 1.5$ and 0.40 $z\sim 7$ galaxies per arcmin$^2$.  These
results suggest that in a ``blank'' field we are $\sim2.2$ times more
like to find a blue high-redshift galaxy than a faint red low-redshift
galaxy.  This ratio of blue high-redshift to red low-redshift galaxies
is even larger in a lensed field because the number density of
high-redshift galaxies intrinsically fainter than 27 mag is
increasing, while the number density of the $z\sim1-2$ faint red
galaxy population is decreasing.  As a consequence, our use of a flat
redshift prior for the specific task of identifying high-redshift
galaxy candidates is conservative and not preferentially selecting
high-redshift galaxies (see also Appendix A from
\citealt{Bouwens2012clash}, who used a similar argument for the use of
a flat prior in the context of $z\sim9$ sources from CLASH).

\section{High-redshift Candidate Selection}
\label{sec:selection}

\subsection{Catalog Construction}

We use the photometric redshift catalog to select high-redshift galaxy
candidates at redshifts $z > 5.5$.  While we do not use a two-color
Lyman-break selection technique to select high-redshift galaxies
\citep[e.g.,][]{Bouwens2011b}, BPZ identifies high-redshift galaxy
candidates primarily based on the presence of the Lyman break feature
in their SED.  BPZ optimally utilizes the photometry in all 16
broadband filters, including UV and optical non-detections or marginal
detections, and provides a quantitative estimate for the redshift
uncertainty.  Traditional LBG color$-$color selections are generally
limited to two (or three) bands plus nondetection in the optical.  The
color$-$color cuts can exclude genuine high-redshift candidates (for
example, the specific Lyman-break color cut chosen necessarily selects
objects at different redshift cutoffs due to intrinsic differences in
object colors), while possibly including more low-redshift
contaminants \citep[e.g., see][]{Finkelstein2010, Dunlop2013r}.

These two alternative approaches to LBG selection have been employed
with great success on deep and ultradeep \HST\ fields such as the
HUDF.  While some groups have used color$-$color selections
\citep[e.g.,][]{Oesch2010a, Bouwens2011b, Bunker2010, Ellis2013,
Schenker2013} and others have used photometric-redshift selections
\citep[e.g.,][]{Finkelstein2010, McLure2010, McLure2013}, the
resulting $z\sim7-8$ galaxy samples are generally in very good
agreement, especially at brighter magnitudes.

To ensure reliable photometric redshifts and to limit the number of
possible contaminants due to photometric scatter, we require that our
high-redshift candidates are detected at $\ge 6\sigma$ in the combined
\JHband\ and \Hband\ bands.  As demonstrated in
\cite{Bouwens2012clash}, our high-redshift galaxy selections would
otherwise be subject to significant contamination ($\ga25\%$) for
sources detected at lower significance levels, especially faintward of
$26.5$ mag, due the effects of noise on the photometry of other
lower-redshift sources.  Further, we investigate any BPZ fit solutions
that give a non-physical result, such as an elliptical galaxy SED
template fit at $z\ga6$.

A few of our candidates lack coverage in one or more of the WFC3/IR
bands because the observations in these filters were obtained at only
one \HST\ orient to accommodate the CLASH supernova search program.
While all of our candidates have coverage in the \Hband\ band, if
coverage in the \JHband\ band is missing, we applied the $6\sigma$
detection threshold to the \Hband\ band plus the next available
reddest WFC3/IR filter.

One galaxy in our sample, MACS0744-0225, is detected at only
$5.5\sigma$ significance in the combined \JHband\ and \Hband\ bands.
Despite this fact, we include this candidate in our $z\sim7$ sample
because the lensing model for MACS0744 suggests that this source and
MACS0744-1695 likely represent a doubly lensed system at $z\sim6.6$,
and hence it is unlikely to be a low-redshift contaminant.

As mentioned in Section~\ref{sec:photometry}, the drizzling procedure
introduces pixel-to-pixel noise correlations that are not reflected in
the output weight or rms maps.  The typical correction for the effects
of correlated noise involves rescaling the rms map by measuring the
empirical noise in ``blank'' areas of size comparable to the observed
galaxies and comparing it with the noise measured from the unscaled
rms maps \citep[e.g.,][]{Trenti2011, Bradley2012b, Guo2013}.  After
correcting the rms maps for the effects of correlated noise using
``empty aperture'' measurements on the data in each of the filters, we
then investigated the effects of this increased noise on the
photometric redshifts.  In particular, without the correlated noise
correction, the potential concern is that the significance of the
UV/optical nondetections, required for the selection of high-redshift
candidates, may be underestimated.

We find that the median correction factors for the noise in the CLASH
UVIS and optical data are relatively small at $\sim1.11$ and
$\sim1.30$, respectively (these values are larger than the simplified
formula presented in \citealt{Casertano2000}).  After re-calculating
the photometric redshifts with rescaled errors, we find only two
galaxies, both in the initial $z\sim6$ sample, that subsequently were
best fit by low-redshift solutions.  Thus, we removed both galaxies
from our high-redshift sample.

\subsection{Resulting High-redshift Candidate Samples}

Using our photometric redshift catalogs, we find 204 $z\sim6$
candidates, 45 $z\sim7$ candidates, and 13 $z\sim8$ candidates, for a
total of 262 lensed high-redshift galaxy candidates
(Table~\ref{tbl:nobj}).  These numbers have been corrected for
multiply imaged lensed systems identified by our fiducial lens models
(see Section~\ref{sec:models}).  The coordinates, photometry, and
photometric redshift estimates for these candidates are presented in
Tables~\ref{tbl:z6}, \ref{tbl:z7}, and \ref{tbl:z8}.  The mean
photometric redshifts for our $z\sim6$, $z\sim7$, and $z\sim8$ samples
are 5.9, 6.7, and 7.8, respectively.

\tabnobj

We have already spectroscopically confirmed a few of these candidates,
including a quintuply lensed $z=6.11$ galaxy behind RXCJ2248
\citep{Balestra2013, Monna2014} and a pair of faint galaxies at
$z=6.387$ behind MACS0717 \citep{Vanzella2014}.  We aim to
spectroscopically confirm even more of these candidates with upcoming
observations at several facilities.

Our sample of LBG candidates is the largest sample of magnified
star-forming galaxies at $z\sim 6-8$ presented to date.  Given the
general lack of high-quality optical and NIR multiband observations of
lensing clusters prior to the CLASH survey, previous studies in this
redshift range have typically focused on one or a few spectacular
lensed candidates or particular clusters for which multiband optical
and NIR data exist.

Such studies include the triply lensed $z\sim6.4$ candidate behind
Abell~2218 \citep{Kneib2004}, the $z\sim7.6$ candidate behind
Abell~1689 \citep{Bradley2008}, a possible $z\sim7$ candidate behind
Abell~2219 and Abell~2667 \citep[][but see also
\citealt{Bouwens2009a}]{Richard2008}, the $z\sim6$ candidate behind
Abell~1703 and two $z\sim6.5$ candidates behind CL0024+16
\citep{Zheng2009}, four $z\sim6$ $i$-dropouts \citep{Bradac2009} and
10 $z\sim7$ candidates \citep{Hall2012} behind the Bullet Cluster
(1E0657$-$56), and seven $z\sim7$ candidates behind Abell~1703
\citep{Bradley2012a}.  Lensed candidates in this redshift range
previously studied with CLASH data include the doubly imaged $z=6.027$
candidate behind A383 \citep{Richard2011}, the quadruply lensed
$z\sim6.2$ candidate behind MACS0329 \citep{Zitrin2012m0329}, a
quintuply lensed $z\sim5.9$ candidate behind RXCJ2248
\citep{Monna2014}, and other studies that briefly mentioned
multiply lensed candidates at $z\sim5.7$ in MACS1206
\citep{Zitrin2012m1206} and $z\sim6$ in MACS0416
\citep{Zitrin2013m0416}.  In total, these studies comprise 31 lensed
high-redshift galaxies at $z\sim 6-8$.  Our CLASH lensed sample of 262
candidates (which includes the five CLASH galaxies from previous
studies) represents nearly an order of magnitude increase in the
number of lensed star-forming galaxies at $z\sim 6-8$.

As an example of a $z\sim6$ candidate, we plot the observed SED in the
16 observed bands along with the best-fit BPZ template for the
$z\sim6.4$ candidate A2261-0754 in Figure~\ref{fig:z6obj}.  We also
show in this figure its posterior photometric redshift probability
distribution, $P(z)$.  The postage stamp images in each of the 16
filters, as well as the total inverse-variance weighted sum of the ACS
and IR images, are also illustrated.  We also present the same set of
plots and images for the $z\sim7.1$ candidate RXJ1532-0844
(Figure~\ref{fig:z7obj}) and the $z\sim7.5$ candidate A2261-0187
(Figure~\ref{fig:z8obj}).

For the candidate MACS1206-1796, we have obtained a spectrum from {\em
VLT}/VIMOS as part of the CLASH {\em VLT} program (PI: P. Rosati).
The longslit 1D and 2D spectra, obtained in a one hour exposure, are
presented in Figure~\ref{fig:z6arc}.  Based on the fiducial lens model
for MACS1206 (see Section~\ref{sec:models}), this relatively bright
candidate (observed \Hband\ magnitude of 23.8 mag) is likely part of a
quadruply lensed system \citep[obj 8.4 in][]{Zitrin2012macs1206} along
with MACS1206-0457, MACS1206-0861, and MACS1206-1135.  Because this
object is magnified by the cluster by only a factor of $\sim2.1$ and
given its long curved arc-like morphology and relative brightness, it
is most likely being additionally magnified by the neighboring
foreground galaxy with $z_{\mathrm{phot}} \sim 1.1$, making it a
probable galaxy$-$galaxy lens candidate.  The spectrum of
MACS1206-1796 is cleanly separated from the bright foreground object
and exhibits a clear emission line at 8146 \AA, corresponding to
Ly$\alpha$ at $z=5.701$.  An alternative possibility is that this
emission line represents \OIIw\ at $z=1.186$.  However, the
photometric redshift of this galaxy, based on the 16 band photometry,
is $z_{\mathrm{phot}} = 5.6$, which differs from the $z=5.701$
spectroscopic hypothesis by only $1.8\%$ (likely demonstrating the
reliability of the photometric redshifts).  Moreover, this probable
quadruple-lens system was predicted to have a redshift of $z=5.7$
based on the lens model \citep{Zitrin2012m1206}.

In Figures~\ref{fig:cl1}, \ref{fig:cl2}, \ref{fig:cl3}, \ref{fig:cl4},
and \ref{fig:cl5}, we indicate the positions of our high-redshift
candidates within the field of view of the cluster images.  In these
figures, we also plot the approximate location of the critical curves
at $z\sim6$ based on the fiducial lens models we constructed for these
clusters (see Section~\ref{sec:models}).

\figmaghist

In Figure~\ref{fig:maghist}, we present histograms of both the observed
and intrinsic (unlensed) rest-frame UV magnitudes at $\sim1750$~\AA\
for our sample of $z\sim6$, $z\sim7$, and $z\sim8$ high-redshift
galaxy candidates identified behind the 17 galaxy clusters (excluding
RXJ1532).  The intrinsic magnitude of each high-redshift candidate has
been calculated using the magnification estimates from the detailed
lens models of each cluster.  The highest magnifications, and hence
the faintest intrinsic magnitudes, usually have the largest
magnification errors due to typical uncertainties in the precise
location of the critical curves (see Section~\ref{sec:muerr}).
Therefore, we separate the intrinsic magnitude histograms for
candidates with magnifications $< 5$ and $\ge 5$.  Because of the
strong lensing effect, the intrinsic magnitudes in the relatively
shallow CLASH survey ($5\sigma$ limiting magnitude of $\sim27.5$ mag
in the \Hband\ band) reach deeper ($> 29.5$ AB mag) than the
ultra-deep HUDF12 observations.  Our observations of the 17 clusters
cover a total area of $\sim22.9$~arcmin$^2$ with magnifications $\mu >
6.3$, needed to surpass the depth of the HUDF12 observations.  This is
a relatively robust measure of area in spite of the model
uncertainties (see Section~\ref{sec:muarea}).

\subsection{Possible Contaminants}
\label{sec:contaminants}

Supernovae, extreme emission-line galaxies (EELG), low-mass stars, and
photometric scatter of red low-redshift galaxies can all be sources of
contamination for high-redshift galaxy selections.  Given that our UV
and optical observations of the cluster fields were obtained over the
same extended time period (typically $\sim2-3$ months) as the WFC3/IR
observations, we can rule out the possibility of contamination from
supernovae.

Because we observe each cluster in 16 overlapping broadband filters
spanning from $0.2$ to $1.7$~$\mu$m, contamination from low-redshift
extreme emission-line galaxies \citep[e.g.,][]{vanderWel2011,
Atek2011} is minimized in our high-redshift samples.  We find two
$z\sim6$ sources, MACS1115-0352 and MACS1720-1114, with very blue SEDs
for which we cannot completely rule out the EELG possibility.  The
EELG hypothesis would require rest-frame equivalent widths of
$\sim2000$~\AA\ in \OIIIww\ and \Ha, but without any substantial
\OIIw\ emission, which falls in our \iband\ and \Iband\ bands where we
have no significant detections (blueward of the Lyman break for a
$z\sim6$ candidate).  This possibility and a more general search for
EELGs in CLASH data is further explored in X. Huang et al. 2014
(submitted).

In principle, the large number of overlapping filters in CLASH also
allows for robust discrimination of low-mass stars, which can be
identified by their distinct colors.  However, the
photometric-redshift code BPZ does not employ stellar templates.  To
further investigate this possible source of contamination for our
unresolved sources, we used the photometric-redshift code LePhare
\citep{Arnouts1999, Ilbert2006}.

LePhare is a SED fitting code that estimates photometric redshifts
with a $\chi^2$ fitting method to fit the observed fluxes with
template spectra.  The code allows us to fit the photometry using
galaxy, QSO, and stellar SED templates.  The resulting galaxy
solutions include the redshift probability distribution function
(PDF($z$)) and also a secondary solution from the PDF($z$), if
available.  For the galaxy templates, we adopt the COSMOS library
\citep{Ilbert2009}, which includes 31 templates of ellipticals,
spirals, and starburst galaxies.  To take into account the extinction
due to the interstellar medium (ISM), we apply the Calzetti
\citep{Calzetti2000} extinction law to the starburst templates and the
Small Magellanic Cloud Prevot law \citep{Prevot1984} to the Sc and Sd
galaxy templates.  We also allow for inclusion of emission lines in
the SED fitting.  For stellar templates, we include the Pickles
stellar library \citep{Pickles1998}, which include all the normal
spectral types plus metal-poor F$-$K dwarfs and G$-$K giants, and cool
M, L, T dwarf star templates from \citep{Cushing2005} and
\cite{Rayner2009}.

In general, it is very difficult to differentiate between extended and
point sources at fainter magnitudes.  Of course high-redshift galaxies
also become smaller and more compact at higher redshifts
\citep[e.g.][]{Ferguson2004, Bouwens2004b, Oesch2010b, Grazian2012,
Ono2013}, and in fact may be unresolved even in lensed images at \HST\
resolution.  For example, the lensed ($\mu = 5.2^{+0.3}_{-0.9}$)
$z_{\mathrm{phot}} \sim 7$ candidate A1703-zD6 behind Abell~1703 is
unresolved in \HST\ WFC3/IR data \citep{Bradley2012a}.  This galaxy
was subsequently confirmed with Keck spectroscopy to be at $z=7.045$
\citep{Schenker2012} and to date remains the highest-redshift lensed
galaxy with a spectroscopic confirmation.

Based on an empirical PSF model constructed from stars in the cluster
fields, we define candidates to be unresolved if they have a FWHM
$<0\farcs22$ in \Hband.  The unresolved candidates are indicated in
Tables~\ref{tbl:z6}, \ref{tbl:z7}, and \ref{tbl:z8}, but most of our
high-redshift candidates appear to be resolved.  The brightest
observed object in our catalog (MACS0416-2028, $\Hband = 23.6$, $\mu
\sim 1.5$) is unresolved and despite having a best-fit photometric
redshift of $z_{\mathrm{phot}} = 7.2$, we suspect this object is most
likely a star.  Based on our current understanding of the $z\sim7$ LF
\citep[e.g.,][]{Bouwens2011b}, the probability of detecting a slightly
magnified $z\sim7$ galaxy with $\Hband = 23.6$ in the small area
covered by the clusters in this paper is exceedingly small.  For
completeness, we include this candidate in the object tables but do
not use it for subsequent analysis given its suspect nature.

Using LePhare, we find that only six of our unresolved candidates have
a good fit with stellar templates, with $\chi^{2}_{\mathrm{star}}$
lower than $\chi^{2}_{\mathrm{galaxy}}$.  Another five candidates have
$\chi^{2}_{\mathrm{star}}$ comparable to $\chi^{2}_{\mathrm{galaxy}}$.
Therefore, we conclude that the contamination rate from low-mass stars
is relatively low at $\la 4\%$, consistent with other studies
\cite[e.g.,][]{Bouwens2011b}.  We note these candidates in
Table~\ref{tbl:z6}.  We also note that LePhare slightly prefers a
low-redshift galaxy solution ($z\sim1$) over the high-redshift
solution for three of these candidates: A2261-0309, MACS1931-0938, and
MACS0647-1670.  Given this and the possible fit with stellar
templates, these candidates should be considered less confident than
the others, even though BPZ prefers a high-redshift solution.

As discussed earlier, the most significant source of contamination to
high-redshift galaxy samples are faint red galaxies at $z\sim 1-2$
that enter the sample due the effects of noise on the photometry.
Based on simulations in which we add photometric errors to a sample of
low-redshift galaxies at $z \sim 1-2$, generate a random realizations
of the photometry within the error bars, and then recalculate the
photometric redshifts using BPZ, we find a low contamination fraction
of $<12\%$ by low-redshift interlopers.  Further evidence of a low
contamination fraction comes from the distribution of rest-frame UV
colors of our $z\sim6-8$ candidate samples, which is much bluer than
one would infer for a $z\sim1-2$ red galaxy sample.

\section{Cluster Lens Models}
\label{sec:models}

In the framework of the CLASH program, detailed lensing models are
being constructed for all 25 CLASH clusters and will eventually be
supplied as high-end science products for the community.  As the lens
modeling is exhaustive and in progress, we use the models available to
date to estimate objects magnifications and to assess the
possibilities of multiply lensed high-redshift candidates.  The
detailed mass models are all constructed using either the modeling
method of \cite{Zitrin2009} (see also \citealt{Broadhurst2005} and
other examples in \citealt{Zitrin2011a383, Zitrin2012m0329}) or using
a second common parameterization of Pseudo Isothermal Elliptical Mass
Distributions (PIEMD) for the galaxies, plus elliptical NFW
distributions for the dark-matter halos \citep[e.g.,][on MACS0416 and
El Gordo]{Zitrin2013m0416, Zitrin2013elgordo}.  The first method
consists of four to six basic free parameters, explained below.  Its
main advantage is that the parameterization allows us to readily find
multiple-image systems physically, using the preliminary mass model,
which is relatively already well constrained.  Once multiple images
are found the model is refined and the best-fit model is obtained
either by a multi-dimensional grid minimization or an Markov Chain
Monte Carlo (MCMC) method.

Briefly, the adopted model parameterization is as follows.  Galaxies
located on the cluster's red sequence are identified as cluster
members and are modeled using a power-law surface mass density, scaled
by their apparent luminosity.  The individual galaxy contributions are
then added to represent the overall galaxy contribution to the total
deflection field.  The superposed mass distribution of the galaxies is
then smoothed, with either a 2D spline interpolation or a Gaussian
kernel, to obtain a light-traces-mass representation of the smooth
dark matter component.  These two components are then added with a
relative scaling to adjust for the relative contribution of galaxies
to the total mass and then the overall added deflection field is
normalized to the corresponding lensing distance.  In addition, it is
often useful to introduce an external shear imitating ellipticity so
that more flexibility is allowed when fitting the location of multiple
images.  For full details on the cluster lens modeling procedure, see
\cite{Zitrin2009}.

The second method is similar in essence to the first method, but the
main difference in the parameterization is that the DM is represented
by an analytical form, specifically an elliptical NFW profile (i.e.,
it is no longer represented by a smooth version of the galaxy light).
Compared to the first parameterization, this method is less coupled to
the light distribution and can give better fits to the data.

Each method has its own advantages and disadvantages and both methods
are well common in the literature.  A more explicit comparison was
discussed in several recent works such as \citep{Zitrin2013m0416,
Zitrin2013elgordo}, modeling MACS0416 and El Gordo, respectively.  The
underlying systematics can be also assessed further by comparing a
wider range different lens models as is now being done in the Frontier
Fields program,\footnote{See
http://archive.stsci.edu/prepds/frontier/lensmodels/.} which includes
our own models.  We typically estimate these systematics at the
10\%$-$20\% level in regions not too close to the critical curves
where the magnification diverges.  We are in the progress of
quantifying the accuracy of these methods in more detail (A. Zitrin et
al. 2014, in preparation).

We have constructed detailed lensing models for 17 of the 18 clusters
in this paper.  Ten of these models were constructed with the ``light
traces mass" technique, which excels in approximating the mass
distribution even with very few constraints because of the underlying
coupling to the light distribution.  The remaining seven were
constructed using the PIEMD parameterization.  We exclude
RXJ1532.9+3021 because we have not been able to clearly identify any
multiply imaged galaxies that are required to constrain the strong
lensing model.  Therefore, we do not further discuss the high-redshift
candidates behind RXJ1532 in this paper.  They are listed in
Appendix~\ref{sec:rxj1532} for reference.

The 17 cluster models include the published mass models for Abell~383
\citep{Zitrin2011a383}, MACS1149 \citep{Zheng2012}, Abell~2261
\citep{Coe2012}, MACS1206 \citep[][also see an alternative CLASH model
in \citealt{Eichner2013}]{Zitrin2012m1206}, MACS0329
\citep{Zitrin2012m0329}, MACS0717 \citep{Medezinski2013}, MACS0647
\citep{Coe2013}, and MACS0416 \citep{Zitrin2013m0416}.  The
unpublished lens models are for the clusters RXJ1347, MACS2129,
MS2137, MACS0744, MACS1115, Abell~611, MACS1720, MACS1931, and
RXCJ2248 (see also \citealt{Monna2014} for an alternate CLASH model).

\figcla
\figclb
\figclc
\figcld
\figcle

\subsection{Magnifications}
\label{sec:mu}

We show the approximate critical lines, where the magnification is
formally infinite, at $z\sim6$ for these clusters as the white
contours in Figures~\ref{fig:cl1} $-$ \ref{fig:cl4}.  Because of the
very small redshift dependence on the angular diameter distance ratio
$d_{ls}/d_{s}$ at $z \ga 6$, the critical curves for $z\sim7$ and
$z\sim8$ are similar in shape, but move slightly outward from those
shown for $z\sim6$.  At $z=0.44$, the mean redshift of the clusters
explored in this paper, the relative distance ratio $d_{ls}/d_{s}$ is
only 2\% higher at $z=8$ than at $z=6$.

We utilize the detailed cluster lens models to estimate the
magnifications of our high-redshift candidates, which are presented in
Tables~\ref{tbl:z6}, \ref{tbl:z7}, and \ref{tbl:z8}.  For the few
candidates that are located outside of the modeled region, we assign a
magnification of $\mu=1.1$, which is typically correct to a few
percent given their large radial distances from the cluster center.
The median magnifications from the models are 4.2, 4.2, and 4.5 for
the $z\sim6$, $z\sim7$, and $z\sim8$ samples, respectively, over an
average area of 4.5 arcmin$^2$ per cluster.  In total, nine of our
high-redshift candidates have estimated magnifications of $\mu > 100$
(i.e., amplifications of $>5$ mag) as a result of their close
proximity to the critical lines.  However, as discussed in
Section~\ref{sec:muerr}, we emphasize that these magnification factors
have enormous uncertainties.  A small change in the precise location
of the critical curve can result in a large but localized inherent
uncertainty in the magnification of a few select objects.

\subsection{Magnification Uncertainties}
\label{sec:muerr}

To estimate the magnification uncertainties, we tested our ability to
accurately measure magnifications given a test case of a simulated
lensing cluster.  The cluster is ``g1'' from the
numerical$-$hydrodynamical simulations discussed in \cite{Saro2006}.
This cluster is also part of the sample investigated in
\cite{Meneghetti2010}, where simulated observations of this cluster
with the SkyLens software \citep{Meneghetti2008} are presented.  Based
on our analysis of the simulated images, we correctly identified the
strongly lensed images of eight background galaxies, spanning the
redshift range 1.1 - 3.7.  We modeled the strong lensing using the
\cite{Zitrin2009} method, then compared our magnification map to the
``true'' magnification map from the simulated lensing
(Figure~\ref{fig:muerr}).

Consistent with previous work \citep{Bradac2009, Maizy2010}, we found
the magnification uncertainties increase, in general, as a function of
magnification.  However, we find larger uncertainties, as the
aforementioned study primarily investigated the uncertainties due to
low-mass cluster substructure not included in the lens models.  We
find that large model magnifications $> 30$ are most likely to be
significantly overestimated ($\ga 1\sigma$), as the lens model
critical curves (regions of formally infinite magnification) are
offset by $\sim3\arcsec$ from their true location, which is impossible
to deduce given the lack of multiple images around these positions.
This affects only a small percentage of our high-redshift candidates,
10\%, 8.5\%, and 15\% of our $z\sim6$, $z\sim7$, and $z\sim8$ samples,
respectively.  In addition, we warrant that while our method assumes
light traces mass, the simulations have a different way of assigning
the light to halos, which renders the presented comparison not pure.
Tests of additional simulated clusters are required to confirm these
levels of uncertainties as a more general result.

\subsection{Total Observed Area as a Function of Magnification}
\label{sec:muarea}

In Figure~\ref{fig:muarea}, we plot the total area over the 17
clusters as a function of the magnification factor.  Because of the
very small redshift dependence on $d_{ls}/d_{s}$ between $z\sim6$ and
$z\sim8$ (see Section~\ref{sec:models}), the areas at $z\sim7$ and
$z\sim8$ are very similar to those at $z\sim6$.  One consequence of
this effect is that the search volumes behind these clusters is
relatively insensitive to redshift, allowing for a differential
determination of UV LF with lower overall uncertainties.
\cite{Bouwens2012clash} use three lensed $z\sim9$ candidates in CLASH
and take advantage of this effect to derive the UV LF at $z\sim9$
based on the well-determined $z\sim8$ LF determined from the
HUDF09+ERS deep fields \citep{Bouwens2011b}.

Further, while local magnifications close to the critical curves can
have large uncertainties, the overall shape of this total area versus
magnification curve, which is critical in deriving lensed UV LFs (see
Section~\ref{sec:nden}), is not significantly affected by
uncertainties in lensing models.  Because the model constraints, which
are the multiple images and their redshifts, are not changed, the area
of high magnification is well known and is not very sensitive to the
exact position of the critical curves.  To surpass the HUDF12 depth,
we need to extend our limiting magnitude by 2 mag, which corresponds
to a magnification $\mu > 6.3$ (shaded area in
Figure~\ref{fig:muarea}).  We estimate that our observations of 17
clusters cover a total area of $22.9$~arcmin$^2$ where $\mu > 6.3$.

Due to the strong lensing effect, several regions in the observed
image plane can map back to the same area in the source plane at high
redshift.  However, the redundant search area reduces the effective
search areas by only $\sim10\%$.

\subsection{Multiply Lensed Systems}

Using the lensing models described above, we find seven likely
multiple image systems between $z\sim 5.5-8.5$ in the clusters
examined here.  These multiple systems are noted in
Tables~\ref{tbl:z6}, \ref{tbl:z7}, and \ref{tbl:z8}.  They include the
spectroscopically confirmed $z=6.027$ system in Abell~383
\citep{Richard2011}, the quadruply lensed galaxies at $z\sim6.2$ in
MACS0329 \citep{Zitrin2012m0329} and at $z=5.701$ in MACS1206
\citep{Zitrin2012m1206}, a doubly lensed galaxy at $z\sim6$ in
MACS0416 \citep{Zitrin2013m0416}, and a quintuply lensed $z\sim5.9$
galaxy in RXCJ2248 \citep{Monna2014}, which have been
spectroscopically confirmed at $z=6.11$ from our CLASH-VLT program
\citep{Balestra2013}.  We have also identified two new doubly lensed
multiple systems:  one at $z\sim6.5$ in MACS0647 and one at $z\sim6.6$
in MACS0744.  This represents the largest sample of multiply imaged
LBGs at $z>5.5$ presented to date.  For completeness, we also note
that the $z\sim10.7$ candidate behind MACS0647 is also a
multiple-image system with three separate images \citep{Coe2013} that
help to make this the most robust candidate at $z>10$.

\section{Observed Number Densities of Star-Forming Galaxies at $\lowercase{z} \sim 6-8$}
\label{sec:nden}

Gravitational lensing allows us to reach much deeper limiting
magnitudes ($\sim$2.0 - 2.5 mag over large areas), thus revealing a
previously unseen population of intrinsically faint star-forming
galaxies.  However, there is an important tradeoff to consider with
lensing searches.  The magnification effect also reduces the effective
source plane area at high-redshift inversely proportional to the
magnification ($A \sim \mu^{-1}$), which in turn reduces the search
volume behind the cluster.  Therefore, the overall efficiency of
cluster lensing searches depends critically on the slope of the galaxy
luminosity function at faint magnitudes.

The trade-off between depth and area is such that the surface density
should be enhanced over the field where the galaxy UV LF is steep
($-d(\log \phi) / d(\log L) > 1$), and reduced where the LF is
shallower \citep{Broadhurst1995}.  The effective slope at the bright
end (the exponential cutoff region for $L > \Lstar$) of the $z\sim6-8$
LFs is sufficiently steep such that the surface density of bright
high-redshift candidates should be higher behind lensing clusters than
in field surveys.  This effect is dependent on magnification such that
higher magnification regions should exhibit higher number densities
than lower magnification regions.

At fainter magnitudes, in the power-law regime of the UV LF (i.e., $L
< \Lstar$), one would expect the number counts to be diminished
because of the reduction in effective volume at high redshift.
However, the observed faint-end slopes recently derived for UV LFs at
$z\sim6 - 8$ \citep{Bouwens2011b, Oesch2012, Bradley2012b,
Schenker2013, McLure2013} are very steep, e.g., $\alpha = 1.98 \pm
0.2$ at $z\sim8$ \citep{Bradley2012b} corresponding to a $-d(\log
\phi) / d(\log L)$ effective slope of $\sim1$.  The consequence of the
very steep faint-end slopes is that the number densities behind
lensing clusters at faint magnitudes should be very similar or
slightly higher than those found in blank fields to the same limiting
magnitude.

\figmuerr
\figmuarea
\figlf

Using our sample of high-redshift lensed candidates and detailed
cluster lensing models, we can compare our number densities with
expectations for lensed fields.  The UV LF has been robustly derived
for ``blank'' fields at $z\sim6-7$ from deep {\HST} observations of
the GOODS-S, HUDF09, HUDF12, ERS, and CANDELS fields.  Here we utilize
the results from \cite{Bouwens2007b} and \cite{Bouwens2011b} based on
the GOODS-S, HUDF, and ERS fields.  At $z\sim6$, \cite{Bouwens2007b}
derive a UV LF with a normalization $\phi^{*} = 1.4^{+0.6}_{-0.4}
\times 10^{-3}$ Mpc$^{-3}$, a characteristic rest-frame UV absolute
magnitude of $M^{*}_{UV} = -20.24 \pm 0.19$, and a faint-end slope of
$\alpha = -1.74 \pm 0.16$.  At $z\sim7$, \cite{Bouwens2011b} find
$\phi^{*} = 0.86^{+0.7}_{-0.39} \times 10^{-3}$ Mpc$^{-3}$,
$M^{*}_{UV} = -20.14 \pm 0.26$, and $\alpha = -2.01 \pm 0.21$.

Combining the \HST\ WFC3 pure-parallel BoRG observations, which
constrain the bright end of the UV LF, with the deeper HUDF09+ERS
data, \cite{Bradley2012b} derived the $z\sim8$ UV LF over a very wide
dynamic range in magnitude.  The combined data sets are well fitted by
a Schechter function with $\phi^{*} =4.3^{+3.5}_{-2.1} \times 10^{-4}$
Mpc$^{-3}$, $M^{*}_{UV} = -20.26^{+0.29}_{-0.34}$, and $\alpha =
-1.98^{+0.23}_{-0.22}$.

In Figure~\ref{fig:lf}, we plot these field UV LFs with their
uncertainties as the red curves.  The expected field number counts are
derived using the total area covered by the lensing clusters.  Using
the SExtractor object segmentation maps, we exclude the search area
behind clusters lost as a result of intervening foreground sources.
We find a total area of 76.9 arcmin$^2$ over the 17 clusters.

We then convolve the UV LFs at $z\sim6$, $z\sim7$, and $z\sim8$
through our strong lens models of the 17 clusters.  This procedure
accounts for both effects of brightening the sources and the reduction
in search area with magnification.  These effects are encapsulated by
the total observed area in each magnification bin (see
Figure~\ref{fig:muarea}), which is relatively insensitive to the
magnification uncertainties.  The resulting expected lensed field
number counts are plotted with their uncertainties in
Figure~\ref{fig:lf} as the blue curves.  We include the lens model
uncertainties, estimated from our analysis of simulated lensing
(Section \ref{sec:muerr}), which we find to be subdominant to Poisson
uncertainties.

As a result of the apparent increased steepness of the faint-end slope
of the UV luminosity function at $z\sim7-8$ with $\alpha \sim -2.0$,
we find that lensing clusters are more efficient than blank field
surveys in searching for $z\ga7$ galaxies down to at least 29 AB mag.
At $z\sim6$, where the faint-end slope is relatively shallower with
$\alpha \sim -1.75$, it appears that lensing clusters are more
efficient than field surveys at the bright end down to $\sim27$ AB
mag.  Fainter than 27 AB mag, field surveys at $z\sim6$ appear to be
only marginally more efficient than lensing surveys down to at least
29 AB mag.

In Figure~\ref{fig:lf}, we also plot our observed number counts, with
68\% ($1\sigma$) confidence intervals for a Poisson distribution, for
our lensed galaxy samples at $z\sim6$, $z\sim7$, and $z\sim8$.  Our
number counts have been corrected for an estimated contamination
fraction of $12\%$ (see Section~\ref{sec:contaminants}).  To estimate
the effects of photometric scatter on the photometric-redshift
completeness of our samples, we performed Monte Carlo simulations by
generating 1000 random realizations of each high-redshift galaxy in
our sample within its $1\sigma$ photometric errors in each band.  We
then run BPZ on the randomly generated photometry and determine the
derived photometric redshifts to estimate our completeness as a
function of observed magnitude.

As can be seen in the figure, the observed number counts of our lensed
high-redshift sample are roughly consistent with the expected lensed
number counts down to $\sim27$ mag, where we begin to suffer
significant ($>50$\%) and dramatically increasing incompleteness.  In
particular, in all three redshift bins we find a higher number density
at brighter observed magnitudes than the field predictions.  The
observed number counts for our $z\sim6$ sample, where we have good
statistics, are overall in excellent agreement with the lensed
expectations down to 27 AB mag.  The $z\sim7$ and $z\sim8$ samples
likely suffer from the effects of small sample statistics, but the
lensing effect is still clearly evident at bright magnitudes where the
lensing effect is most pronounced because of the steepness of the LF.
A more detailed exploration of incompleteness and contamination of our
high-redshift candidates will be explored in an upcoming study to
derive accurate effective volumes behind the clusters and lensed LFs
in a Bayesian framework (L. A. Moustakas et al., in preparation)

\section{Summary}
\label{sec:summary}

We have analyzed the 16 band \HST\ observations of 18 lensing clusters
obtained as part of the CLASH MCT program to search for $z\sim6-8$
galaxies.  Using purely photometric redshift selections, we find 204,
45, and 13 high-redshift LBG candidates at $z\sim6$, $z\sim7$, and
$z\sim8$, respectively.  Our large sample of magnified star-forming
galaxies at these redshifts represents the largest sample presented to
date, nearly an order of magnitude largest that previous lensed
samples.  The accurate photometric redshift selections obtained here
are enabled by our observations of these $z\sim 6-8$ LBG candidates in
seven ACS optical and all five WFC3/IR broadband filters.

We constructed detailed lensing models for 17 of the 18 clusters
(excluding RXJ1532) searched in this paper.  We utilize these models
to both estimate object magnifications and to identify two new
multiply lensed $z\ga6$ candidates.  The median magnifications
provided by these 17 clusters are 4.2, 4.2, and 4.5 for the $z\sim6$,
$z\sim7$, and $z\sim8$ samples, respectively, over an average area of
4.5 arcmin$^2$ per cluster.  We note that the highest magnifications
have the largest magnification errors due to inherent uncertainties in
the precise location of the critical curves, as discussed in
Section~\ref{sec:muerr}.

The intrinsic magnitudes in the relatively shallow CLASH survey reach
deeper ($>29.5$ AB mag) than the ultra-deep HUDF12 observations thanks
to the strong lensing effect.  Our observations of the 17 clusters
cover a total area of $\sim22.9$~arcmin$^2$ with magnifications $\mu >
6.3$, needed to surpass the depth of the HUDF12 observations.  This is
a relatively robust measure of area in spite of the model
uncertainties (see Section~\ref{sec:muarea}).

Utilizing our detailed lensing models, we identified seven likely
multiple image systems over the 17 clusters explored in this paper.
Five of them have been previously found in CLASH data: the
spectroscopically confirmed $z=6.027$ system in Abell~383
\citep{Richard2011}, the quadruply lensed galaxy at $z\sim6.2$ in
MACS0329 \citep{Zitrin2012m0329}, the spectroscopically confirmed
quadruple system at $z=5.701$ in MACS1206 \citep{Zitrin2012m1206}, a
doubly lensed galaxy at $z\sim6$ in MACS0416 \citep{Zitrin2013m0416},
and a quintuply lensed $z\sim5.9$ galaxy in RXCJ2248
\citep{Monna2014}, which have been spectroscopically confirmed at
$z=6.11$ from our CLASH-VLT program \citep{Balestra2013}.  We find two
new multiply lensed systems, one at $z\sim6.5$ in MACS0647 and one at
$z\sim6.6$ in MACS0744.  In total, this represents the largest sample
of multiply imaged LBGs at $z>5.5$ presented to date.

Finally, we compare ``blank'' field UV LFs with their lensed
counterparts and our observed number counts with expectations based on
convolving ``blank'' field UV LFs with the 17 cluster lens models.  We
find that lensing clusters are more efficient than blank field surveys
in searching for $z\ga7$ galaxies down to at least 29 AB mag.  This
result follows from the apparent increased steepness of the faint-end
slope of the UV luminosity function at $z\sim7-8$ with $\alpha \sim
-2.0$ \citep[e.g.,][]{Bouwens2011b, Oesch2012, Bradley2012b,
Schenker2013, McLure2013}.  At $z\sim6$, we find that lensing clusters
are more efficient than field surveys at the bright end down to
$\sim26$ AB mag due to the relatively shallower faint-end slope of
$\alpha \sim -1.75$ \citep{Bouwens2011b}.

The observed number counts of our lensed high-redshift sample are
approximately consistent with the expected lensed number counts down
to $\sim27$ mag, where we begin to suffer significant incompleteness.
Our number counts have been corrected for a small ($\sim12\%$)
contamination from low-redshift red galaxies (see
Section~\ref{sec:contaminants}).  Where we have our best statistics at
$z\sim6$, we find our observed number counts overall to be in
excellent agreement with the lensed expectations down to 27 AB mag.
For the $z\sim7$ and $z\sim8$ samples, which likely suffer from the
effects of small number statistics, the lensing effect is also clearly
evident at bright magnitudes where the lensing effect is most
pronounced.  In all three redshift bins, our observed number densities
are higher than one would expect from a ``blank'' field survey at the
brightest magnitudes, where the lensing effects are most significant
because of the steepness of the LF.

This large new sample of lensed star-forming galaxies at $z \ga 5.5$
provides a wealth of information on galaxies in the reionization epoch
of the universe.  Because these galaxies are brighter than typical
field surveys, a sample of high-redshift CLASH candidates have also
been detected and studied with \Spitzer/IRAC at $3.6\mu$m and
$4.5\mu$m, elucidating their stellar masses and specific
star-formation rates \citep{Zitrin2012m0329, Zheng2012, Coe2013,
Smit2014} and nebular emission-line strengths \citep{Smit2014}.
Future work will focus on additional spectroscopic followup
observations, investigating their rest-frame UV colors, deriving
accurate effective volumes behind the clusters and lensed UV LFs in a
Bayesian framework (L. A. Moustakas et al., in preparation), and
studying their intrinsic sizes and morphologies.

With the recent exciting CLASH discoveries of two $z\sim9$ candidates
\citep{Bouwens2012clash}, a $z\sim9.6$ candidate \citep{Zheng2012},
and a triply lensed candidate at $z\sim10.7$ behind MACS0647
\citep{Coe2013}, lensing clusters have proven to be a powerful tool in
the discovery and study of high-redshift galaxies.  This technique
will continue to be availed with the HFF campaign, which will obtain
ultra-deep ACS and WFC3/IR observations of four to six lensing
clusters (four of which are presented in this paper) to unprecedented
depths.

\acknowledgments

We thank the anonymous referee for very helpful feedback that improved
the paper.  We are especially grateful to our program coordinator Beth
Perrillo for her expert assistance in implementing the \HST\
observations in this program.  We thank Jay Anderson and Norman Grogin
for providing the ACS CTE and bias striping correction algorithms used
in our data pipeline.  Finally, we are indebted to the hundreds of
people who have labored many years to plan, develop, manufacture,
install, repair, and calibrate the WFC3 and ACS instruments as well as
to all those who maintain and operate the {\em Hubble Space
Telescope}.

The CLASH Multi-Cycle Treasury Program (GO-12065) is based on
observations made with the NASA/ESA Hubble Space Telescope.  The Space
Telescope Science Institute is operated by the Association of
Universities for Research in Astronomy, Inc. under NASA contract NAS
5-26555.  A.Z. is supported by contract research ``Internationale
Spitzenforschung II/2-6'' of the Baden W\"urttemberg Stiftung.  The
work of L.A.M., J.M., and M.M. was carried out at Jet Propulsion
Laboratory, California Institute of Technology, under a contract with
NASA.  The Dark Cosmology Centre is funded by the DNRF.

%------------------------------------------------------------------------------
% Bibliography
%------------------------------------------------------------------------------
%\bibliographystyle{apj}
%\bibliographystyle{astroads}
%\bibliography{highz}

\begin{thebibliography}{104}
\expandafter\ifx\csname natexlab\endcsname\relax\def\natexlab#1{#1}\fi
\expandafter\ifx\csname href\endcsname\relax
  \def\href#1#2{}\fi
\expandafter\ifx\csname urllinklabel\endcsname\relax
  \def\urllinklabel{[URL]}\fi
\expandafter\ifx\csname adsurllinklabel\endcsname\relax
  \def\adsurllinklabel{[ADS]}\fi

\bibitem[{{Arnouts} {et~al.}(1999){Arnouts}, {Cristiani}, {Moscardini}, {Matarrese}, {Lucchin}, {Fontana}, \& {Giallongo}}]{Arnouts1999}
{Arnouts}, S., {Cristiani}, S., {Moscardini}, L., {Matarrese}, S., {Lucchin}, F., {Fontana}, A., \& {Giallongo}, E. 1999, \mnras, 310, 540 \href{http://adsabs.harvard.edu/abs/1999MNRAS.310..540A}{\adsurllinklabel}

\bibitem[{{Atek} {et~al.}(2011){Atek}, {Siana}, {Scarlata}, {Malkan}, {McCarthy}, {Teplitz}, {Henry}, {Colbert}, {Bridge}, {Bunker}, {Dressler}, {Fosbury}, {Hathi}, {Martin}, {Ross}, \& {Shim}}]{Atek2011}
{Atek}, H., {Siana}, B., {Scarlata}, C., {et al.} 2011, \apj, 743, 121 \href{http://adsabs.harvard.edu/abs/2011ApJ...743..121A}{\adsurllinklabel}

\bibitem[{{Balestra} {et~al.}(2013){Balestra}, {Vanzella}, {Rosati}, {Monna}, {Grillo}, {Nonino}, {Mercurio}, {Biviano}, {Bradley}, {Coe}, {Fritz}, {Postman}, {Seitz}, {Scodeggio}, {Tozzi}, {Zheng}, {Ziegler}, {Zitrin}, {Annunziatella}, {Bartelmann}, {Benitez}, {Broadhurst}, {Bouwens}, {Czoske}, {Donahue}, {Ford}, {Girardi}, {Infante}, {Jouvel}, {Kelson}, {Koekemoer}, {Kuchner}, {Lemze}, {Lombardi}, {Maier}, {Medezinski}, {Melchior}, {Meneghetti}, {Merten}, {Molino}, {Moustakas}, {Presotto}, {Smit}, \& {Umetsu}}]{Balestra2013}
{Balestra}, I., {Vanzella}, E., {Rosati}, P., {et al.} 2013, \aap, 559, L9 \href{http://adsabs.harvard.edu/abs/2013A%26A...559L...9B}{\adsurllinklabel}

\bibitem[{{Ben{\'{\i}}tez}(2000){Ben{\'{\i}}tez}}]{Benitez2000}
{Ben{\'{\i}}tez}, N. 2000, \apj, 536, 571 \href{http://adsabs.harvard.edu/abs/2000ApJ...536..571B}{\adsurllinklabel}

\bibitem[{{Ben{\'{\i}}tez} {et~al.}(2004){Ben{\'{\i}}tez}, {Ford}, {Bouwens}, {Menanteau}, {Blakeslee}, {Gronwall}, {Illingworth}, {Meurer}, {Broadhurst}, {Clampin}, {Franx}, {Hartig}, {Magee}, {Sirianni}, {Ardila}, {Bartko}, {Brown}, {Burrows}, {Cheng}, {Cross}, {Feldman}, {Golimowski}, {Infante}, {Kimble}, {Krist}, {Lesser}, {Levay}, {Martel}, {Miley}, {Postman}, {Rosati}, {Sparks}, {Tran}, {Tsvetanov}, {White}, \& {Zheng}}]{Benitez2004}
{Ben{\'{\i}}tez}, N., {Ford}, H., {Bouwens}, R., {et al.} 2004, \apjs, 150, 1 \href{http://adsabs.harvard.edu/abs/2004ApJS..150....1B}{\adsurllinklabel}

\bibitem[{{Bertin} \& {Arnouts}(1996){Bertin} \& {Arnouts}}]{Bertin1996}
{Bertin}, E. \& {Arnouts}, S. 1996, \aaps, 117, 393 \href{http://adsabs.harvard.edu/abs/1996A%26AS..117..393B}{\adsurllinklabel}

\bibitem[{{Bouwens} {et~al.}(2012{\natexlab{a}}){Bouwens}, {Bradley}, {Zitrin}, {Coe}, {Franx}, {Zheng}, {Smit}, {Host}, {Postman}, {Moustakas}, {Labbe}, {Carrasco}, {Molino}, {Donahue}, {Kelson}, {Meneghetti}, {Jha}, {Benitez}, {Lemze}, {Umetsu}, {Broadhurst}, {Moustakas}, {Rosati}, {Bartelmann}, {Ford}, {Graves}, {Grillo}, {Infante}, {Jiminez-Teja}, {Jouvel}, {Lahav}, {Maoz}, {Medezinski}, {Melchior}, {Merten}, {Nonino}, {Ogaz}, \& {Seitz}}]{Bouwens2012clash}
{Bouwens}, R., {Bradley}, L., {Zitrin}, A., {et al.} 2012{\natexlab{a}}, ArXiv e-prints, 1211.2230 \href{http://adsabs.harvard.edu/abs/2012arXiv1211.2230B}{\adsurllinklabel}

\bibitem[{{Bouwens} {et~al.}(2004){Bouwens}, {Illingworth}, {Blakeslee}, {Broadhurst}, \& {Franx}}]{Bouwens2004b}
{Bouwens}, R.~J., {Illingworth}, G.~D., {Blakeslee}, J.~P., {Broadhurst}, T.~J., \& {Franx}, M. 2004, \apjl, 611, L1 \href{http://adsabs.harvard.edu/abs/2004ApJ...611L...1B}{\adsurllinklabel}

\bibitem[{{Bouwens} {et~al.}(2009){Bouwens}, {Illingworth}, {Bradley}, {Ford}, {Franx}, {Zheng}, {Broadhurst}, {Coe}, \& {Jee}}]{Bouwens2009a}
{Bouwens}, R.~J., {Illingworth}, G.~D., {Bradley}, L.~D., {et al.} 2009, \apj, 690, 1764 \href{http://adsabs.harvard.edu/abs/2009ApJ...690.1764B}{\adsurllinklabel}

\bibitem[{{Bouwens} {et~al.}(2007){Bouwens}, {Illingworth}, {Franx}, \& {Ford}}]{Bouwens2007b}
{Bouwens}, R.~J., {Illingworth}, G.~D., {Franx}, M., \& {Ford}, H. 2007, \apj, 670, 928 \href{http://adsabs.harvard.edu/abs/2007ApJ...670..928B}{\adsurllinklabel}

\bibitem[{{Bouwens} {et~al.}(2012{\natexlab{b}}){Bouwens}, {Illingworth}, {Oesch}, {Franx}, {Labb{\'e}}, {Trenti}, {van Dokkum}, {Carollo}, {Gonz{\'a}lez}, {Smit}, \& {Magee}}]{Bouwens2012beta}
{Bouwens}, R.~J., {Illingworth}, G.~D., {Oesch}, P.~A., {et al.} 2012{\natexlab{b}}, \apj, 754, 83 \href{http://adsabs.harvard.edu/abs/2012ApJ...754...83B}{\adsurllinklabel}

\bibitem[{{Bouwens} {et~al.}(2011){Bouwens}, {Illingworth}, {Oesch}, {Labb{\'e}}, {Trenti}, {van Dokkum}, {Franx}, {Stiavelli}, {Carollo}, {Magee}, \& {Gonzalez}}]{Bouwens2011b}
{Bouwens}, R.~J., {Illingworth}, G.~D., {Oesch}, P.~A., {et al.} 2011, \apj, 737, 90 \href{http://adsabs.harvard.edu/abs/2011ApJ...737...90B}{\adsurllinklabel}

\bibitem[{{Bouwens} {et~al.}(2013){Bouwens}, {Illingworth}, {Oesch}, {Labbe}, {van Dokkum}, {Trenti}, {Franx}, {Smit}, {Gonzalez}, \& {Magee}}]{Bouwens2013beta}
{Bouwens}, R.~J., {Illingworth}, G.~D., {Oesch}, P.~A., {et al.} 2013, ArXiv e-prints, 1306.2950 \href{http://adsabs.harvard.edu/abs/2013arXiv1306.2950B}{\adsurllinklabel}

\bibitem[{{Brada{\v c}} {et~al.}(2009){Brada{\v c}}, {Treu}, {Applegate}, {Gonzalez}, {Clowe}, {Forman}, {Jones}, {Marshall}, {Schneider}, \& {Zaritsky}}]{Bradac2009}
{Brada{\v c}}, M., {Treu}, T., {Applegate}, D., {et al.} 2009, \apj, 706, 1201 \href{http://adsabs.harvard.edu/abs/2009ApJ...706.1201B}{\adsurllinklabel}

\bibitem[{{Brada{\v c}} {et~al.}(2012){Brada{\v c}}, {Vanzella}, {Hall}, {Treu}, {Fontana}, {Gonzalez}, {Clowe}, {Zaritsky}, {Stiavelli}, \& {Cl{\'e}ment}}]{Bradac2012}
{Brada{\v c}}, M., {Vanzella}, E., {Hall}, N., {et al.} 2012, \apjl, 755, L7 \href{http://adsabs.harvard.edu/abs/2012ApJ...755L...7B}{\adsurllinklabel}

\bibitem[{{Bradley} {et~al.}(2008){Bradley}, {Bouwens}, {Ford}, {Illingworth}, {Jee}, {Ben{\'{\i}}tez}, {Broadhurst}, {Franx}, {Frye}, {Infante}, {Motta}, {Rosati}, {White}, \& {Zheng}}]{Bradley2008}
{Bradley}, L.~D., {Bouwens}, R.~J., {Ford}, H.~C., {et al.} 2008, \apj, 678, 647 \href{http://adsabs.harvard.edu/abs/2008ApJ...678..647B}{\adsurllinklabel}

\bibitem[{{Bradley} {et~al.}(2012{\natexlab{a}}){Bradley}, {Bouwens}, {Zitrin}, {Smit}, {Coe}, {Ford}, {Zheng}, {Illingworth}, {Ben{\'{\i}}tez}, \& {Broadhurst}}]{Bradley2012a}
{Bradley}, L.~D., {Bouwens}, R.~J., {Zitrin}, A., {et al.} 2012{\natexlab{a}}, \apj, 747, 3 \href{http://adsabs.harvard.edu/abs/2012ApJ...747....3B}{\adsurllinklabel}

\bibitem[{{Bradley} {et~al.}(2012{\natexlab{b}}){Bradley}, {Trenti}, {Oesch}, {Stiavelli}, {Treu}, {Bouwens}, {Shull}, {Holwerda}, \& {Pirzkal}}]{Bradley2012b}
{Bradley}, L.~D., {Trenti}, M., {Oesch}, P.~A., {et al.} 2012{\natexlab{b}}, \apj, 760, 108 \href{http://adsabs.harvard.edu/abs/2012ApJ...760..108B}{\adsurllinklabel}

\bibitem[{{Broadhurst} {et~al.}(2005){Broadhurst}, {Ben{\'{\i}}tez}, {Coe}, {Sharon}, {Zekser}, {White}, {Ford}, {Bouwens}, {Blakeslee}, {Clampin}, {Cross}, {Franx}, {Frye}, {Hartig}, {Illingworth}, {Infante}, {Menanteau}, {Meurer}, {Postman}, {Ardila}, {Bartko}, {Brown}, {Burrows}, {Cheng}, {Feldman}, {Golimowski}, {Goto}, {Gronwall}, {Herranz}, {Holden}, {Homeier}, {Krist}, {Lesser}, {Martel}, {Miley}, {Rosati}, {Sirianni}, {Sparks}, {Steindling}, {Tran}, {Tsvetanov}, \& {Zheng}}]{Broadhurst2005}
{Broadhurst}, T., {Ben{\'{\i}}tez}, N., {Coe}, D., {et al.} 2005, \apj, 621, 53 \href{http://adsabs.harvard.edu/abs/2005ApJ...621...53B}{\adsurllinklabel}

\bibitem[{{Broadhurst} {et~al.}(1995){Broadhurst}, {Taylor}, \& {Peacock}}]{Broadhurst1995}
{Broadhurst}, T.~J., {Taylor}, A.~N., \& {Peacock}, J.~A. 1995, \apj, 438, 49 \href{http://adsabs.harvard.edu/abs/1995ApJ...438...49B}{\adsurllinklabel}

\bibitem[{{Bunker} {et~al.}(2010){Bunker}, {Wilkins}, {Ellis}, {Stark}, {Lorenzoni}, {Chiu}, {Lacy}, {Jarvis}, \& {Hickey}}]{Bunker2010}
{Bunker}, A.~J., {Wilkins}, S., {Ellis}, R.~S., {et al.} 2010, \mnras, 409, 855 \href{http://adsabs.harvard.edu/abs/2010MNRAS.409..855B}{\adsurllinklabel}

\bibitem[{{Calzetti} {et~al.}(2000){Calzetti}, {Armus}, {Bohlin}, {Kinney}, {Koornneef}, \& {Storchi-Bergmann}}]{Calzetti2000}
{Calzetti}, D., {Armus}, L., {Bohlin}, R.~C., {Kinney}, A.~L., {Koornneef}, J., \& {Storchi-Bergmann}, T. 2000, \apj, 533, 682 \href{http://adsabs.harvard.edu/abs/2000ApJ...533..682C}{\adsurllinklabel}

\bibitem[{{Casertano} {et~al.}(2000){Casertano}, {de Mello}, {Dickinson}, {Ferguson}, {Fruchter}, {Gonzalez-Lopezlira}, {Heyer}, {Hook}, {Levay}, {Lucas}, {Mack}, {Makidon}, {Mutchler}, {Smith}, {Stiavelli}, {Wiggs}, \& {Williams}}]{Casertano2000}
{Casertano}, S., {de Mello}, D., {Dickinson}, M., {et al.} 2000, \aj, 120, 2747 \href{http://adsabs.harvard.edu/abs/2000AJ....120.2747C}{\adsurllinklabel}

\bibitem[{{Cimatti} {et~al.}(2002){Cimatti}, {Daddi}, {Mignoli}, {Pozzetti}, {Renzini}, {Zamorani}, {Broadhurst}, {Fontana}, {Saracco}, {Poli}, {Cristiani}, {D'Odorico}, {Giallongo}, {Gilmozzi}, \& {Menci}}]{Cimatti2002}
{Cimatti}, A., {Daddi}, E., {Mignoli}, M., {et al.} 2002, \aap, 381, L68 \href{http://adsabs.harvard.edu/abs/2002A%26A...381L..68C}{\adsurllinklabel}

\bibitem[{{Coe} {et~al.}(2006){Coe}, {Ben{\'{\i}}tez}, {S{\'a}nchez}, {Jee}, {Bouwens}, \& {Ford}}]{Coe2006}
{Coe}, D., {Ben{\'{\i}}tez}, N., {S{\'a}nchez}, S.~F., {Jee}, M., {Bouwens}, R., \& {Ford}, H. 2006, \aj, 132, 926 \href{http://adsabs.harvard.edu/abs/2006AJ....132..926C}{\adsurllinklabel}

\bibitem[{{Coe} {et~al.}(2012){Coe}, {Umetsu}, {Zitrin}, {Donahue}, {Medezinski}, {Postman}, {Carrasco}, {Anguita}, {Geller}, {Rines}, {Diaferio}, {Kurtz}, {Bradley}, {Koekemoer}, {Zheng}, {Nonino}, {Molino}, {Mahdavi}, {Lemze}, {Infante}, {Ogaz}, {Melchior}, {Host}, {Ford}, {Grillo}, {Rosati}, {Jim{\'e}nez-Teja}, {Moustakas}, {Broadhurst}, {Ascaso}, {Lahav}, {Bartelmann}, {Ben{\'{\i}}tez}, {Bouwens}, {Graur}, {Graves}, {Jha}, {Jouvel}, {Kelson}, {Moustakas}, {Maoz}, {Meneghetti}, {Merten}, {Riess}, {Rodney}, \& {Seitz}}]{Coe2012}
{Coe}, D., {Umetsu}, K., {Zitrin}, A., {et al.} 2012, \apj, 757, 22 \href{http://adsabs.harvard.edu/abs/2012ApJ...757...22C}{\adsurllinklabel}

\bibitem[{{Coe} {et~al.}(2013){Coe}, {Zitrin}, {Carrasco}, {Shu}, {Zheng}, {Postman}, {Bradley}, {Koekemoer}, {Bouwens}, {Broadhurst}, {Monna}, {Host}, {Moustakas}, {Ford}, {Moustakas}, {van der Wel}, {Donahue}, {Rodney}, {Ben{\'{\i}}tez}, {Jouvel}, {Seitz}, {Kelson}, \& {Rosati}}]{Coe2013}
{Coe}, D., {Zitrin}, A., {Carrasco}, M., {et al.} 2013, \apj, 762, 32 \href{http://adsabs.harvard.edu/abs/2013ApJ...762...32C}{\adsurllinklabel}

\bibitem[{{Cushing} {et~al.}(2005){Cushing}, {Rayner}, \& {Vacca}}]{Cushing2005}
{Cushing}, M.~C., {Rayner}, J.~T., \& {Vacca}, W.~D. 2005, \apj, 623, 1115 \href{http://adsabs.harvard.edu/abs/2005ApJ...623.1115C}{\adsurllinklabel}

\bibitem[{{Dunlop}(2013){Dunlop}}]{Dunlop2013r}
{Dunlop}, J.~S. 2013, Astrophysics and Space Science Library, 396, 223 \href{http://adsabs.harvard.edu/abs/2013ASSL..396..223D}{\adsurllinklabel}

\bibitem[{{Dunlop} {et~al.}(2013){Dunlop}, {Rogers}, {McLure}, {Ellis}, {Robertson}, {Koekemoer}, {Dayal}, {Curtis-Lake}, {Wild}, {Charlot}, {Bowler}, {Schenker}, {Ouchi}, {Ono}, {Cirasuolo}, {Furlanetto}, {Stark}, {Targett}, \& {Schneider}}]{Dunlop2013}
{Dunlop}, J.~S., {Rogers}, A.~B., {McLure}, R.~J., {et al.} 2013, \mnras, 432, 3520 \href{http://adsabs.harvard.edu/abs/2013MNRAS.432.3520D}{\adsurllinklabel}

\bibitem[{{Ebeling} {et~al.}(2007){Ebeling}, {Barrett}, {Donovan}, {Ma}, {Edge}, \& {van Speybroeck}}]{Ebeling2007}
{Ebeling}, H., {Barrett}, E., {Donovan}, D., {Ma}, C.-J., {Edge}, A.~C., \& {van Speybroeck}, L. 2007, \apjl, 661, L33 \href{http://adsabs.harvard.edu/abs/2007ApJ...661L..33E}{\adsurllinklabel}

\bibitem[{{Ebeling} {et~al.}(2001){Ebeling}, {Edge}, \& {Henry}}]{Ebeling2001}
{Ebeling}, H., {Edge}, A.~C., \& {Henry}, J.~P. 2001, \apj, 553, 668 \href{http://adsabs.harvard.edu/abs/2001ApJ...553..668E}{\adsurllinklabel}

\bibitem[{{Ebeling} {et~al.}(2010){Ebeling}, {Edge}, {Mantz}, {Barrett}, {Henry}, {Ma}, \& {van Speybroeck}}]{Ebeling2010}
{Ebeling}, H., {Edge}, A.~C., {Mantz}, A., {Barrett}, E., {Henry}, J.~P., {Ma}, C.~J., \& {van Speybroeck}, L. 2010, \mnras, 407, 83 \href{http://adsabs.harvard.edu/abs/2010MNRAS.407...83E}{\adsurllinklabel}

\bibitem[{{Egami} {et~al.}(2005){Egami}, {Kneib}, {Rieke}, {Ellis}, {Richard}, {Rigby}, {Papovich}, {Stark}, {Santos}, {Huang}, {Dole}, {Le Floc'h}, \& {P{\'e}rez-Gonz{\'a}lez}}]{Egami2005}
{Egami}, E., {Kneib}, J.-P., {Rieke}, G.~H., {et al.} 2005, \apjl, 618, L5 \href{http://adsabs.harvard.edu/abs/2005ApJ...618L...5E}{\adsurllinklabel}

\bibitem[{{Eichner} {et~al.}(2013){Eichner}, {Seitz}, {Suyu}, {Halkola}, {Umetsu}, {Zitrin}, {Coe}, {Monna}, {Rosati}, {Grillo}, {Balestra}, {Postman}, {Koekemoer}, {Zheng}, {H{\o}st}, {Lemze}, {Broadhurst}, {Moustakas}, {Bradley}, {Molino}, {Nonino}, {Mercurio}, {Scodeggio}, {Bartelmann}, {Benitez}, {Bouwens}, {Donahue}, {Infante}, {Jouvel}, {Kelson}, {Lahav}, {Medezinski}, {Melchior}, {Merten}, \& {Riess}}]{Eichner2013}
{Eichner}, T., {Seitz}, S., {Suyu}, S.~H., {et al.} 2013, \apj, 774, 124 \href{http://adsabs.harvard.edu/abs/2013ApJ...774..124E}{\adsurllinklabel}

\bibitem[{{Ellis} {et~al.}(2013){Ellis}, {McLure}, {Dunlop}, {Robertson}, {Ono}, {Schenker}, {Koekemoer}, {Bowler}, {Ouchi}, {Rogers}, {Curtis-Lake}, {Schneider}, {Charlot}, {Stark}, {Furlanetto}, \& {Cirasuolo}}]{Ellis2013}
{Ellis}, R.~S., {McLure}, R.~J., {Dunlop}, J.~S., {et al.} 2013, \apjl, 763, L7 \href{http://adsabs.harvard.edu/abs/2013ApJ...763L...7E}{\adsurllinklabel}

\bibitem[{{Ferguson} {et~al.}(2004){Ferguson}, {Dickinson}, {Giavalisco}, {Kretchmer}, {Ravindranath}, {Idzi}, {Taylor}, {Conselice}, {Fall}, {Gardner}, {Livio}, {Madau}, {Moustakas}, {Papovich}, {Somerville}, {Spinrad}, \& {Stern}}]{Ferguson2004}
{Ferguson}, H.~C., {Dickinson}, M., {Giavalisco}, M., {et al.} 2004, \apjl, 600, L107 \href{http://adsabs.harvard.edu/abs/2004ApJ...600L.107F}{\adsurllinklabel}

\bibitem[{{Finkelstein} {et~al.}(2010){Finkelstein}, {Papovich}, {Giavalisco}, {Reddy}, {Ferguson}, {Koekemoer}, \& {Dickinson}}]{Finkelstein2010}
{Finkelstein}, S.~L., {Papovich}, C., {Giavalisco}, M., {Reddy}, N.~A., {Ferguson}, H.~C., {Koekemoer}, A.~M., \& {Dickinson}, M. 2010, \apj, 719, 1250 \href{http://adsabs.harvard.edu/abs/2010ApJ...719.1250F}{\adsurllinklabel}

\bibitem[{{Fioc} \& {Rocca-Volmerange}(1997){Fioc} \& {Rocca-Volmerange}}]{Fioc1997}
{Fioc}, M. \& {Rocca-Volmerange}, B. 1997, \aap, 326, 950 \href{http://adsabs.harvard.edu/abs/1997A%26A...326..950F}{\adsurllinklabel}

\bibitem[{{Franx} {et~al.}(1997){Franx}, {Illingworth}, {Kelson}, {van Dokkum}, \& {Tran}}]{Franx1997}
{Franx}, M., {Illingworth}, G.~D., {Kelson}, D.~D., {van Dokkum}, P.~G., \& {Tran}, K.-V. 1997, \apjl, 486, L75 \href{http://adsabs.harvard.edu/abs/1997ApJ...486L..75F}{\adsurllinklabel}

\bibitem[{{Giallongo} {et~al.}(2005){Giallongo}, {Salimbeni}, {Menci}, {Zamorani}, {Fontana}, {Dickinson}, {Cristiani}, \& {Pozzetti}}]{Giallongo2005}
{Giallongo}, E., {Salimbeni}, S., {Menci}, N., {Zamorani}, G., {Fontana}, A., {Dickinson}, M., {Cristiani}, S., \& {Pozzetti}, L. 2005, \apj, 622, 116 \href{http://adsabs.harvard.edu/abs/2005ApJ...622..116G}{\adsurllinklabel}

\bibitem[{{Grazian} {et~al.}(2012){Grazian}, {Castellano}, {Fontana}, {Pentericci}, {Dunlop}, {McLure}, {Koekemoer}, {Dickinson}, {Faber}, {Ferguson}, {Galametz}, {Giavalisco}, {Grogin}, {Hathi}, {Kocevski}, {Lai}, {Newman}, \& {Vanzella}}]{Grazian2012}
{Grazian}, A., {Castellano}, M., {Fontana}, A., {et al.} 2012, \aap, 547, A51 \href{http://adsabs.harvard.edu/abs/2012A%26A...547A..51G}{\adsurllinklabel}

\bibitem[{{Grogin} {et~al.}(2011){Grogin}, {Kocevski}, {Faber}, {Ferguson}, {Koekemoer}, {Riess}, {Acquaviva}, {Alexander}, {Almaini}, {Ashby}, {Barden}, {Bell}, {Bournaud}, {Brown}, {Caputi}, {Casertano}, {Cassata}, {Castellano}, {Challis}, {Chary}, {Cheung}, {Cirasuolo}, {Conselice}, {Roshan Cooray}, {Croton}, {Daddi}, {Dahlen}, {Dav{\'e}}, {de Mello}, {Dekel}, {Dickinson}, {Dolch}, {Donley}, {Dunlop}, {Dutton}, {Elbaz}, {Fazio}, {Filippenko}, {Finkelstein}, {Fontana}, {Gardner}, {Garnavich}, {Gawiser}, {Giavalisco}, {Grazian}, {Guo}, {Hathi}, {H{\"a}ussler}, {Hopkins}, {Huang}, {Huang}, {Jha}, {Kartaltepe}, {Kirshner}, {Koo}, {Lai}, {Lee}, {Li}, {Lotz}, {Lucas}, {Madau}, {McCarthy}, {McGrath}, {McIntosh}, {McLure}, {Mobasher}, {Moustakas}, {Mozena}, {Nandra}, {Newman}, {Niemi}, {Noeske}, {Papovich}, {Pentericci}, {Pope}, {Primack}, {Rajan}, {Ravindranath}, {Reddy}, {Renzini}, {Rix}, {Robaina}, {Rodney}, {Rosario}, {Rosati}, {Salimbeni}, {Scarlata}, {Siana}, {Simard}, {Smidt}, {Somerville}, {Spinrad}, {Straughn}, {Strolger}, {Telford}, {Teplitz}, {Trump}, {van der Wel}, {Villforth}, {Wechsler}, {Weiner}, {Wiklind}, {Wild}, {Wilson}, {Wuyts}, {Yan}, \& {Yun}}]{Grogin2011}
{Grogin}, N.~A., {Kocevski}, D.~D., {Faber}, S.~M., {et al.} 2011, \apjs, 197, 35 \href{http://adsabs.harvard.edu/abs/2011ApJS..197...35G}{\adsurllinklabel}

\bibitem[{{Guo} {et~al.}(2013){Guo}, {Ferguson}, {Giavalisco}, {Barro}, {Willner}, {Ashby}, {Dahlen}, {Donley}, {Faber}, {Fontana}, {Galametz}, {Grazian}, {Huang}, {Kocevski}, {Koekemoer}, {Koo}, {McGrath}, {Peth}, {Salvato}, {Wuyts}, {Castellano}, {Cooray}, {Dickinson}, {Dunlop}, {Fazio}, {Gardner}, {Gawiser}, {Grogin}, {Hathi}, {Hsu}, {Lee}, {Lucas}, {Mobasher}, {Nandra}, {Newman}, \& {van der Wel}}]{Guo2013}
{Guo}, Y., {Ferguson}, H.~C., {Giavalisco}, M., {et al.} 2013, \apjs, 207, 24 \href{http://adsabs.harvard.edu/abs/2013ApJS..207...24G}{\adsurllinklabel}

\bibitem[{{Hall} {et~al.}(2012){Hall}, {Brada{\v c}}, {Gonzalez}, {Treu}, {Clowe}, {Jones}, {Stiavelli}, {Zaritsky}, {Cuby}, \& {Cl{\'e}ment}}]{Hall2012}
{Hall}, N., {Brada{\v c}}, M., {Gonzalez}, A.~H., {et al.} 2012, \apj, 745, 155 \href{http://adsabs.harvard.edu/abs/2012ApJ...745..155H}{\adsurllinklabel}

\bibitem[{{Ilbert} {et~al.}(2006){Ilbert}, {Arnouts}, {McCracken}, {Bolzonella}, {Bertin}, {Le F{\`e}vre}, {Mellier}, {Zamorani}, {Pell{\`o}}, {Iovino}, {Tresse}, {Le Brun}, {Bottini}, {Garilli}, {Maccagni}, {Picat}, {Scaramella}, {Scodeggio}, {Vettolani}, {Zanichelli}, {Adami}, {Bardelli}, {Cappi}, {Charlot}, {Ciliegi}, {Contini}, {Cucciati}, {Foucaud}, {Franzetti}, {Gavignaud}, {Guzzo}, {Marano}, {Marinoni}, {Mazure}, {Meneux}, {Merighi}, {Paltani}, {Pollo}, {Pozzetti}, {Radovich}, {Zucca}, {Bondi}, {Bongiorno}, {Busarello}, {de La Torre}, {Gregorini}, {Lamareille}, {Mathez}, {Merluzzi}, {Ripepi}, {Rizzo}, \& {Vergani}}]{Ilbert2006}
{Ilbert}, O., {Arnouts}, S., {McCracken}, H.~J., {et al.} 2006, \aap, 457, 841 \href{http://adsabs.harvard.edu/abs/2006A%26A...457..841I}{\adsurllinklabel}

\bibitem[{{Ilbert} {et~al.}(2009){Ilbert}, {Capak}, {Salvato}, {Aussel}, {McCracken}, {Sanders}, {Scoville}, {Kartaltepe}, {Arnouts}, {Le Floc'h}, {Mobasher}, {Taniguchi}, {Lamareille}, {Leauthaud}, {Sasaki}, {Thompson}, {Zamojski}, {Zamorani}, {Bardelli}, {Bolzonella}, {Bongiorno}, {Brusa}, {Caputi}, {Carollo}, {Contini}, {Cook}, {Coppa}, {Cucciati}, {de la Torre}, {de Ravel}, {Franzetti}, {Garilli}, {Hasinger}, {Iovino}, {Kampczyk}, {Kneib}, {Knobel}, {Kovac}, {Le Borgne}, {Le Brun}, {F{\`e}vre}, {Lilly}, {Looper}, {Maier}, {Mainieri}, {Mellier}, {Mignoli}, {Murayama}, {Pell{\`o}}, {Peng}, {P{\'e}rez-Montero}, {Renzini}, {Ricciardelli}, {Schiminovich}, {Scodeggio}, {Shioya}, {Silverman}, {Surace}, {Tanaka}, {Tasca}, {Tresse}, {Vergani}, \& {Zucca}}]{Ilbert2009}
{Ilbert}, O., {Capak}, P., {Salvato}, M., {et al.} 2009, \apj, 690, 1236 \href{http://adsabs.harvard.edu/abs/2009ApJ...690.1236I}{\adsurllinklabel}

\bibitem[{{Jouvel} {et~al.}(2014){Jouvel}, {Host}, {Lahav}, {Seitz}, {Molino}, {Coe}, {Postman}, {Moustakas}, {Ben{\`i}tez}, {Rosati}, {Balestra}, {Grillo}, {Bradley}, {Fritz}, {Kelson}, {Koekemoer}, {Lemze}, {Medezinski}, {Mercurio}, {Moustakas}, {Nonino}, {Scodeggio}, {Zheng}, {Zitrin}, {Bartelmann}, {Bouwens}, {Broadhurst}, {Donahue}, {Ford}, {Graves}, {Infante}, {Jimenez-Teja}, {Lazkoz}, {Melchior}, {Meneghetti}, {Merten}, {Ogaz}, \& {Umetsu}}]{Jouvel2014}
{Jouvel}, S., {Host}, O., {Lahav}, O., {et al.} 2014, \aap, 562, A86 \href{http://adsabs.harvard.edu/abs/2014A%26A...562A..86J}{\adsurllinklabel}

\bibitem[{{Kneib} {et~al.}(2004){Kneib}, {Ellis}, {Santos}, \& {Richard}}]{Kneib2004}
{Kneib}, J.-P., {Ellis}, R.~S., {Santos}, M.~R., \& {Richard}, J. 2004, \apj, 607, 697 \href{http://adsabs.harvard.edu/abs/2004ApJ...607..697K}{\adsurllinklabel}

\bibitem[{{Koekemoer} {et~al.}(2011){Koekemoer}, {Faber}, {Ferguson}, {Grogin}, {Kocevski}, {Koo}, {Lai}, {Lotz}, {Lucas}, {McGrath}, {Ogaz}, {Rajan}, {Riess}, {Rodney}, {Strolger}, {Casertano}, {Castellano}, {Dahlen}, {Dickinson}, {Dolch}, {Fontana}, {Giavalisco}, {Grazian}, {Guo}, {Hathi}, {Huang}, {van der Wel}, {Yan}, {Acquaviva}, {Alexander}, {Almaini}, {Ashby}, {Barden}, {Bell}, {Bournaud}, {Brown}, {Caputi}, {Cassata}, {Challis}, {Chary}, {Cheung}, {Cirasuolo}, {Conselice}, {Roshan Cooray}, {Croton}, {Daddi}, {Dav{\'e}}, {de Mello}, {de Ravel}, {Dekel}, {Donley}, {Dunlop}, {Dutton}, {Elbaz}, {Fazio}, {Filippenko}, {Finkelstein}, {Frazer}, {Gardner}, {Garnavich}, {Gawiser}, {Gruetzbauch}, {Hartley}, {H{\"a}ussler}, {Herrington}, {Hopkins}, {Huang}, {Jha}, {Johnson}, {Kartaltepe}, {Khostovan}, {Kirshner}, {Lani}, {Lee}, {Li}, {Madau}, {McCarthy}, {McIntosh}, {McLure}, {McPartland}, {Mobasher}, {Moreira}, {Mortlock}, {Moustakas}, {Mozena}, {Nandra}, {Newman}, {Nielsen}, {Niemi}, {Noeske}, {Papovich}, {Pentericci}, {Pope}, {Primack}, {Ravindranath}, {Reddy}, {Renzini}, {Rix}, {Robaina}, {Rosario}, {Rosati}, {Salimbeni}, {Scarlata}, {Siana}, {Simard}, {Smidt}, {Snyder}, {Somerville}, {Spinrad}, {Straughn}, {Telford}, {Teplitz}, {Trump}, {Vargas}, {Villforth}, {Wagner}, {Wandro}, {Wechsler}, {Weiner}, {Wiklind}, {Wild}, {Wilson}, {Wuyts}, \& {Yun}}]{Koekemoer2011}
{Koekemoer}, A.~M., {Faber}, S.~M., {Ferguson}, H.~C., {et al.} 2011, \apjs, 197, 36 \href{http://adsabs.harvard.edu/abs/2011ApJS..197...36K}{\adsurllinklabel}

\bibitem[{{Koekemoer} {et~al.}(2003){Koekemoer}, {Fruchter}, {Hook}, \& {Hack}}]{Koekemoer2003}
{Koekemoer}, A.~M., {Fruchter}, A.~S., {Hook}, R.~N., \& {Hack}, W. 2003, in The 2002 HST Calibration Workshop : Hubble after the Installation of the ACS and the NICMOS Cooling System, ed. {S.~Arribas, A.~Koekemoer, \& B.~Whitmore}, 337

\bibitem[{{Kron}(1980){Kron}}]{Kron1980}
{Kron}, R.~G. 1980, \apjs, 43, 305 \href{http://adsabs.harvard.edu/abs/1980ApJS...43..305K}{\adsurllinklabel}

\bibitem[{{Labb{\'e}} {et~al.}(2010){Labb{\'e}}, {Gonz{\'a}lez}, {Bouwens}, {Illingworth}, {Franx}, {Trenti}, {Oesch}, {van Dokkum}, {Stiavelli}, {Carollo}, {Kriek}, \& {Magee}}]{Labbe2010b}
{Labb{\'e}}, I., {Gonz{\'a}lez}, V., {Bouwens}, R.~J., {et al.} 2010, \apjl, 716, L103 \href{http://adsabs.harvard.edu/abs/2010ApJ...716L.103L}{\adsurllinklabel}

\bibitem[{{Labb{\'e}} {et~al.}(2013){Labb{\'e}}, {Oesch}, {Bouwens}, {Illingworth}, {Magee}, {Gonz{\'a}lez}, {Carollo}, {Franx}, {Trenti}, {van Dokkum}, \& {Stiavelli}}]{Labbe2013}
{Labb{\'e}}, I., {Oesch}, P.~A., {Bouwens}, R.~J., {et al.} 2013, \apjl, 777, L19 \href{http://adsabs.harvard.edu/abs/2013ApJ...777L..19L}{\adsurllinklabel}

\bibitem[{{Lorenzoni} {et~al.}(2011){Lorenzoni}, {Bunker}, {Wilkins}, {Stanway}, {Jarvis}, \& {Caruana}}]{Lorenzoni2011}
{Lorenzoni}, S., {Bunker}, A.~J., {Wilkins}, S.~M., {Stanway}, E.~R., {Jarvis}, M.~J., \& {Caruana}, J. 2011, \mnras, 414, 1455 \href{http://adsabs.harvard.edu/abs/2011MNRAS.414.1455L}{\adsurllinklabel}

\bibitem[{{Madau}(1995){Madau}}]{Madau1995}
{Madau}, P. 1995, \apj, 441, 18 \href{http://adsabs.harvard.edu/abs/1995ApJ...441...18M}{\adsurllinklabel}

\bibitem[{{Maizy} {et~al.}(2010){Maizy}, {Richard}, {de Leo}, {Pell{\'o}}, \& {Kneib}}]{Maizy2010}
{Maizy}, A., {Richard}, J., {de Leo}, M.~A., {Pell{\'o}}, R., \& {Kneib}, J.~P. 2010, \aap, 509, A105 \href{http://adsabs.harvard.edu/abs/2010A%26A...509A.105M}{\adsurllinklabel}

\bibitem[{{Mann} \& {Ebeling}(2012){Mann} \& {Ebeling}}]{Mann2012}
{Mann}, A.~W. \& {Ebeling}, H. 2012, \mnras, 420, 2120 \href{http://adsabs.harvard.edu/abs/2012MNRAS.420.2120M}{\adsurllinklabel}

\bibitem[{{McLure} {et~al.}(2013){McLure}, {Dunlop}, {Bowler}, {Curtis-Lake}, {Schenker}, {Ellis}, {Robertson}, {Koekemoer}, {Rogers}, {Ono}, {Ouchi}, {Charlot}, {Wild}, {Stark}, {Furlanetto}, {Cirasuolo}, \& {Targett}}]{McLure2013}
{McLure}, R.~J., {Dunlop}, J.~S., {Bowler}, R.~A.~A., {et al.} 2013, \mnras, 432, 2696 \href{http://adsabs.harvard.edu/abs/2013MNRAS.432.2696M}{\adsurllinklabel}

\bibitem[{{McLure} {et~al.}(2010){McLure}, {Dunlop}, {Cirasuolo}, {Koekemoer}, {Sabbi}, {Stark}, {Targett}, \& {Ellis}}]{McLure2010}
{McLure}, R.~J., {Dunlop}, J.~S., {Cirasuolo}, M., {Koekemoer}, A.~M., {Sabbi}, E., {Stark}, D.~P., {Targett}, T.~A., \& {Ellis}, R.~S. 2010, \mnras, 403, 960 \href{http://adsabs.harvard.edu/abs/2010MNRAS.403..960M}{\adsurllinklabel}

\bibitem[{{McLure} {et~al.}(2011){McLure}, {Dunlop}, {de Ravel}, {Cirasuolo}, {Ellis}, {Schenker}, {Robertson}, {Koekemoer}, {Stark}, \& {Bowler}}]{McLure2011}
{McLure}, R.~J., {Dunlop}, J.~S., {de Ravel}, L., {et al.} 2011, \mnras, 418, 2074 \href{http://adsabs.harvard.edu/abs/2011MNRAS.418.2074M}{\adsurllinklabel}

\bibitem[{{Medezinski} {et~al.}(2013){Medezinski}, {Umetsu}, {Nonino}, {Merten}, {Zitrin}, {Broadhurst}, {Donahue}, {Sayers}, {Waizmann}, {Koekemoer}, {Coe}, {Molino}, {Melchior}, {Mroczkowski}, {Czakon}, {Postman}, {Meneghetti}, {Lemze}, {Ford}, {Grillo}, {Kelson}, {Bradley}, {Moustakas}, {Bartelmann}, {Ben{\'{\i}}tez}, {Biviano}, {Bouwens}, {Golwala}, {Graves}, {Infante}, {Jim{\'e}nez-Teja}, {Jouvel}, {Lahav}, {Moustakas}, {Ogaz}, {Rosati}, {Seitz}, \& {Zheng}}]{Medezinski2013}
{Medezinski}, E., {Umetsu}, K., {Nonino}, M., {et al.} 2013, \apj, 777, 43 \href{http://adsabs.harvard.edu/abs/2013ApJ...777...43M}{\adsurllinklabel}

\bibitem[{{Meneghetti} {et~al.}(2008){Meneghetti}, {Melchior}, {Grazian}, {De Lucia}, {Dolag}, {Bartelmann}, {Heymans}, {Moscardini}, \& {Radovich}}]{Meneghetti2008}
{Meneghetti}, M., {Melchior}, P., {Grazian}, A., {et al.} 2008, \aap, 482, 403 \href{http://adsabs.harvard.edu/abs/2008A%26A...482..403M}{\adsurllinklabel}

\bibitem[{{Meneghetti} {et~al.}(2010){Meneghetti}, {Rasia}, {Merten}, {Bellagamba}, {Ettori}, {Mazzotta}, {Dolag}, \& {Marri}}]{Meneghetti2010}
{Meneghetti}, M., {Rasia}, E., {Merten}, J., {Bellagamba}, F., {Ettori}, S., {Mazzotta}, P., {Dolag}, K., \& {Marri}, S. 2010, \aap, 514, A93 \href{http://adsabs.harvard.edu/abs/2010A%26A...514A..93M}{\adsurllinklabel}

\bibitem[{{Monna} {et~al.}(2014){Monna}, {Seitz}, {Greisel}, {Eichner}, {Drory}, {Postman}, {Zitrin}, {Coe}, {Halkola}, {Suyu}, {Grillo}, {Rosati}, {Lemze}, {Balestra}, {Snigula}, {Bradley}, {Umetsu}, {Koekemoer}, {Kuchner}, {Moustakas}, {Bartelmann}, {Ben{\'{\i}}tez}, {Bouwens}, {Broadhurst}, {Donahue}, {Ford}, {Host}, {Infante}, {Jimenez-Teja}, {Jouvel}, {Kelson}, {Lahav}, {Medezinski}, {Melchior}, {Meneghetti}, {Merten}, {Molino}, {Moustakas}, {Nonino}, \& {Zheng}}]{Monna2014}
{Monna}, A., {Seitz}, S., {Greisel}, N., {et al.} 2014, \mnras, 438, 1417 \href{http://adsabs.harvard.edu/abs/2014MNRAS.438.1417M}{\adsurllinklabel}

\bibitem[{{Oesch} {et~al.}(2010{\natexlab{a}}){Oesch}, {Bouwens}, {Carollo}, {Illingworth}, {Trenti}, {Stiavelli}, {Magee}, {Labb{\'e}}, \& {Franx}}]{Oesch2010b}
{Oesch}, P.~A., {Bouwens}, R.~J., {Carollo}, C.~M., {et al.} 2010{\natexlab{a}}, \apjl, 709, L21 \href{http://adsabs.harvard.edu/abs/2010ApJ...709L..21O}{\adsurllinklabel}

\bibitem[{{Oesch} {et~al.}(2010{\natexlab{b}}){Oesch}, {Bouwens}, {Illingworth}, {Carollo}, {Franx}, {Labb{\'e}}, {Magee}, {Stiavelli}, {Trenti}, \& {van Dokkum}}]{Oesch2010a}
{Oesch}, P.~A., {Bouwens}, R.~J., {Illingworth}, G.~D., {et al.} 2010{\natexlab{b}}, \apjl, 709, L16 \href{http://adsabs.harvard.edu/abs/2010ApJ...709L..16O}{\adsurllinklabel}

\bibitem[{{Oesch} {et~al.}(2012){Oesch}, {Bouwens}, {Illingworth}, {Gonzalez}, {Trenti}, {van Dokkum}, {Franx}, {Labb{\'e}}, {Carollo}, \& {Magee}}]{Oesch2012}
{Oesch}, P.~A., {Bouwens}, R.~J., {Illingworth}, G.~D., {et al.} 2012, \apj, 759, 135 \href{http://adsabs.harvard.edu/abs/2012ApJ...759..135O}{\adsurllinklabel}

\bibitem[{{Oke}(1974){Oke}}]{Oke1974}
{Oke}, J.~B. 1974, \apjs, 27, 21 \href{http://adsabs.harvard.edu/abs/1974ApJS...27...21O}{\adsurllinklabel}

\bibitem[{{Ono} {et~al.}(2013){Ono}, {Ouchi}, {Curtis-Lake}, {Schenker}, {Ellis}, {McLure}, {Dunlop}, {Robertson}, {Koekemoer}, {Bowler}, {Rogers}, {Schneider}, {Charlot}, {Stark}, {Shimasaku}, {Furlanetto}, \& {Cirasuolo}}]{Ono2013}
{Ono}, Y., {Ouchi}, M., {Curtis-Lake}, E., {et al.} 2013, \apj, 777, 155 \href{http://adsabs.harvard.edu/abs/2013ApJ...777..155O}{\adsurllinklabel}

\bibitem[{{Pickles}(1998){Pickles}}]{Pickles1998}
{Pickles}, A.~J. 1998, \pasp, 110, 863 \href{http://adsabs.harvard.edu/abs/1998PASP..110..863P}{\adsurllinklabel}

\bibitem[{{Postman} {et~al.}(2012){Postman}, {Coe}, {Ben{\'{\i}}tez}, {Bradley}, {Broadhurst}, {Donahue}, {Ford}, {Graur}, {Graves}, {Jouvel}, {Koekemoer}, {Lemze}, {Medezinski}, {Molino}, {Moustakas}, {Ogaz}, {Riess}, {Rodney}, {Rosati}, {Umetsu}, {Zheng}, {Zitrin}, {Bartelmann}, {Bouwens}, {Czakon}, {Golwala}, {Host}, {Infante}, {Jha}, {Jimenez-Teja}, {Kelson}, {Lahav}, {Lazkoz}, {Maoz}, {McCully}, {Melchior}, {Meneghetti}, {Merten}, {Moustakas}, {Nonino}, {Patel}, {Reg{\"o}s}, {Sayers}, {Seitz}, \& {Van der Wel}}]{Postman2012}
{Postman}, M., {Coe}, D., {Ben{\'{\i}}tez}, N., {et al.} 2012, \apjs, 199, 25 \href{http://adsabs.harvard.edu/abs/2012ApJS..199...25P}{\adsurllinklabel}

\bibitem[{{Prevot} {et~al.}(1984){Prevot}, {Lequeux}, {Prevot}, {Maurice}, \& {Rocca-Volmerange}}]{Prevot1984}
{Prevot}, M.~L., {Lequeux}, J., {Prevot}, L., {Maurice}, E., \& {Rocca-Volmerange}, B. 1984, \aap, 132, 389 \href{http://adsabs.harvard.edu/abs/1984A%26A...132..389P}{\adsurllinklabel}

\bibitem[{{Rayner} {et~al.}(2009){Rayner}, {Cushing}, \& {Vacca}}]{Rayner2009}
{Rayner}, J.~T., {Cushing}, M.~C., \& {Vacca}, W.~D. 2009, \apjs, 185, 289 \href{http://adsabs.harvard.edu/abs/2009ApJS..185..289R}{\adsurllinklabel}

\bibitem[{{Richard} {et~al.}(2011){Richard}, {Kneib}, {Ebeling}, {Stark}, {Egami}, \& {Fiedler}}]{Richard2011}
{Richard}, J., {Kneib}, J.-P., {Ebeling}, H., {Stark}, D.~P., {Egami}, E., \& {Fiedler}, A.~K. 2011, \mnras, 414, L31 \href{http://adsabs.harvard.edu/abs/2011MNRAS.414L..31R}{\adsurllinklabel}

\bibitem[{{Richard} {et~al.}(2008){Richard}, {Stark}, {Ellis}, {George}, {Egami}, {Kneib}, \& {Smith}}]{Richard2008}
{Richard}, J., {Stark}, D.~P., {Ellis}, R.~S., {George}, M.~R., {Egami}, E., {Kneib}, J.-P., \& {Smith}, G.~P. 2008, \apj, 685, 705 \href{http://adsabs.harvard.edu/abs/2008ApJ...685..705R}{\adsurllinklabel}

\bibitem[{{Saro} {et~al.}(2006){Saro}, {Borgani}, {Tornatore}, {Dolag}, {Murante}, {Biviano}, {Calura}, \& {Charlot}}]{Saro2006}
{Saro}, A., {Borgani}, S., {Tornatore}, L., {Dolag}, K., {Murante}, G., {Biviano}, A., {Calura}, F., \& {Charlot}, S. 2006, \mnras, 373, 397 \href{http://adsabs.harvard.edu/abs/2006MNRAS.373..397S}{\adsurllinklabel}

\bibitem[{{Schenker} {et~al.}(2013){Schenker}, {Robertson}, {Ellis}, {Ono}, {McLure}, {Dunlop}, {Koekemoer}, {Bowler}, {Ouchi}, {Curtis-Lake}, {Rogers}, {Schneider}, {Charlot}, {Stark}, {Furlanetto}, \& {Cirasuolo}}]{Schenker2013}
{Schenker}, M.~A., {Robertson}, B.~E., {Ellis}, R.~S., {et al.} 2013, \apj, 768, 196 \href{http://adsabs.harvard.edu/abs/2013ApJ...768..196S}{\adsurllinklabel}

\bibitem[{{Schenker} {et~al.}(2012){Schenker}, {Stark}, {Ellis}, {Robertson}, {Dunlop}, {McLure}, {Kneib}, \& {Richard}}]{Schenker2012}
{Schenker}, M.~A., {Stark}, D.~P., {Ellis}, R.~S., {Robertson}, B.~E., {Dunlop}, J.~S., {McLure}, R.~J., {Kneib}, J.-P., \& {Richard}, J. 2012, \apj, 744, 179 \href{http://adsabs.harvard.edu/abs/2012ApJ...744..179S}{\adsurllinklabel}

\bibitem[{{Schlegel} {et~al.}(1998){Schlegel}, {Finkbeiner}, \& {Davis}}]{Schlegel1998}
{Schlegel}, D.~J., {Finkbeiner}, D.~P., \& {Davis}, M. 1998, \apj, 500, 525 \href{http://adsabs.harvard.edu/abs/1998ApJ...500..525S}{\adsurllinklabel}

\bibitem[{{Sharon} {et~al.}(2012){Sharon}, {Gladders}, {Rigby}, {Wuyts}, {Koester}, {Bayliss}, \& {Barrientos}}]{Sharon2012}
{Sharon}, K., {Gladders}, M.~D., {Rigby}, J.~R., {Wuyts}, E., {Koester}, B.~P., {Bayliss}, M.~B., \& {Barrientos}, L.~F. 2012, \apj, 746, 161 \href{http://adsabs.harvard.edu/abs/2012ApJ...746..161S}{\adsurllinklabel}

\bibitem[{{Smit} {et~al.}(2014){Smit}, {Bouwens}, {Labb{\'e}}, {Zheng}, {Bradley}, {Donahue}, {Lemze}, {Moustakas}, {Umetsu}, {Zitrin}, {Coe}, {Postman}, {Gonzalez}, {Bartelmann}, {Ben{\'{\i}}tez}, {Broadhurst}, {Ford}, {Grillo}, {Infante}, {Jimenez-Teja}, {Jouvel}, {Kelson}, {Lahav}, {Maoz}, {Medezinski}, {Melchior}, {Meneghetti}, {Merten}, {Molino}, {Moustakas}, {Nonino}, {Rosati}, \& {Seitz}}]{Smit2014}
{Smit}, R., {Bouwens}, R.~J., {Labb{\'e}}, I., {et al.} 2014, \apj, 784, 58 \href{http://adsabs.harvard.edu/abs/2014ApJ...784...58S}{\adsurllinklabel}

\bibitem[{{Steidel} {et~al.}(1996){Steidel}, {Giavalisco}, {Dickinson}, \& {Adelberger}}]{Steidel1996a}
{Steidel}, C.~C., {Giavalisco}, M., {Dickinson}, M., \& {Adelberger}, K.~L. 1996, \aj, 112, 352 \href{http://adsabs.harvard.edu/abs/1996AJ....112..352S}{\adsurllinklabel}

\bibitem[{{Swinbank} {et~al.}(2009){Swinbank}, {Webb}, {Richard}, {Bower}, {Ellis}, {Illingworth}, {Jones}, {Kriek}, {Smail}, {Stark}, \& {van Dokkum}}]{Swinbank2009}
{Swinbank}, A.~M., {Webb}, T.~M., {Richard}, J., {et al.} 2009, \mnras, 400, 1121 \href{http://adsabs.harvard.edu/abs/2009MNRAS.400.1121S}{\adsurllinklabel}

\bibitem[{{Trenti} {et~al.}(2011){Trenti}, {Bradley}, {Stiavelli}, {Oesch}, {Treu}, {Bouwens}, {Shull}, {MacKenty}, {Carollo}, \& {Illingworth}}]{Trenti2011}
{Trenti}, M., {Bradley}, L.~D., {Stiavelli}, M., {et al.} 2011, \apjl, 727, L39 \href{http://adsabs.harvard.edu/abs/2011ApJ...727L..39T}{\adsurllinklabel}

\bibitem[{{Trenti} {et~al.}(2012{\natexlab{a}}){Trenti}, {Bradley}, {Stiavelli}, {Shull}, {Oesch}, {Bouwens}, {Mu{\~n}oz}, {Romano-Diaz}, {Treu}, {Shlosman}, \& {Carollo}}]{Trenti2012a}
{Trenti}, M., {Bradley}, L.~D., {Stiavelli}, M., {et al.} 2012{\natexlab{a}}, \apj, 746, 55 \href{http://adsabs.harvard.edu/abs/2012ApJ...746...55T}{\adsurllinklabel}

\bibitem[{{Trenti} {et~al.}(2012{\natexlab{b}}){Trenti}, {Perna}, {Levesque}, {Shull}, \& {Stocke}}]{Trenti2012b}
{Trenti}, M., {Perna}, R., {Levesque}, E.~M., {Shull}, J.~M., \& {Stocke}, J.~T. 2012{\natexlab{b}}, \apjl, 749, L38 \href{http://adsabs.harvard.edu/abs/2012ApJ...749L..38T}{\adsurllinklabel}

\bibitem[{{van der Wel} {et~al.}(2011){van der Wel}, {Straughn}, {Rix}, {Finkelstein}, {Koekemoer}, {Weiner}, {Wuyts}, {Bell}, {Faber}, {Trump}, {Koo}, {Ferguson}, {Scarlata}, {Hathi}, {Dunlop}, {Newman}, {Dickinson}, {Jahnke}, {Salmon}, {de Mello}, {Kocevski}, {Lai}, {Grogin}, {Rodney}, {Guo}, {McGrath}, {Lee}, {Barro}, {Huang}, {Riess}, {Ashby}, \& {Willner}}]{vanderWel2011}
{van der Wel}, A., {Straughn}, A.~N., {Rix}, H.-W., {et al.} 2011, \apj, 742, 111 \href{http://adsabs.harvard.edu/abs/2011ApJ...742..111V}{\adsurllinklabel}

\bibitem[{{Vanzella} {et~al.}(2014){Vanzella}, {Fontana}, {Zitrin}, {Coe}, {Bradley}, {Postman}, {Grazian}, {Castellano}, {Pentericci}, {Giavalisco}, {Rosati}, {Nonino}, {Smit}, {Balestra}, {Bouwens}, {Cristiani}, {Giallongo}, {Zheng}, {Infante}, {Cusano}, \& {Speziali}}]{Vanzella2014}
{Vanzella}, E., {Fontana}, A., {Zitrin}, A., {et al.} 2014, \apjl, 783, L12 \href{http://adsabs.harvard.edu/abs/2014ApJ...783L..12V}{\adsurllinklabel}

\bibitem[{{Wilkins} {et~al.}(2011){Wilkins}, {Bunker}, {Stanway}, {Lorenzoni}, \& {Caruana}}]{Wilkins2011}
{Wilkins}, S.~M., {Bunker}, A.~J., {Stanway}, E., {Lorenzoni}, S., \& {Caruana}, J. 2011, \mnras, 417, 717 \href{http://adsabs.harvard.edu/abs/2011MNRAS.417..717W}{\adsurllinklabel}

\bibitem[{{Windhorst} {et~al.}(2011){Windhorst}, {Cohen}, {Hathi}, {McCarthy}, {Ryan}, {Yan}, {Baldry}, {Driver}, {Frogel}, {Hill}, {Kelvin}, {Koekemoer}, {Mechtley}, {O'Connell}, {Robotham}, {Rutkowski}, {Seibert}, {Straughn}, {Tuffs}, {Balick}, {Bond}, {Bushouse}, {Calzetti}, {Crockett}, {Disney}, {Dopita}, {Hall}, {Holtzman}, {Kaviraj}, {Kimble}, {MacKenty}, {Mutchler}, {Paresce}, {Saha}, {Silk}, {Trauger}, {Walker}, {Whitmore}, \& {Young}}]{Windhorst2011}
{Windhorst}, R.~A., {Cohen}, S.~H., {Hathi}, N.~P., {et al.} 2011, \apjs, 193, 27 \href{http://adsabs.harvard.edu/abs/2011ApJS..193...27W}{\adsurllinklabel}

\bibitem[{{Wuyts} {et~al.}(2008){Wuyts}, {Labb{\'e}}, {Schreiber}, {Franx}, {Rudnick}, {Brammer}, \& {van Dokkum}}]{Wuyts2008}
{Wuyts}, S., {Labb{\'e}}, I., {Schreiber}, N.~M.~F., {Franx}, M., {Rudnick}, G., {Brammer}, G.~B., \& {van Dokkum}, P.~G. 2008, \apj, 682, 985 \href{http://adsabs.harvard.edu/abs/2008ApJ...682..985W}{\adsurllinklabel}

\bibitem[{{Yan} {et~al.}(2012){Yan}, {Finkelstein}, {Huang}, {Ryan}, {Ferguson}, {Koekemoer}, {Grogin}, {Dickinson}, {Newman}, {Somerville}, {Dav{\'e}}, {Faber}, {Papovich}, {Guo}, {Giavalisco}, {Lee}, {Reddy}, {Cooray}, {Siana}, {Hathi}, {Fazio}, {Ashby}, {Weiner}, {Lucas}, {Dekel}, {Pentericci}, {Conselice}, {Kocevski}, \& {Lai}}]{Yan2012}
{Yan}, H., {Finkelstein}, S.~L., {Huang}, K.-H., {et al.} 2012, \apj, 761, 177 \href{http://adsabs.harvard.edu/abs/2012ApJ...761..177Y}{\adsurllinklabel}

\bibitem[{{Yan} {et~al.}(2011){Yan}, {Yan}, {Zamojski}, {Windhorst}, {McCarthy}, {Fan}, {R{\"o}ttgering}, {Koekemoer}, {Robertson}, {Dav{\'e}}, \& {Cai}}]{Yan2011a}
{Yan}, H., {Yan}, L., {Zamojski}, M.~A., {et al.} 2011, \apjl, 728, L22 \href{http://adsabs.harvard.edu/abs/2011ApJ...728L..22Y}{\adsurllinklabel}

\bibitem[{{Zheng} {et~al.}(2009){Zheng}, {Bradley}, {Bouwens}, {Ford}, {Illingworth}, {Ben{\'{\i}}tez}, {Broadhurst}, {Frye}, {Infante}, {Jee}, {Motta}, {Shu}, \& {Zitrin}}]{Zheng2009}
{Zheng}, W., {Bradley}, L.~D., {Bouwens}, R.~J., {et al.} 2009, \apj, 697, 1907 \href{http://adsabs.harvard.edu/abs/2009ApJ...697.1907Z}{\adsurllinklabel}

\bibitem[{{Zheng} {et~al.}(2012){Zheng}, {Postman}, {Zitrin}, {Moustakas}, {Shu}, {Jouvel}, {H{\o}st}, {Molino}, {Bradley}, {Coe}, {Moustakas}, {Carrasco}, {Ford}, {Ben{\'{\i}}tez}, {Lauer}, {Seitz}, {Bouwens}, {Koekemoer}, {Medezinski}, {Bartelmann}, {Broadhurst}, {Donahue}, {Grillo}, {Infante}, {Jha}, {Kelson}, {Lahav}, {Lemze}, {Melchior}, {Meneghetti}, {Merten}, {Nonino}, {Ogaz}, {Rosati}, {Umetsu}, \& {van der Wel}}]{Zheng2012}
{Zheng}, W., {Postman}, M., {Zitrin}, A., {et al.} 2012, \nat, 489, 406 \href{http://adsabs.harvard.edu/abs/2012Natur.489..406Z}{\adsurllinklabel}

\bibitem[{{Zitrin} {et~al.}(2012{\natexlab{a}}){Zitrin}, {Broadhurst}, {Bartelmann}, {Rephaeli}, {Oguri}, {Ben{\'{\i}}tez}, {Hao}, \& {Umetsu}}]{Zitrin2012sdss}
{Zitrin}, A., {Broadhurst}, T., {Bartelmann}, M., {Rephaeli}, Y., {Oguri}, M., {Ben{\'{\i}}tez}, N., {Hao}, J., \& {Umetsu}, K. 2012{\natexlab{a}}, \mnras, 423, 2308 \href{http://adsabs.harvard.edu/abs/2012MNRAS.423.2308Z}{\adsurllinklabel}

\bibitem[{{Zitrin} {et~al.}(2011){Zitrin}, {Broadhurst}, {Coe}, {Umetsu}, {Postman}, {Ben{\'{\i}}tez}, {Meneghetti}, {Medezinski}, {Jouvel}, {Bradley}, {Koekemoer}, {Zheng}, {Ford}, {Merten}, {Kelson}, {Lahav}, {Lemze}, {Molino}, {Nonino}, {Donahue}, {Rosati}, {Van der Wel}, {Bartelmann}, {Bouwens}, {Graur}, {Graves}, {Host}, {Infante}, {Jha}, {Jimenez-Teja}, {Lazkoz}, {Maoz}, {McCully}, {Melchior}, {Moustakas}, {Ogaz}, {Patel}, {Regoes}, {Riess}, {Rodney}, \& {Seitz}}]{Zitrin2011a383}
{Zitrin}, A., {Broadhurst}, T., {Coe}, D., {et al.} 2011, \apj, 742, 117 \href{http://adsabs.harvard.edu/abs/2011ApJ...742..117Z}{\adsurllinklabel}

\bibitem[{{Zitrin} {et~al.}(2009){Zitrin}, {Broadhurst}, {Umetsu}, {Coe}, {Ben{\'{\i}}tez}, {Ascaso}, {Bradley}, {Ford}, {Jee}, {Medezinski}, {Rephaeli}, \& {Zheng}}]{Zitrin2009}
{Zitrin}, A., {Broadhurst}, T., {Umetsu}, K., {et al.} 2009, \mnras, 396, 1985 \href{http://adsabs.harvard.edu/abs/2009MNRAS.396.1985Z}{\adsurllinklabel}

\bibitem[{{Zitrin} {et~al.}(2013{\natexlab{a}}){Zitrin}, {Menanteau}, {Hughes}, {Coe}, {Barrientos}, {Infante}, \& {Mandelbaum}}]{Zitrin2013elgordo}
{Zitrin}, A., {Menanteau}, F., {Hughes}, J.~P., {Coe}, D., {Barrientos}, L.~F., {Infante}, L., \& {Mandelbaum}, R. 2013{\natexlab{a}}, \apjl, 770, L15 \href{http://adsabs.harvard.edu/abs/2013ApJ...770L..15Z}{\adsurllinklabel}

\bibitem[{{Zitrin} {et~al.}(2013{\natexlab{b}}){Zitrin}, {Meneghetti}, {Umetsu}, {Broadhurst}, {Bartelmann}, {Bouwens}, {Bradley}, {Carrasco}, {Coe}, {Ford}, {Kelson}, {Koekemoer}, {Medezinski}, {Moustakas}, {Moustakas}, {Nonino}, {Postman}, {Rosati}, {Seidel}, {Seitz}, {Sendra}, {Shu}, {Vega}, \& {Zheng}}]{Zitrin2013m0416}
{Zitrin}, A., {Meneghetti}, M., {Umetsu}, K., {et al.} 2013{\natexlab{b}}, \apjl, 762, L30 \href{http://adsabs.harvard.edu/abs/2013ApJ...762L..30Z}{\adsurllinklabel}

\bibitem[{{Zitrin} {et~al.}(2012{\natexlab{b}}){Zitrin}, {Moustakas}, {Bradley}, {Coe}, {Moustakas}, {Postman}, {Shu}, {Zheng}, {Ben{\'{\i}}tez}, {Bouwens}, {Broadhurst}, {Ford}, {Host}, {Jouvel}, {Koekemoer}, {Meneghetti}, {Rosati}, {Donahue}, {Grillo}, {Kelson}, {Lemze}, {Medezinski}, {Molino}, {Nonino}, \& {Ogaz}}]{Zitrin2012m0329}
{Zitrin}, A., {Moustakas}, J., {Bradley}, L., {et al.} 2012{\natexlab{b}}, \apjl, 747, L9 \href{http://adsabs.harvard.edu/abs/2012ApJ...747L...9Z}{\adsurllinklabel}

\bibitem[{{Zitrin} {et~al.}(2012{\natexlab{c}}){Zitrin}, {Rosati}, {Nonino}, {Grillo}, {Postman}, {Coe}, {Seitz}, {Eichner}, {Broadhurst}, {Jouvel}, {Balestra}, {Mercurio}, {Scodeggio}, {Ben{\'{\i}}tez}, {Bradley}, {Ford}, {Host}, {Jimenez-Teja}, {Koekemoer}, {Zheng}, {Bartelmann}, {Bouwens}, {Czoske}, {Donahue}, {Graur}, {Graves}, {Infante}, {Jha}, {Kelson}, {Lahav}, {Lazkoz}, {Lemze}, {Lombardi}, {Maoz}, {McCully}, {Medezinski}, {Melchior}, {Meneghetti}, {Merten}, {Molino}, {Moustakas}, {Ogaz}, {Patel}, {Regoes}, {Riess}, {Rodney}, {Umetsu}, \& {Van der Wel}}]{Zitrin2012m1206}
{Zitrin}, A., {Rosati}, P., {Nonino}, M., {et al.} 2012{\natexlab{c}}, \apj, 749, 97 \href{http://adsabs.harvard.edu/abs/2012ApJ...749...97Z}{\adsurllinklabel}

\bibitem[{{Zitrin} {et~al.}(2012{\natexlab{d}}){Zitrin}, {Rosati}, {Nonino}, {Grillo}, {Postman}, {Coe}, {Seitz}, {Eichner}, {Broadhurst}, {Jouvel}, {Balestra}, {Mercurio}, {Scodeggio}, {Ben{\'{\i}}tez}, {Bradley}, {Ford}, {Host}, {Jimenez-Teja}, {Koekemoer}, {Zheng}, {Bartelmann}, {Bouwens}, {Czoske}, {Donahue}, {Graur}, {Graves}, {Infante}, {Jha}, {Kelson}, {Lahav}, {Lazkoz}, {Lemze}, {Lombardi}, {Maoz}, {McCully}, {Medezinski}, {Melchior}, {Meneghetti}, {Merten}, {Molino}, {Moustakas}, {Ogaz}, {Patel}, {Regoes}, {Riess}, {Rodney}, {Umetsu}, \& {Van der Wel}}]{Zitrin2012macs1206}
---. 2012{\natexlab{d}}, \apj, 749, 97 \href{http://adsabs.harvard.edu/abs/2012ApJ...749...97Z}{\adsurllinklabel}

\end{thebibliography}

%------------------------------------------------------------------------------
% Long Tables
%------------------------------------------------------------------------------
\appendix

\ifemulateapj\LongTables
\clearpage
\tabzsix
\clearpage
\clearpage
\tabzseven
\clearpage
\tabzeight

\section{High-redshift Candidates Behind RXJ1532.9+3021}
\label{sec:rxj1532}

In Table~\ref{tbl:rxj1532}, we list the high-redshifts candidates
identified behind RXJ1532.9+3021.  For this cluster, we have not been
able to clearly identify any multiply imaged galaxies that are
required to constrain the strong lensing model.  However, we have
generated an approximate model of this cluster constrained based
solely on the light distribution \citep[e.g.,][]{Zitrin2012sdss},
without any information of multiple images as constraints.  Such
models, utilizing the useful parameterizations of this modeling
method, were found to be accurate at the level of $\sim20\%$ on the
mass and location of the critical curves, but magnification errors can
be higher \citep{Zitrin2012sdss}.  Using this model, we include our
magnification estimates for these candidates, but the important
caveats about this lens model need to be taken into consideration.

\ifemulateapj\LongTables
\clearpage
\tabrxjonefive

\end{document}